\documentclass[12pt,preprint]{aastex}
\newcommand{\msun}{M$_{\odot}$\,}

\newcommand{\noprint}[1]{}
\newcommand{\figsetstart}{{\bf Fig. Set} }
\newcommand{\figsetend}{}
\newcommand{\figsetgrpstart}{}
\newcommand{\figsetgrpend}{}
\newcommand{\figsetnum}[1]{{\bf #1.}}
\newcommand{\figsettitle}[1]{ {\bf #1} }
\newcommand{\figsetgrpnum}[1]{\noprint{#1}}
\newcommand{\figsetgrptitle}[1]{\noprint{#1}}
\newcommand{\figsetplot}[1]{\noprint{#1}}
\newcommand{\figsetgrpnote}[1]{\noprint{#1}}

\shorttitle{A Disk of Young Stars at the Galactic Center}
\shortauthors{Lu et al.}

\begin{document}

\title{A Disk of Young Stars at the Galactic Center as Determined by 
Individual Stellar Orbits}

\author{
J. R. Lu\altaffilmark{1},
A. M. Ghez\altaffilmark{1,2},
S. D. Hornstein\altaffilmark{1,3},
M. R. Morris\altaffilmark{1},
E. E. Becklin\altaffilmark{1}, 
K. Matthews\altaffilmark{4}
}
\email{jlu@astro.ucla.edu, ghez@astro.ucla.edu, seth.hornstein@colorado.edu,
morris@astro.ucla.edu, becklin@astro.ucla.edu, kym@caltech.edu}
\altaffiltext{1}{UCLA Department of Physics and Astronomy,
Los Angeles, CA 90095-1562}
\altaffiltext{2}{UCLA Institute of Geophysics and Planetary Physics,
Los Angeles, CA 90095-1565}
\altaffiltext{3}{Center for Astrophysics \& Space Astronomy, 
Department of Astrophysical and Planetary Sciences, 
University of Colorado, Boulder, CO 80309}
\altaffiltext{4}{Caltech Optical Observatories, California Institute of
Technology, MS 320-47, Pasadena, CA 91125}

\begin{abstract}
We present new proper motions from the 10 m Keck telescopes 
for a puzzling population of massive, young stars located within 3\farcs5
(0.14 pc) of the supermassive black hole at the Galactic Center.
Our proper motion measurements have uncertainties of only 
0.07 mas yr$^{-1}$ (3 km s$^{-1}$), which is 
$\gtrsim$ 7 times better than previous 
proper motion measurements for these stars, and
enables us to measure accelerations as low as 0.2 mas yr$^{-2}$ 
(7 km s$^{-1}$ yr$^{-1}$).
Using these measurements, line-of-sight velocities from the literature, 
and 3D velocities for additional young stars in the central parsec, we
constrain the true orbit of each individual star
and directly test the hypothesis that the massive stars 
reside in two stellar disks as has been previously proposed.
Analysis of the stellar orbits reveals only one of the previously 
proposed disks of young stars
using a method that is capable of detecting disks containing at least 7 stars. 
The detected disk contains 50\% of the young stars, is inclined
by $\sim115^\circ$ from the plane of the sky, and is oriented at
a position angle of $\sim100^\circ$ East of North. 
Additionally, the on-disk and off-disk populations have similar K-band 
luminosity functions and radial distributions that decrease at larger 
projected radii as $\propto r^{-2}$. 
The disk has an out-of-the-disk velocity dispersion
of 28 $\pm$ 6 km s$^{-1}$, which corresponds to a half-opening angle of 
$7^\circ \pm 2^\circ$, and several candidate disk members have
eccentricities greater than 0.2. 
Our findings suggest that the young stars may have formed {\it in situ}
but in a more complex geometry than a simple, thin circular disk.
\end{abstract}

\keywords{black hole physics -- Galaxy:center --
infrared:stars -- techniques:high angular resolution}

\section{Introduction}
\label{sec:intro}

The center of our Galaxy harbors not only a supermassive black hole 
\citep[Sgr A*, $M_\bullet \sim 4 \times 10^6$ \msun;][]{
eckart96,genzel96,ghez98pm,ghez00nat,ghez03spec,ghez05orbits,
schodel02,schodel03,eisenhauer06}, but also 
a population of massive (10-120 \msun), young ($\lesssim$10-100 Myr) stars
whose existence is a puzzle. 
The origin of such young stars has been difficult to explain since
the gas densities observed today are orders of magnitude too low 
for a gas clump to overcome the extreme tidal forces and
collapse to form stars 
\citep[e.g.][for reviews]{sanders92,morris93,ghez05orbits,alexander05review}. 
And yet, within the central parsec of our Galaxy, nearly 100 stars have
been classified as OB main-sequence stars, 
more luminous OB giants and supergiants, and post-main-sequence 
Wolf-Rayet stars \citep{allen90,krabbe91,blum95heI,krabbe95,tamblyn96,
najarro97,ghez03spec,paumard06}, with the more evolved massive stars having
ages as young as 6$\pm$2 Myr \citep{paumard06}.
Populations of young stars have also been observed in the nuclei of other
galaxies, such as M31 \citep{bender05}, suggesting that star formation
near a supermassive black hole may be a common, but not understood,
phenomenon in galaxy evolution. The close proximity 
of the black hole at the center of the Milky Way provides a unique laboratory
for studying this ''paradox of youth''
\citep[e.g.][]{ghez03spec,ghez05orbits,schodel03,eisenhauer06}.

Proposed resolutions to the paradox of youth
can be grouped into several broad categories, including
(1) rejuvenation of an older population such that older stars appear young, 
(2) dynamical migration from larger radii, and
(3) {\it in situ} formation. 
Rejuvenation scenarios include stripping \citep{davies98,davies05} 
or tidal heating of the atmospheres of old stars \citep{alexander03}, 
or combining multiple low mass stars via collisional mergers
to form a higher-mass hot star akin to a ``blue straggler''
\citep{lee96gcmergers,morris93,genzel03cusp}. 
Although these processes may be candidates for explaining the 
closest young stars within the central arcsecond,
they cannot account for the OB giants, OB supergiants, and 
Wolf-Rayet stars that are located at larger radii (1\arcsec-14\arcsec),
since the rate of collisions is too low to produce 
the observed total numbers.
Thus, it appears that these massive young stars must have formed,
or were deposited, in the central region within the last 4-8 Myr.
Dynamical migration scenarios attempt to resolve the paradox of youth 
with the formation of a massive star cluster at larger distances from 
the black hole (3-30 pc). 
Such a cluster would spiral in due to dynamical friction
and deposit stars at smaller radii where they
are observed today \citep{gerhard01}. However,
for a cluster to reach the central parsec in only a few million
years, it must be very massive and centrally concentrated 
\citep{kim03,pzwart03irs16,mcmillan03,gurkan05}, 
and it may even require the existence of an intermediate-mass 
black hole (IMBH) as an anchor in the cluster core \citep{hansen03,kim04}.
{\it In situ} star formation scenarios can resolve the paradox
of youth if a massive, self-gravitating gas disk
was once present around the black hole \citep{levin03}. 
Such a disk would be sufficiently dense to overcome the strong tidal
forces, and gravitational instabilities would then lead to 
fragmentation and the formation of stars,
as has been suggested in the context of both the Galactic Center
circumnuclear disk and AGN accretion disks in other galaxies
\citep[e.g.][]{kolykhalov80,shlosman89,morris96,sanders98,goodman03,nayakshinCuadra05}.

Insight into the origins of the massive, young stars may be obtained
through observations of the spatial distribution and stellar dynamics of 
this population. 
Already, high-resolution infrared imaging and spectroscopy have shown
that the young stars between 0\farcs5 and 14\arcsec (0.02-0.6 pc) 
exhibit coherent rotation \citep{genzel00}.
Analyses of the statistical properties of the 
three-dimensional velocity vectors for these stars
suggest that they may reside in two disks.
The first proposed disk has a clockwise sense of rotation, as
projected onto the plane of the sky
\citep[][hereafter: clockwise-rotating or CW disk]{levin03},
while the second proposed disk is
counter-clockwise-rotating \citep[CCW][]{genzel03cusp} and is nearly 
perpendicular to the first. The proposed disks extend
from $\sim$0\farcs8 to at least 7\arcsec \citep{paumard06}.
Other velocity vector analyses show that there are possible 
co-moving groups or clusters of stars, including the IRS 13
cluster, which is proposed to lie within the putative CCW
disk \citep{maillard04irs13,schodel05}, and the IRS 16SW co-moving
group, which are also consistent with the proposed CW disk \citep{lu05irs16sw}.
The two proposed disks are inferred
to be oriented with an inclination and angle to the ascending node of 
[$i_{CW}$=127$^\circ \pm$ 2$^\circ$, $\Omega_{CW}$=99$^\circ \pm$ 2$^\circ$] and 
[$i_{CCW}$=24$^\circ \pm$ 4$^\circ$, $\Omega_{CCW}$=167$^\circ \pm$ 7$^\circ$] and 
to have a finite angular thickness of 
$\Delta\theta_{CW} \sim 14^\circ$ and 
$\Delta\theta_{CW} \sim 19^\circ$ where $\Delta\theta$ is the
standard deviation of the orbital inclinations distributed
normally about the disk plane \citep{paumard06}. 
The thickness of the stellar disks has been attributed to 
thickening as a result of gravitational interactions between the two disks,
which provides an estimate of the disk masses \citep{nayakshin06thick}. 
The derived mass is smaller than the mass inferred from the number of 
observed young stars, assuming a Salpeter initial mass function (IMF); 
accordingly, \citet{nayakshin06thick}
suggest that the disks have a top-heavy mass function.
Both {\it in situ} gas disk and in-spiraling star cluster 
formation scenarios have been used to explain
the kinematics of this young star population and to
predict that the stars should lie in a common orbital plane. 
However, the presence of two stellar disks with similarly aged
populations requires either two nearly concurrent gas disks or two 
infalling star clusters; and both of these scenarios are difficult to produce.
Therefore, to understand the recent star formation history,
it is critical to measure the orbital planes of individual
stars in order to confirm the existence of the two stellar disks previously 
derived from a statistical analysis of velocity vectors alone.

The {\it in situ} gas disk and inspiraling star cluster 
formation scenarios predict different structures and evolutions
for the resulting stellar disk, particularly with respect to the
eccentricities and radial distribution of stars within the disk.
Early models of a self-gravitating gas disk around the supermassive 
black hole at the center of the Milky Way produce stars with 
a steep radial profile in the disk surface density, 
$\Sigma \propto r^{\alpha}$, with $\alpha \sim -2$ \citep{linPringle87,levin06}. 
These models typically result in stars on
circular orbits as would be the case for the slow build up of a
gas disk that is circularized before there is sufficient mass for
gravitational instabilities to set in
\citep{milosav04,nayakshinCuadra05,levin06}.
The stellar eccentricities of an initially circular disk can relax 
to higher eccentricities up to $e_{rms} = \sqrt{<e^2>} \sim $0.15 for a normal 
IMF or $e_{rms} \sim $0.3 for a top-heavy IMF \citep{alexander07imf,cuadra08}.
More recent models have also shown that star formation can occur rapidly
before circularization in an initially eccentric 
disk as might result from the infall of a single massive molecular cloud
or a cloud-cloud collision \citep{sanders98,nayakshin07sims,alexander08}.
These eccentric self-gravitating accretion disk models typically produce
a more top-heavy IMF than initially circular disks.
On the other hand, an inspiraling star cluster would dissolve into a 
disk of stars with a flatter radial profile 
\citep[$\Sigma \propto r^{-0.75}$;][]{berukoff06}
whose orbital eccentricities would reflect the 
eccentricity of the cluster's orbit, which could be either circular 
or eccentric \citep{pzwart03irs16,mcmillan03,kim03,kim04,gurkan05,berukoff06}. 
Previous measurements of the radial distribution of young stars
yields a steep radial profile consistent with {\it in situ} formation
\citep{paumard06}.
Also, the eccentricities of the stars have previously been estimated from 
observations by assuming that the stars orbit in a disk; 
however, there are conflicting
results claiming that the stars in the clockwise-rotating disk are on
nearly circular orbits \citep{paumard06} or on eccentric orbits 
\citep{beloborodov06}. 
Determining the radial profile and stellar eccentricities of stars in a 
disk may provide observational constraints on the origin of the young stars.
 
We present an improved proper motion study that yields an order 
of magnitude more precise proper motions and 
the first measurement of accelerations in the 
plane of the sky for stars outside the central arcsecond.
By combining the stellar positions, proper motions, radial velocities,
and accelerations, we estimate stellar orbital parameters 
and test whether the young stars reside on one or two stellar disks in a more
direct manner than previous methods using only velocity information.
This provides a {\it direct} test of the existence, membership, and properties
of these disks. The observations are described in 
\S\ref{sec:obs} and the astrometric analysis procedure and results are detailed 
in \S\ref{sec:astrometry}. Orbit analysis and results are presented in 
\S\ref{sec:orbitAnalysis} and \S\ref{sec:orbitResults} and a discussion 
of the implications for the origin of the massive, young stars
at the Galactic Center is presented in \S\ref{sec:discussion}.

\section{Observations}
\label{sec:obs}

This study utilizes 29 epochs of high-resolution, infrared images of the 
Galaxy's central stellar cluster, which were taken from 1995 to 2005 
using both speckle and laser guide star adaptive optics (LGS AO) observing 
techniques on the W.~M. Keck 10 m telescopes.
These data sets are listed in Table \ref{tab:obs} and all but the
additional LGS AO observation from 2005 are described in detail in earlier
papers \citep{ghez98pm,ghez00nat,ghez05orbits,lu05irs16sw,rafelski07}.
Columns 3 and 4 list the individual exposure times and the 
total number of frames for each epoch of data. 
All 27 speckle imaging observations were taken using the 
facility near-infrared camera, NIRC \citep{NIRC,NIRCs}, which has a
plate scale of $\sim$20 mas per pixel, and 
a 5\farcs22 $\times$ 5\farcs22 field of view. 
The two adaptive optics imaging observations used the facility LGS AO 
system \citep{wizinowich06,vanDam06} and the near-infrared camera, 
NIRC2 (PI: K. Matthews) with a plate scale of 
9.963 $\pm$ 0.006 mas per pixel \citep{ghez08}
and a 10\farcs2 $\times$ 10\farcs2 field of view. 
While the laser guide star is used to correct most of the 
atmospheric aberrations, the low-order, tip-tilt terms were corrected 
using visible observations of USNO 0600-28577051 (R = 13.7 mag and
$\Delta r_{SgrA*}$ = 19\arcsec).

In addition to the 27 speckle observation and the 2004 LGS AO observations
described in previous works, a new LGS AO data set was obtained in 
2005 June.   This data set was taken using two different
narrow-band filters, 
K$_{CO}$ ($\lambda_{o}$=2.289 \micron, $\Delta\lambda$=0.027 \micron) and 
K$_{cont}$ ($\lambda_{o}$=2.270 \micron, $\Delta\lambda$=0.030 \micron),
rather than the K' broadband filter used for the 2004 LGS AO observations.
For each filter, images were taken in a 5 position pattern around a 
4\farcs0 box with exposure times of 36 s 
(t$_{exp}$ = 7.2 s, 5 coadds) and 59.5 s (t$_{exp}$ = 11.9 s, 5 coadds)
for the K$_{CO}$ and K$_{cont}$ filters, respectively.
The choice of narrow-band filters was driven by a different project
and the data sets from the two filters were combined together
for the present study (see \S\ref{sec:images}).
Resulting Strehl ratios were $\sim$0.25-0.35 in the individual frames.

\section{Astrometric Data Analysis and Results}
\label{sec:astrometry}

The goal of this analysis is to obtain high precision astrometry for
a sample of young stars that are candidate disk members and 
have existing radial velocity measurements. 
Based on spectroscopic identification, there are currently 90 known
young stars with radial velocity measurements listed in \citet{paumard06} 
based on high quality (``quality 1 or 2'') spectral classifications.
We define a {\it primary sample} that includes those known young stars found
in our astrometric data sets that 
have projected radii between 0\farcs8 and 3\farcs5. The inner
radius is set by the proposed inner edge of the clockwise disk of young
stars and young stars interior to this radius are on more randomly
oriented orbits \citep{ghez05orbits,eisenhauer06}.
The outer radius is set by the field of view of the speckle data sets. 
Over this region, \citet{paumard06} note that all young
stars brighter than K=13.5 should be identified, which includes
OB giants and supergiants.
A total of 32 such young stars are in our 11 year 
astrometric data set and comprise the sample for this study. Of the 
32 stars in our sample, 23 are among the 36 stars thought to be part 
of the clockwise disk, 2 are among the 12 candidate members of the
counter-clockwise disk, and the remaining 7 are among the 42
stars not assigned to either disk by \citet{paumard06}.

We also define an {\it extended sample} that includes both the primary
sample of 32 stars and an additional 41 young stars found by 
\citet{paumard06} at larger radii that are outside the field of view
of our astrometric measurements. The astrometry for the additional
41 stars is taken from \citet{paumard06}\footnote{We note that there
are 4 additional young stars at larger radii that are not included
in our extended sample since they do not have proper motions
listed in \citet{paumard06}.},
which has an order of magnitude lower precision and lacks any constraints
on the accelerations. 
However, we use the extended sample to explore the kinematics of the
young stars at larger radii with the same analysis techniques used
on the primary sample. We also note that the spectroscopic observations
used to identify the young stars at larger radii were taken in a different 
setup than in the central regions, with lower spectral resolution 
and lower Strehl; thus the 
completeness limit may be somewhat brighter in this region. 
However, any difference is statistically insignificant given that
a two-sample KS-test yields a 50\% probability that
the primary sample and those additional stars added to the extended
sample have the same K-band luminosity function. The extended sample
is used only to supplement our analysis; therefore, to avoid confusion,
all analysis and results are reported for the 
primary sample, which has more precise proper motions and accelerations,
unless specifically noted otherwise. 

Astrometric positions for the young stars in the primary sample 
are extracted from the imaging
data sets listed in Table \ref{tab:obs} using similar techniques to those
described in \citet{ghez98pm,ghez00nat}, \citet{lu05irs16sw},
and \citet{ghez05orbits}, with the
following key changes:
(1) geometric distortion is corrected in the speckle images using
an improved distortion solution (see \S\ref{sec:images}, 
Appendix \ref{app:speckDistort}), 
(2) speckle images are combined with an improved algorithm developed 
and implemented by \citet{sethThesis}, and
(3) image coordinates are transformed between data sets with more
degrees of freedom (see \S\ref{sec:coords}).
Sections \ref{sec:images} and \ref{sec:coords} describe the analysis in
detail and Section \ref{sec:astroResults} presents the astrometric results.

\subsection{Image Processing}
\label{sec:images}

To achieve precise astrometry, the basic image reduction steps, 
particularly geometric distortion correction, must be carefully implemented.
First, both speckle and LGS AO individual exposures are processed using 
standard techniques of sky subtraction, flat-fielding, and bad pixel correction.
Next, the images are transformed to correct for optical distortion.
For the LGS AO/NIRC2 images, optical distortions are well characterized at the
$\sim$2 milli-arcsecond level over 2'' 
\citep[][Appendix \ref{app:speckDistort}]{ghez08}
by the pre-ship review distortion 
coefficients\footnote{http://www2.keck.hawaii.edu/inst/nirc2/}
and the distortions are removed from the images using the IRAF routine, 
{\it Drizzle} \citep{drizzle}. 
The speckle images, obtained with NIRC, have a known off-axis 
distortion that can be corrected as described in \citet{ghez98pm}. 
However, this distortion solution 
does not account for any distortion introduced by the 
additional optics in the NIRC reimager, which magnifies the image scale
by a factor of $\sim$7 from 
seeing limited sampling to diffraction limited sampling. 
Speckle data sets were acquired in such a way as to minimize
the effects of this residual distortion in the center of the field
of view and have resulting residual distortion errors that are 
smaller than the typical centroiding error, which is $\sim$2 mas,
for stars at radii $<$ 0\farcs5.
However, astrometric uncertainties for stars outside this region
are dominated by the uncorrected distortion, which grows to
$\sim$6 mas near the field edge at a radius of 2\farcs5 \citep{ghez05orbits}. 
In order to characterize the residual distortion in NIRC, simultaneous images
of the Galactic Center were obtained with both NIRC and NIRC2 with the NIRC2
images serving as a reference coordinate system 
(see Appendix \ref{app:speckDistort}). The speckle image distortion
is mapped by comparing stars' positions in both NIRC and NIRC2 images.
As shown in Appendix \ref{app:speckDistort}, the resulting NIRC to NIRC2
transformation is characterized at the $\sim$2 mas level over the entire
field of view.

After distortion correction, individual exposures are combined into a final 
diffraction-limited image using different methods for speckle and LGS AO data 
sets. 
Speckle images are produced by first rejecting the low Strehl ratio frames 
(typically 75\% of frames are rejected)
and then stacking the remaining frames using a 
weighted shift-and-add (SAA) routine \citep{sethThesis}.
The resulting combined images have a point-spread function (PSF)
composed of a diffraction-limited core 
(FWHM$\sim$0\farcs055) on top of a broad seeing halo (FWHM$\sim$0\farcs4).
The improved image combination algorithm attempts to maximize the
signal-to-noise ratio (SNR) of the final image while preserving
the highest spatial resolution. Quantitatively, the weighted SAA method
doubles the fraction of light contained in the diffraction-limited
core (from 3.5\% to 7.0\%) over the standard SAA scheme with no weighting 
and no frame rejection \citep{sethThesis}. 
The LGS AO individual exposures are all of similar
quality and are thus all averaged together, without weighting, in order
to produce the final high-resolution image for each data set. 
Although the 2005 June data were taken in two different filters 
(K$_{CO}$ and K$_{cont}$), all the images
were combined together to increase the final SNR.
While photometry from this epoch is marginally impacted, 
the astrometry is comparable to other epochs.
Each data set was also sub-divided to produce three equivalent
quality (randomized in time) subsets to make three images used for determining
photometric and astrometric uncertainties. 
The resulting images are summarized in Table \ref{tab:obs},
including the achieved spatial resolution (FWHM) and 
the Strehl ratio.

\subsection{Stellar Positions and Coordinate Transformations}
\label{sec:coords}

In order to extract astrometric information for the sample of young stars, the
coordinate system from each data set is transformed into a common
reference frame using the stars in each image to determine the 
transformation parameters. Since the 
accuracy of this transformation relies on the assumption that there 
is no net rotation of the sample, we use all stars
detected in each data set, not just the young stars, in this analysis.
The steps for (1) measuring stars' positions in each epoch, 
(2) transforming to a common (relative) reference frame, and 
(3) determining the absolute coordinate system are 
described below and utilize all stars detected in the data sets; 
then as a final step, the young star sample is extracted.

\begin{figure}
\begin{center}
\plotone{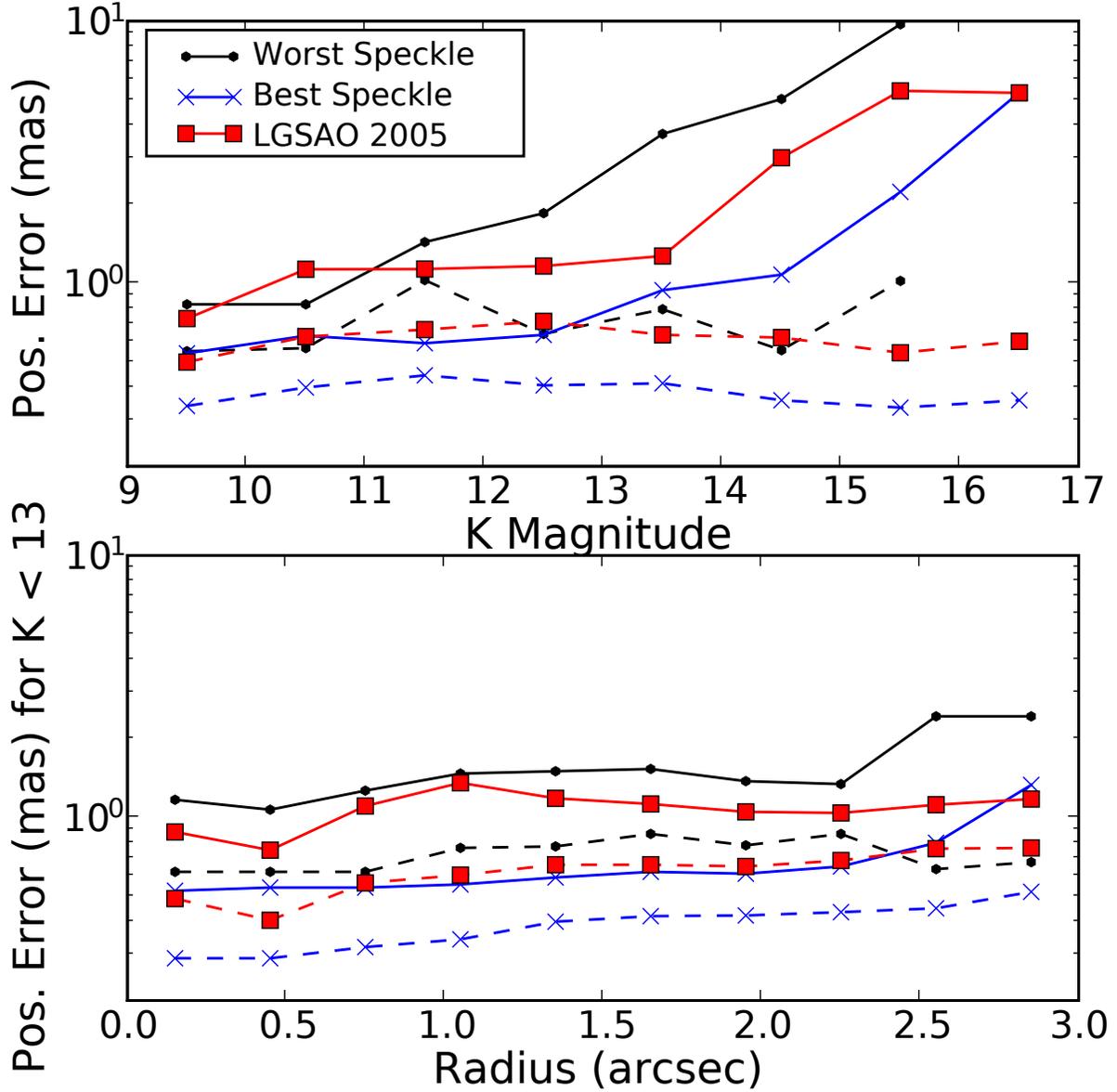}
\end{center}
\caption{
Positional uncertainties for stars as a function of stellar brightness
{\it (top)} and distance from the black hole, Sgr A*, which is 
near the center of the field of view {\it (bottom)}. 
To show the full range of possible values, the
centroiding {\it (solid)} and the alignment {\it (dashed)} uncertainties
are shown for the best (1999.559) and
worst (1996.485) speckle epochs and one of the LGS AO epochs (2004.567). 
The uncertainties are the median values of all stars within
magnitude bins of $\Delta$K = 1 or radius bins of $\Delta r$ = 0\farcs3.
Note that alignment uncertainties are small compared to centroid 
uncertainties.
}
\label{fig:posError}
\end{figure}

In each data set, stars are identified and their positions measured
using the IDL point-spread function fitting routine ``StarFinder'' 
\citep{starfinder}. 
StarFinder generates a PSF from several bright stars in the field
and cross-correlates the resulting PSF with the image. 
The PSF was iteratively constructed using 
IRS 16C and IRS 16NW for the speckle maps and 
IRS 16C, 16NW, 16NE, 16SW, 33E, 33W, 7, 29N, and GEN+2.33+4.60 
for the LGS AO images. 
Candidate stars are those for which StarFinder correlation 
peaks have a correlation value higher than 0.8 and positions
and fluxes are extracted by fitting the PSF to each correlation peak. 
From the candidate star list, spurious detections are then eliminated by
requiring that each star be detected in all three of the subset-images
with a correlation of higher than 0.6. 
The positional centroiding uncertainties for each candidate star are estimated 
from the rms of their locations in the three subset-images, and an additional
systematic error term of 0.88 mas is added in quadrature to all stars in
LGS AO epochs to account for residual distortion in the central 5'' of
NIRC2 \citep{ghez08}.
The candidate stars are flux calibrated using the apparent 
magnitudes of the non-variable stars, 
IRS 16C, IRS 16SW-E, S2-17, S1-23, S1-3, S1-4, S2-22, S2-5, S1-68, S0-13, 
and S1-25, as measured by \citet{rafelski07}. 
The star detections from each epoch are cross-identified 
with stars from all other epochs and those stars that are detected in 
at least 16 out of 29 epochs are used to create a master star list. 
The threshold of 16 or more epochs is used in order to insure
high astrometric precision; for a threshold of less than 16 epochs,
the number of detected stars rises dramatically as does the number of
sources showing significant ($\gtrsim$3$\sigma$) accelerations in 
non-physical directions, indicating a high frequency of false detections
(see \S\ref{sec:astroResults} for further discussion).
Stars in the master list are also examined for source confusion, which may occur
when two stars pass close enough to each other such that StarFinder
only detects a single source with biased astrometry rather than detecting
both stars. Source measurements from individual epochs are rejected if two 
stars pass within 55 mas ($\sim$1 spatial resolution element) of each other and
only one source is detected by StarFinder. 
The results of this stage of the analysis are summarized in 
Table \ref{tab:obs}, which provides for 
each data set the average centroiding error for the brightest stars
(K$<$13; also see Figure \ref{fig:posError}) 
and the sensitivity as estimated 
by the peak in a histogram of the K-band magnitudes (bins = 0.1 mag) 
of all the stars in the data set.
Averaged over all stars in all maps, the centroiding uncertainties have a
mean value of 1.6 mas for the brightest stars (K $\leq$ 13 mag) and 
3.4 mas for the fainter stars (13 $<$ K $<$ 16 mag). 

The coordinate system for each image is transformed to a common
local reference frame defined by the 2004 July LGS AO/NIRC2 image's 
coordinates and pixel scale. 
This particular LGS AO epoch was chosen as the reference because
the NIRC speckle distortion solution is tied to this epoch, thus
providing a smooth transition between speckle and LGS AO data sets.
The procedure for deriving the coordinate transformation for
all of the data sets is non-trivial, since the stars in the
images have detectable motions.
Optimal alignment is achieved by minimizing the error-weighted, net 
displacement for all the stars as described by \citet{ghez98pm}
while allowing for translation, rotation, and two magnifications in
arbitrary, but perpendicular, directions. This is a higher order
transformation than was used in our earlier astrometric works, which
only allowed for translation and rotation. 
The new transformation equations have the form
\begin{eqnarray}
x_{pix} = a_0 + a_1 x'_{pix} + a_2 y'_{pix} \\
y_{pix} = b_0 + b_1 y'_{pix} + b_2 x'_{pix}
\end{eqnarray}
where $x'_{pix}$ and $y'_{pix}$ are the input detector coordinates in pixels 
and $x_{pix}$ and $y_{pix}$ are the output coordinates for each star, 
and all other variables are free parameters that are common across all stars
in the alignment fit. 
As in \citet{ghez05orbits}, stars within 0\farcs5 of Sgr A* are excluded 
from the transformation as they exhibit large non-linear motions.
Additionally, all spectroscopically identified young stars are excluded
from the transformation as they have a known net rotation \citep{genzel00}.
Initially, each image is aligned to the reference image by assuming the 
stars have no proper motions and finding the best-fit values for the
free parameters of the transformation, $a_0, a_1, a_2, b_0, b_1, b_2$, for
that image. However, after a first pass at the alignment of all the images, 
proper motions are estimated and 
used to refine the alignment solutions in a second pass. 
Sources with estimated proper motions higher than 
1.5 mas yr$^{-1}$ (600 km s$^{-1}$)
are excluded from the transformation resulting in the elimination of 
2 sources that are near the edge of the speckle field-of-view and suffer
from edge effects.
Alignment uncertainties are estimated by a half-sample bootstrap method
\citep{astrostats,ghez05orbits}
and are small ($\sim$0.2 mas for
stars at $r<$2\arcsec) compared to the centroiding uncertainties 
(see Figure \ref{fig:posError}). 
Alignment and centroiding 
uncertainties are added in quadrature to produce a final relative
positional uncertainty for each star at each epoch. 
The resulting astrometric data set contains stellar positions
and uncertainties for all epochs, transformed into the 2004 July NIRC2 pixel 
coordinate system ($x_{pix}, y_{pix}$). 

The relative positions and uncertainties are transformed into J2000 
absolute astrometric coordinates defined by radio observations of 
SiO masers and Sgr A*. Using observations of the SiO masers in the infrared,
a set of {\it infrared absolute astrometric standards} are defined in
a process described in detail in an 
appendix of \citet{ghez08}.
These astrometric standards are used to derive the transformation from
2004 July NIRC2 pixel coordinates into absolute coordinates. 
A statistically insignificant adjustment is made to place the origin at 
the dynamical center of S0-2's orbit, which is known to high precision,
by offseting from the radio position of Sgr A* by 1 mas to the East
and 5 mas to the South. This offset is well within the absolute astrometric 
uncertainty of $\sim$6 mas for Sgr A* \citep{ghez08}.
The stellar positions in all epochs are thus expressed in arcseconds
offset from the dynamical center with $+x$ increasing East and $+y$ 
increasing North and can be converted
into celestial coordinates using 
(x, y) = ($\cos{\delta}\; \Delta\alpha$, $\Delta\delta$)
\footnote{When converting from (x, y) to ($\Delta\alpha$, $\Delta\delta$),
higher order terms are negligible (0.06 mas over 5\arcsec) because the 
celestial sphere is sufficiently flat over our field of view.
}.
Positional uncertainties are taken as the quadratic sum of the relative
errors, which dominate, and the absolute error from uncertainties in the
plate scale and position angle.
Errors in the relative position of Sgr A* ($\sim$2 mas) are incorporated later 
during the orbit analysis stage as a parameter of the potential of the 
supermassive black hole
(see \S\ref{sec:orbitAnalysis}).
From the resulting absolute astrometric data set, the sample of young 
stars is extracted.

\subsection{Proper Motions and Acceleration Results}
\label{sec:astroResults}

\begin{figure}
\epsscale{0.6}
\plotone{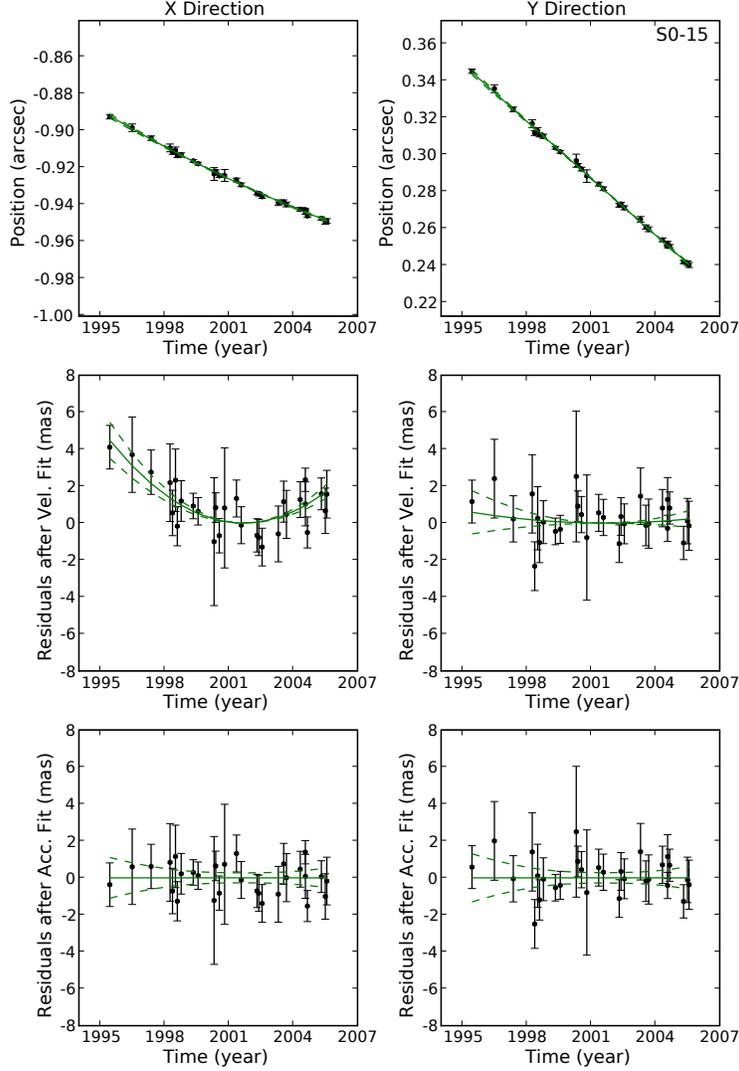}
\caption{
Measured positions and residuals as a function of time for S0-15,
a source with a significant, non-zero acceleration measurement,
in X and Y ({\it top}). Positions are reported relative
to Sgr A* and do not include the uncertainties in the
transformation to the absolute coordinate system (i.e. plate scale, 
position angle, and position of Sgr A*). 
The best fit quadratic polynomial modeling the velocity and acceleration
of the source is shown ({\it green solid}) with the 1$\sigma$ errorbars
({\it green dashed}). 
Also plotted are the X and Y residuals after subtracting off the best fit
velocity ({\it middle}) and the best fit acceleration
curve ({\it bottom}). 
The X (East-West) and Y (North-South) position plots ({\it top}) have 
a (y-axis) range of 0\farcs16 and residual plots ({\it middle, bottom}) have
a (y-axis) range of $\pm$8 mas.
}
\label{fig:posTime_S0-15}
\end{figure}

\begin{figure}
\epsscale{0.6}
\plotone{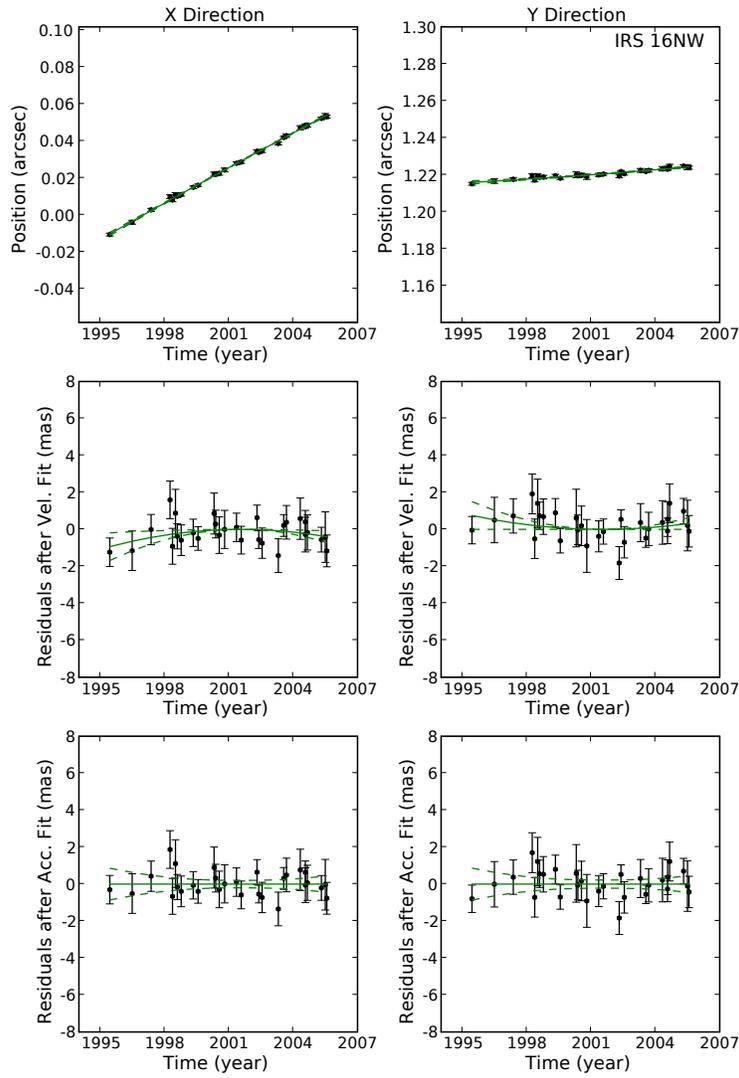}
\caption{
Measured positions and residuals as a function of time for
IRS 16NW, a source that has an acceleration consistent with zero,
but significantly below the maximum possible acceleration. 
See the caption in Figure \ref{fig:posTime_S0-15} for more information.
}
\label{fig:posTime_irs16NW}
\end{figure}

\begin{figure}
\epsscale{1.0}
\plotone{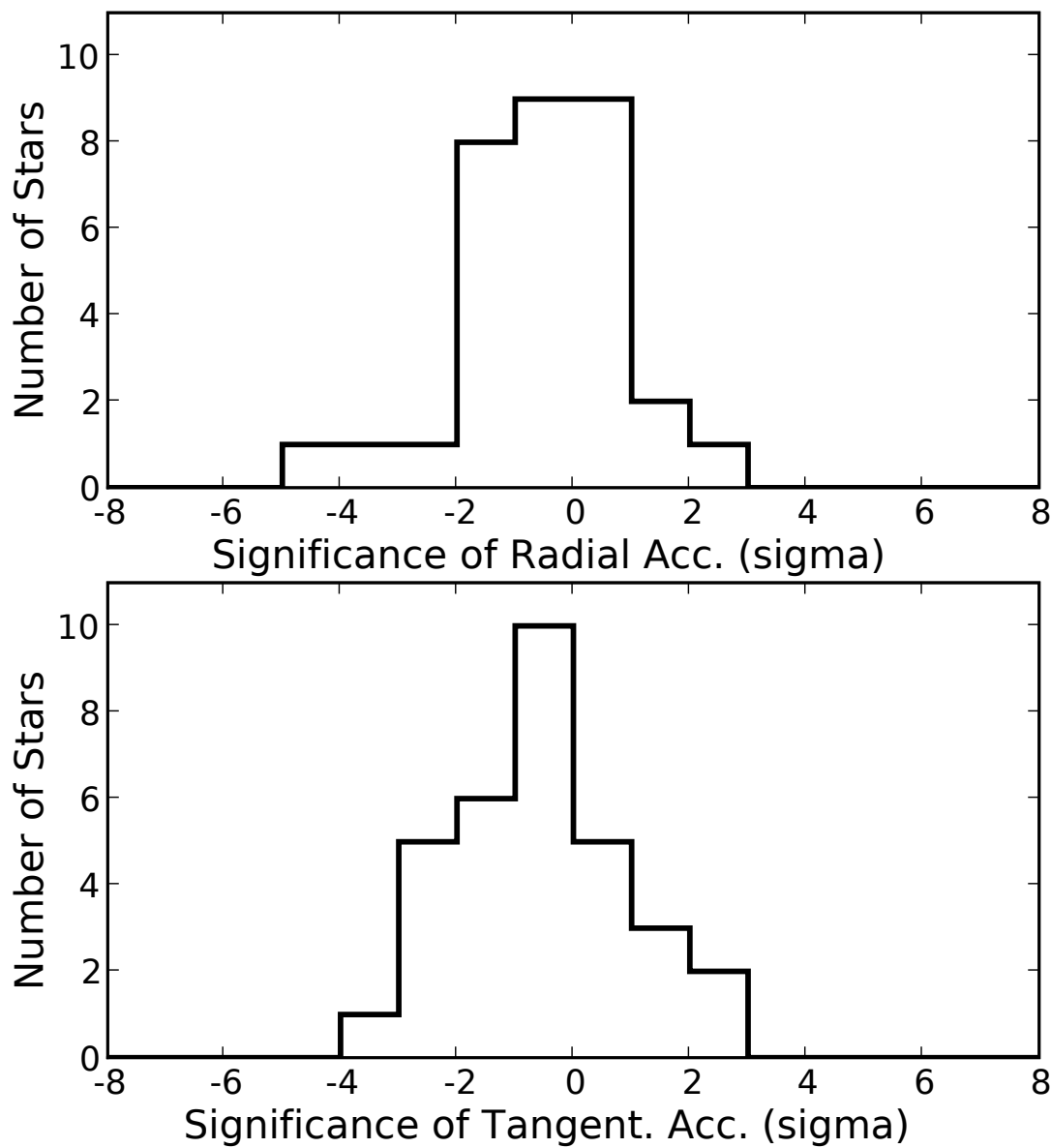}
\caption{Histograms of $a_\rho$/$\sigma_{a_\rho}$ showing the significance of the
acceleration measurements in both the radial ({\it top}) and tangential
({\it bottom}) directions. While, the distributions show an offset from zero
indicating a possible bias due to systematic errors, such as residual
distortion, that are not well characterized, it appears that any biases
are limited to the $\sim$1$\sigma$ level. 
The only star with significant negative radial acceleration 
($\gtrsim$4$\sigma$) is S0-15 and it is assumed to be a real acceleration due
to the gravity of the supermassive black hole.
}
\label{fig:histAccel}
\end{figure}

For each of the young stars in the sample, positions, velocities, and 
accelerations in the plane of the sky are derived by fitting second-order 
polynomials to the star's
position as a function of time, weighted by the positional uncertainties.
The polynomials are fit independently in $x$ and $y$ 
coordinates and have the form
\begin{eqnarray}
x(t) = x_{ref} + v_{x,ref} (t - t_{ref}) + \frac{1}{2} a_{x,ref} (t - t_{ref})^2 \\ 
y(t) = y_{ref} + v_{y,ref} (t - t_{ref}) + \frac{1}{2} a_{y,ref} (t - t_{ref})^2
\end{eqnarray}
where $t$ is the time in years, $t_{ref}$ is a reference time
taken to be the mean of the time of all epochs weighted by the positional
uncertainties for each star,
$x_{ref}$ and $y_{ref}$ are the positions at the reference time, $v_{ref}$ is the 
velocity at the reference time, and $a_{ref}$ is the acceleration at the 
reference time. Uncertainties in the fit
parameters are determined from the covariance matrix.
Figures \ref{fig:posTime_S0-15} and \ref{fig:posTime_irs16NW} show the 
polynomial fits for two example stars and 
the resulting values for the kinematic variables for all stars are reported
in Table \ref{tab:yng_pm_table}. 
Since the stars' motions
are assumed to be dominated by the central force from the black hole,
we convert $a_{x,ref}$ and $a_{y,ref}$ into radial and tangential 
accelerations \footnote{This assumption 
may not hold for stars in a gravitationally bound cluster, such as may be the 
case for the 4 stars in the extended sample that make up the IRS 13 
co-moving group; however, the deviations from the potential assumed above 
should result in only $5-10$\% changes in the velocity vectors.}. 
All tangential accelerations and 
positive radial accelerations are non-physical and therefore provide
a check on the systematic errors of the acceleration measurements. 
Figure \ref{fig:histAccel} shows a histogram of the 
significance of the acceleration measurements both in the radial and
tangential directions for the young stars in our primary sample. 
While the tangential and positive radial distributions are 
slightly offset (0.6$\sigma$) from zero and 
broader (1.5$\sigma$ vs. 1$\sigma$) than is expected 
for a normal distribution, 
any systematic errors appear to impact the results at the
$\lesssim 1\sigma$ level.

The resulting velocity measurements for the young star sample outside
the central arcsecond are improved by at least a factor of 7 when compared
with our previous work \citep{ghez98pm,lu05irs16sw} and other recently 
reported Galactic Center proper motions \citep[e.g.][]{genzel00,ott03thesis}. 
The absolute uncertainties in our proper motions
are typically $\sim$0.06 mas yr$^{-1}$ ($\sim$2 km s$^{-1}$), 
although stars detected in fewer epochs have somewhat higher values
(0.1 - 0.5 mas yr$^{-1}$; 4 - 20 km s$^{-1}$).
Figures \ref{fig:posTime_S0-15} and \ref{fig:posTime_irs16NW}
show examples of the measurements for two stars in our sample (S0-15 
and IRS 16NW), and their corresponding proper motion fits
with 1$\sigma$ errorbars.

\begin{figure}
\epsscale{1.0}
\plottwo{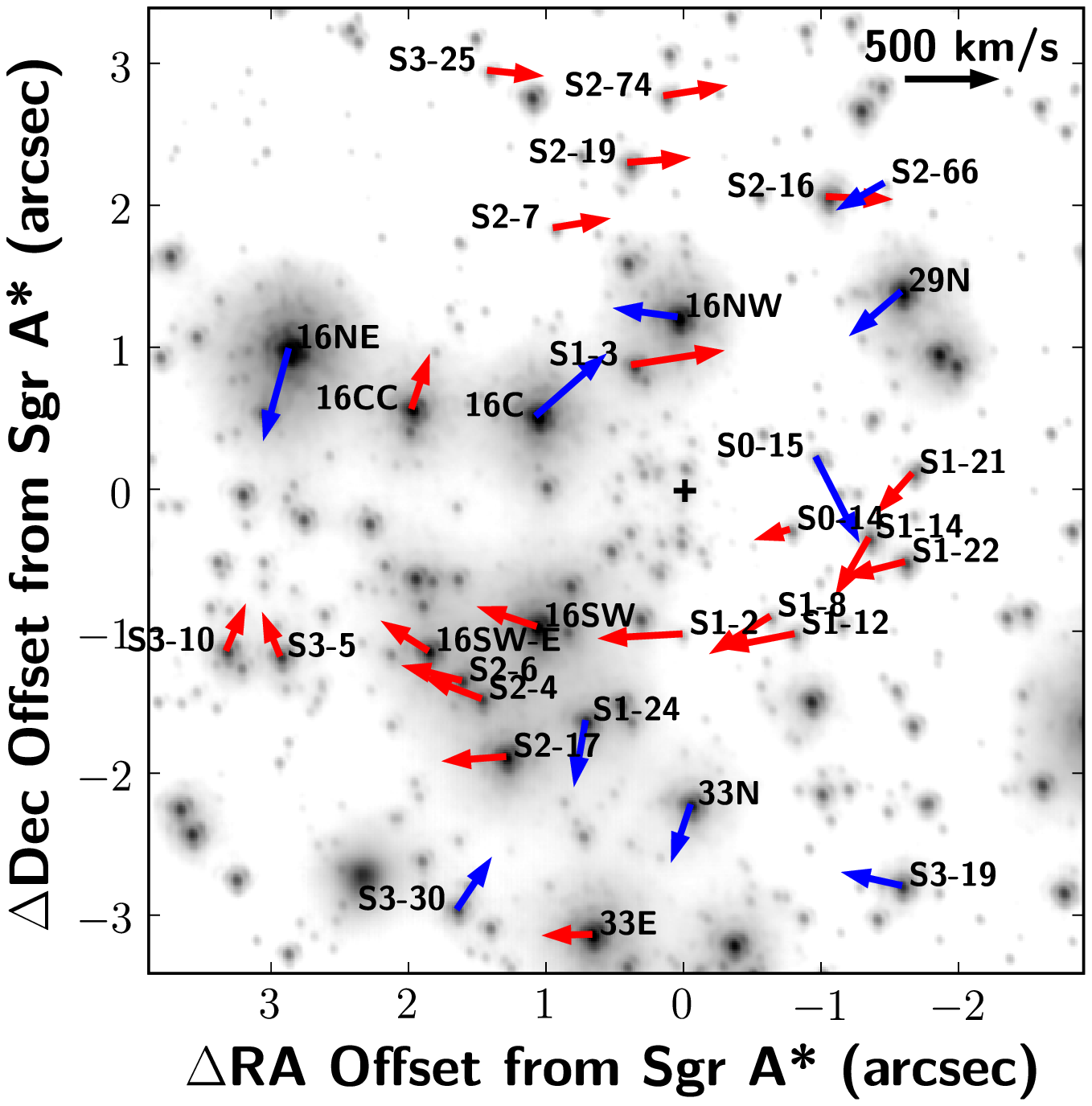}{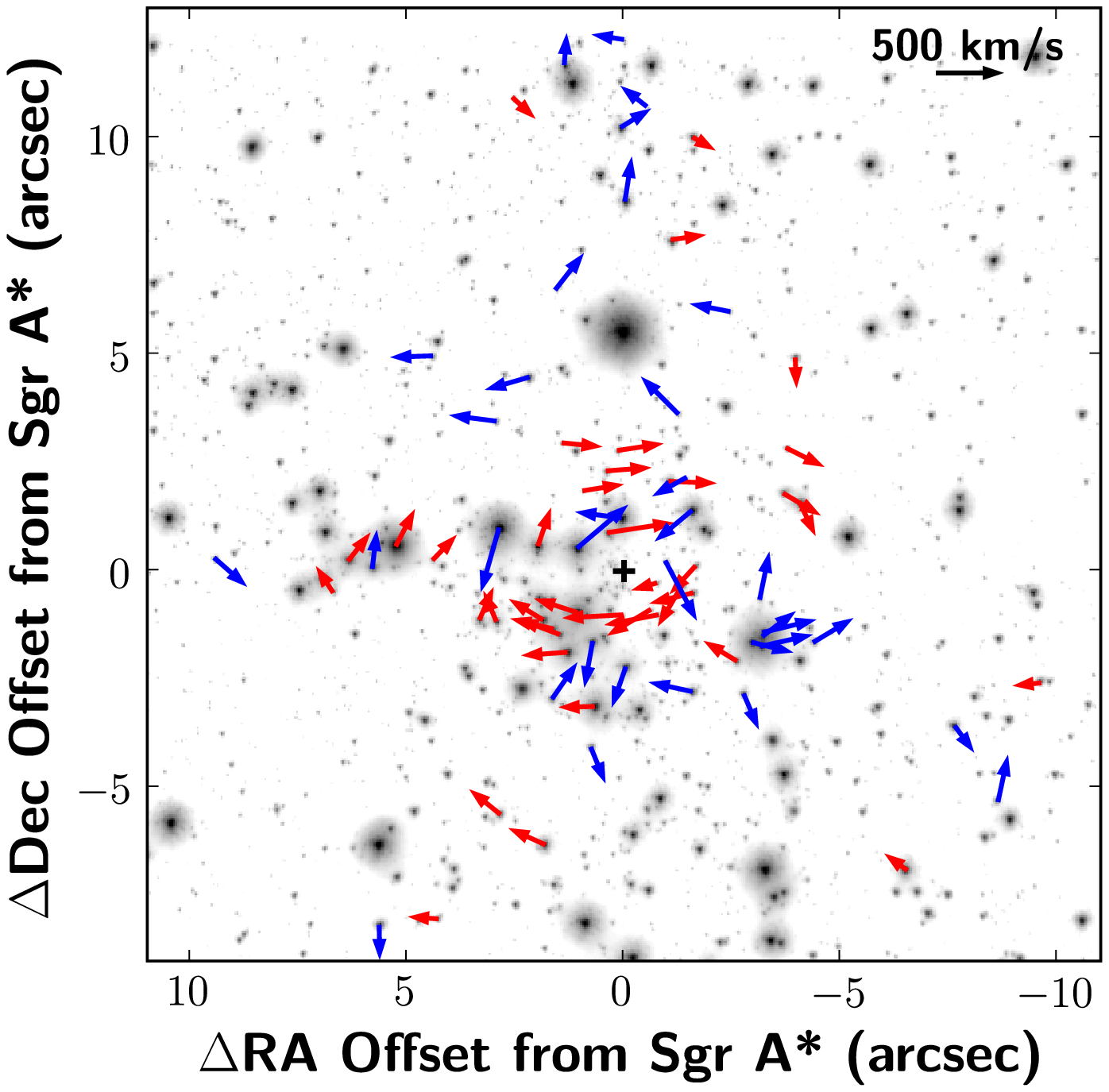}
\caption{
Positions and proper motion vectors of the young stars in our sample.
Candidate disk members are shown in red and non-disk members are
shown in blue over-plotted on an LGS AO image in grey-scale. The names of
the stars in the primary sample are shown in the {\it left} panel and 
the complete extended sample is shown 
in a zoomed-out view in the {\it right} panel. The position of Sgr A*
is marked with a black cross.
}
\label{fig:PMimage}
\end{figure}

In the young star sample, significant ($>$3$\sigma$) acceleration, 
or curvature, in the plane of the sky is detected only for S0-15 
(Figure \ref{fig:posTime_S0-15}).
This star has the second smallest projected separation from Sgr A*
in our sample,
at $\rho$ = 1\farcs0 (0.04 pc), and has a projected radial acceleration of 
-0.21 $\pm$ 0.05 mas yr$^{-2}$ or, equivalently, 
-9.6 $\pm$ 2.0 km s$^{-1}$ yr$^{-1}$ (see Figure \ref{fig:PMimage}).
S0-15 is more than twice as far from Sgr A*, in projection, than the seven
stars with previously detected accelerations, 
which were all within a projected radius of less than 0\farcs4
\citep[0.016 pc, ][]{ghez00nat,eckart02,ghez05orbits,eisenhauer06}.

The detection of acceleration is important in that it allows us to 
solve for the line-of-sight distance, and thus the 
three-dimensional position of a star relative to the black hole. 
For a star in the gravitational potential well of a supermassive black hole,
the plane-of-the-sky acceleration, at a three-dimensional distance r, 
in cylindrical coordinates is
\begin{equation}
\label{eqn:a2z}
a_{\rho} = \frac{-GM  \rho}{r^3} = \frac{-GM \rho}{(\rho^2 + z^2)^{3/2}}
\end{equation}
where $\rho$ is the plane-of-the-sky radial coordinate and z is the
coordinate along the line of sight relative to Sgr A*. 
The magnitude of the line-of-sight distance from Sgr A*, z,
can be solved for by adopting a black hole mass of
$M_{\bullet} = 4.4 \times 10^6$ \msun and a distance of
$R_{\circ} = 8.0$ kpc \citep[see \S\ref{sec:orbitAnalysis};][]{ghez08};
it is important to note that there is a remaining sign ambiguity for z. 
The resulting line-of-sight distance from Sgr A*
for S0-15 is $|0.045 \pm 0.004|$ pc bringing the total separation between
S0-15 and Sgr A* to 0.060 pc. 

\begin{figure}
\epsscale{1.0}
\plotone{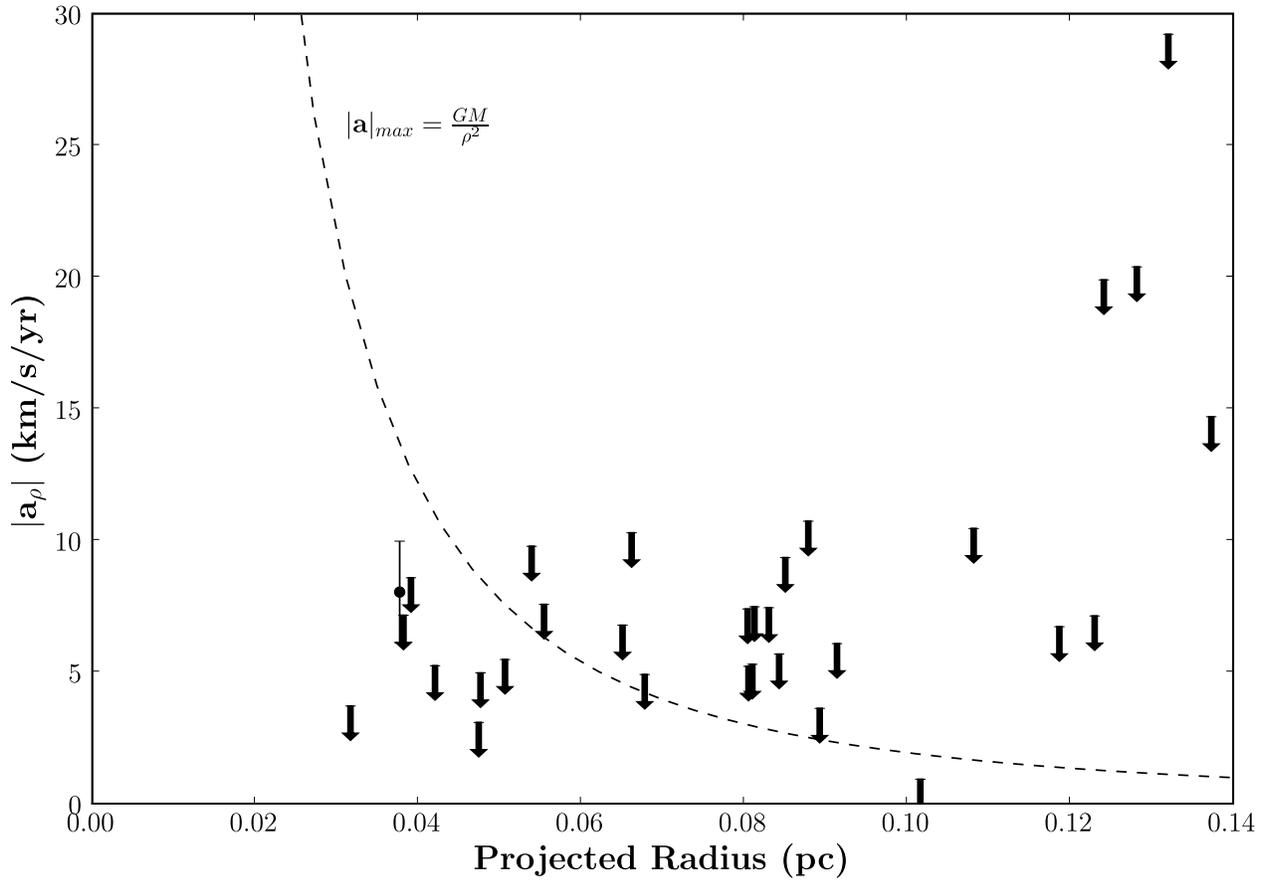}
\caption{The significance of observed limits for the plane-of-the-sky
acceleration. For a given projected radius, there is a maximum
allowed acceleration (dashed line).
If the measured accelerations from polynomial fitting, shown as 
3$\sigma$ upper limits on the y-axis, are less than the maximum allowed
acceleration (below the dashed line), then significant constraints can be 
placed on the line-of-sight distance, z, and subsequently the orbital 
parameters of the star. S0-15 has a significant detection of non-zero 
acceleration and is plotted with its 1$\sigma$ errorbars.}
\label{fig:accSignificance2}
\end{figure}

The remaining stars in our sample have acceleration measurements that
constrain the line-of-sight distance. 
While the lower limits of these acceleration magnitudes are not 
significantly different
from zero at the 3$\sigma$ level, their upper limits are 
smaller than the maximum allowed acceleration.
The maximum possible magnitude of the acceleration for a star at a 
given $\rho$ occurs when z = 0. When the measured acceleration limits are 
below this value, they provide a 
lower limit on the star's line-of-sight distance to the SMBH. 
Figure \ref{fig:accSignificance2} compares the measured acceleration limits
with the maximum possible acceleration for each star. 
Any 3$\sigma$ acceleration limits below the maximum allowed value
gives useful constraints on the line-of-sight distances.
In addition to our explicit measurement for S0-15, 
our high precision astrometric measurements are now yielding
3$\sigma$ acceleration limits with a median value of 
-0.19 mas yr$^{-2}$ (-7.3 km s$^{-1}$ yr$^{-1}$)
that can significantly constrain the line-of-sight
distance for nine stars in our sample that are located as far as 
1\farcs7 (0.07 pc), in projection, away from the black hole. 

\section{Orbit Analysis}
\label{sec:orbitAnalysis}

For a known point-mass Newtonian gravitational potential, 
a star's orbital elements can be fully determined from 
the measurement of only six kinematic variables.
For this analysis, we assume that the central point mass is a
black hole with characteristics determined by analysis of the
orbit of the star S0-2, which has been observed for nearly one
complete revolution 
\citep{eisenhauer06,ghez08}. 
Our proper motion analysis (\S\ref{sec:astroResults}) yields information 
on five kinematic variables, including two positions, two velocities, and one 
acceleration. 
The sixth kinematic variable comes from radial velocities measured by
\citet{paumard06}. The reported radial velocities are averaged 
over several years of observations; however, we adopt the same 
reference epoch, $t_{ref}$, as for the proper motion analysis since
any change in the radial velocity due to acceleration along the 
line-of-sight should be well
within the large measurement uncertainties in radial velocity 
($\sigma_{v_{z,ref}}\sim$20-100 km s$^{-1}$).
As described in \S\ref{sec:astroResults}, the plane-of-the-sky acceleration
can be converted into a line-of-sight distance that, when combined with the
projected distance, gives the full three-dimensional position for a star.
Although most of the stars in our sample have plane-of-the-sky accelerations 
that are consistent with zero, the upper limits on the magnitude of the 
acceleration provide valuable information by ruling out small 
line-of-sight distances. We therefore use our best-fit accelerations
and uncertainties as formal measures of the acceleration when converting 
to a line-of-sight distance.
Therefore the six measured quantities can be expressed as a
three-dimensional position and three-dimensional velocity at a certain 
epoch ($t_{ref}$).
Given the properties of the black hole, these kinematic quantities can be
translated directly into 6 standard orbital elements 
(see Appendix \ref{app:orbits}). 

\begin{figure*}
  \epsscale{1.0}
  \plotone{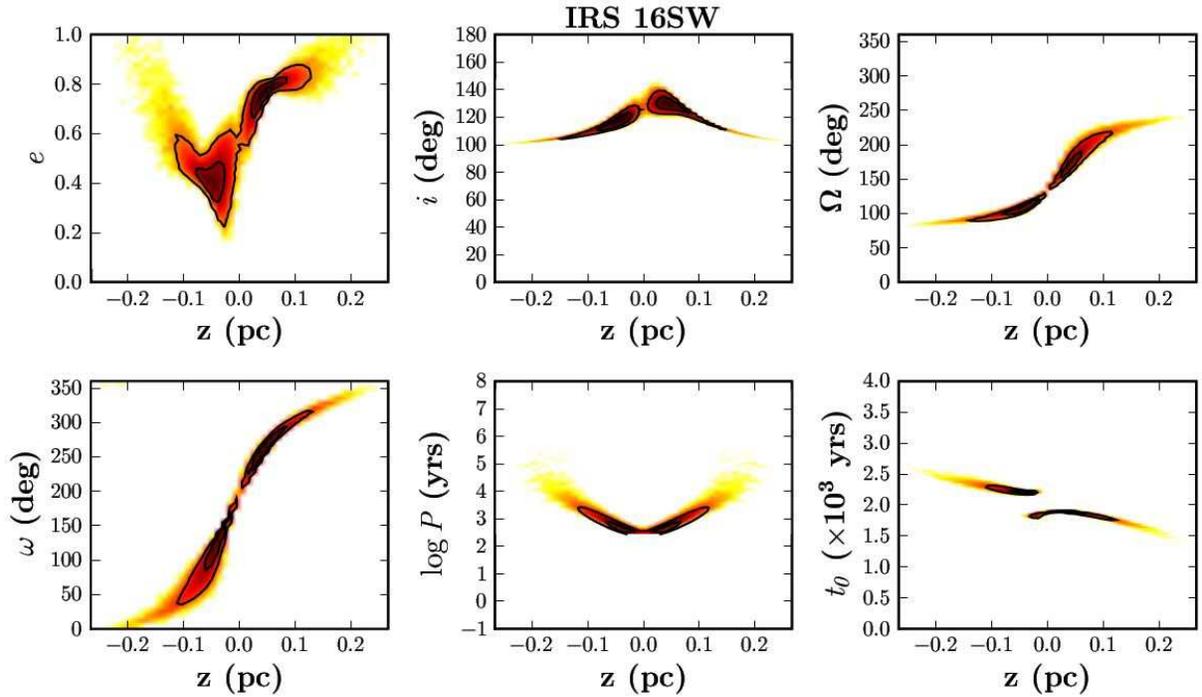}
  \caption{
    The range of allowed orbital parameters for IRS 16SW
    as determined from the observed 
    two-dimensional position in the plane of the sky, 
    the three-dimensional velocity,
    and the acceleration. The probability distribution for
    each orbital parameter is determined 
    by sampling from a gaussian distribution for each of the 
    observed quantities and analytically converting to the standard
    orbital elements.
    High density (dark) regions represent the most probable values for each
    orbital parameter and the resulting 1$\sigma$ and 2$\sigma$ contours
    are shown as black lines.
  }
  \label{fig:pdfParams_irs16SW}
\end{figure*}

A Monte Carlo simulation is carried out to transform
each star's six measured kinematic variables 
($x_{ref}$, $y_{ref}$, $v_{x,ref}$, $v_{y,ref}$, $v_{z,ref}$, $a_{\rho,ref}$)
into six orbital parameters ($i$, $\Omega$, $\omega$, $e$, $P$, $T_o$)
and their uncertainties.
A total of 10$^5$ Monte Carlo trials are run and, in each trial,
$4 + (6 \cdot 32)$ variables are randomly generated; four for the potential 
parameters and six for each of the 32 stars' measured kinematic variables. 
The four potential parameters are pulled from a four-dimensional probability 
density function, PDF(M$_\bullet$, R$_o$, x$_o$, y$_o$), 
based on the orbit of S0-2 derived by \citet{ghez08},
where the black hole's mass and line-of-sight distance are centered on 
$M_{\bullet}=4.4 \times 10^6$ \msun and
$R_{\circ}=8.0$ kpc \footnote{These values correspond to a 12-parameter
orbit model for S0-2 (i.e. $v_z=0$ case) from an early version of 
\citet{ghez08}. In this version, local distortions were
not corrected \citep[][Appendix B]{ghez08}; but the
resulting black hole mass and distance differ by $<1\sigma$ from the final
reported values.}, 
the dynamical center is adopted as the origin
with x$_o$ and y$_o$ defined as zero, and the projected one-dimensional 
probability distributions' RMS errors are [1.0, 1.6] mas for [$x_o$, $y_o$],
$0.3 \times 10^6$ \msun for $M_{\bullet}$, and $0.3$ kpc for $R_\circ$
\footnote{Simulations were also performed using the lower black hole mass and 
distance reported by \citep{eisenhauer06}. Our results on the detection of 
only one stellar disk and on the properties of the disk are all consistent 
within 1$\sigma$ error bars.}.
For each trial, all the stars' orbits are calculated using the same 
potential parameters in order to preserve correlations between the 
potential parameters and the orbital parameters such as eccentricity.
The kinematic variables for each star
are sampled from independent gaussian distributions,
each of which is centered at the best-fit value from Table 
\ref{tab:yng_pm_table} and has a 1$\sigma$ width set to the measurement
uncertainty. 
Any correlations between the measured kinematic variables are negligible
given the small uncertainties in the stars' relative angular positions 
($\lesssim$0.2\%) and velocities ($\lesssim$3\%) in the plane-of-the-sky
as compared to the uncertainties in the black hole mass ($\sim$10\%) 
and the accelerations ($\sim$60\%). 
The distribution for the acceleration, $a_\rho$, is truncated such that only
accelerations of bound orbits are allowed\footnote{The assumption that the
orbits are bound does not effect the results presented in this paper 
discussed in \S\ref{sec:orbitResults} since
all unbound orbital solutions yield high inclination (edge-on) orbits
and large eccentricities (e > 1). Considering only bound orbits 
simplifies the orbit analysis as we need only consider equations for
elliptical orbits rather than hyperbolic or parabolic orbits.
}, which follows from requiring a negative specific orbital energy,
\begin{equation}
E = \frac{v^2}{2} - \frac{GM}{r} < 0
\end{equation}
and substituting from Eq. \ref{eqn:a2z} to give the acceleration constraint
\begin{equation}
|a_\rho| > \frac{\rho v^6}{8(GM)^2}.
\end{equation}
For each trial and each star, the orbital parameters are computed and 
the results of all the trials are combined into a
six-dimensional probability density function (PDF) by dividing up parameter
space into bins, summing the number of trials in each bin, and then
normalizing by the total number of trials.
This Monte Carlo method is a straight-forward way to combine
a star's six measurement PDFs and the
four-dimensional PDF for the central point mass, which shows
strong correlations between $M_\bullet$ and $R_o$,
to produce a six-dimensional PDF for each star's orbital elements,
PDF($i$, $\Omega$, $\omega$, $e$, $P$, $T_o$), which has strong correlations
between the orbital parameters. 
The results of these simulations are plotted for an example star, IRS 16SW,
in Figure \ref{fig:pdfParams_irs16SW} 
to show that $i$ and $\Omega$ are generally well determined and that
$e$, in some cases, can be usefully constrained.
Similar figures of the orbital parameters for every star are
shown in Figure Set 7, which is available online in the electronic edition
of this manuscript.

\begin{figure}
\epsscale{0.8}
\plotone{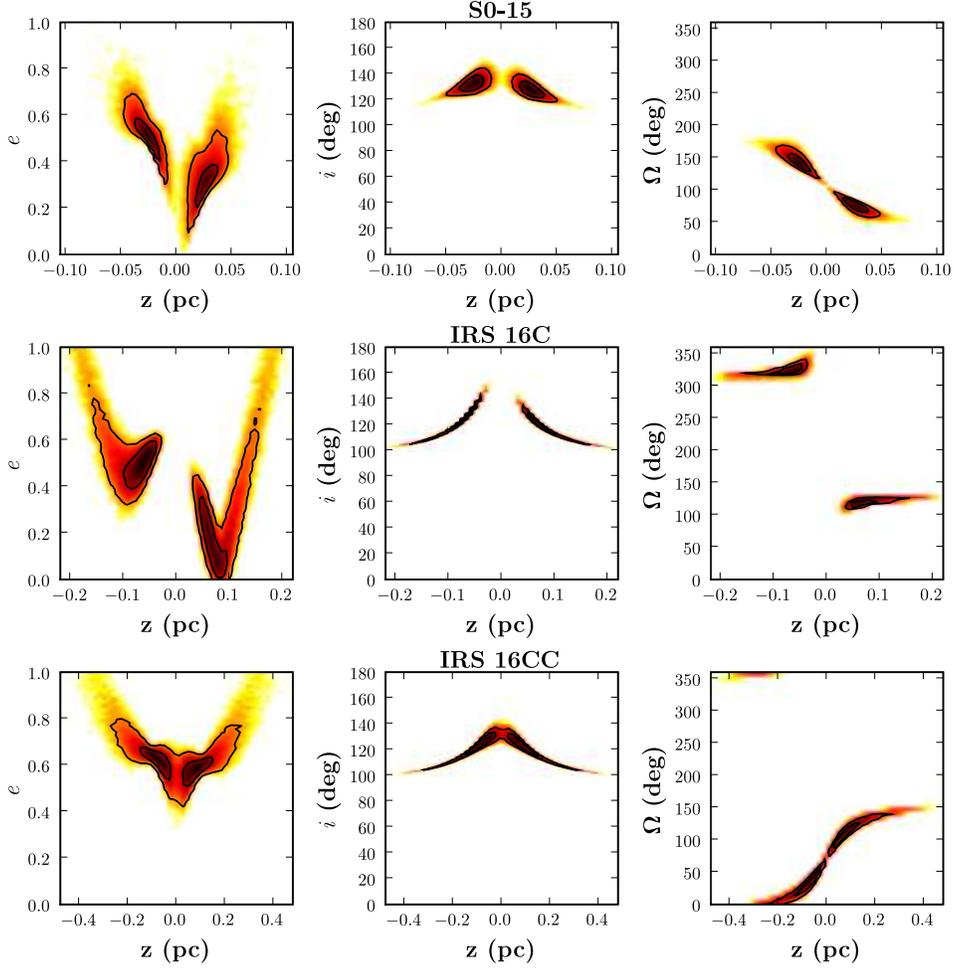}
\caption{The range of allowed eccentricities ($e$), inclinations ($i$),
and angles to ascending nodes ($\Omega$) 
as determined by our orbit analysis for three example stars. 
The range of z values (horizontal axis) extends to all possible 
bound orbits for the star. The probability density function is shown
in color with the 1$\sigma$ and 2$\sigma$ contours drawn shown as black lines.
S0-15 has a measured acceleration that is significantly different from zero.
IRS 16C has an acceleration upper limit that is less
than the maximum allowed acceleration, and thus a lower limit
on the line-of-sight distance, $|z|$. IRS 16CC has no significant
acceleration limit, but has a high velocity that is always larger than
the circular velocity, thus prohibiting circular orbits. Also, by assuming
the star is bound, the direction of the normal vector to 
IRS 16CC's orbital plane
is restricted to a low inclination. 
}
\label{fig:pdfParams_example}
\end{figure}

The resulting stellar orbital parameters are constrained by several different 
factors.
First, a measured acceleration that is significantly different from zero, 
such as for S0-15, yields the best
determined orbit since the line-of-sight distance is confined to a small
range of values (Figure \ref{fig:pdfParams_example}, {\it top}).
Secondly, each star has a maximum allowed acceleration,
$a_{{\rho}, max} = |-GM/\rho^2|$, at the closest 
possible distance set by the observed projected radius. 
Stars with measured accelerations more than 3$\sigma$ below
the maximum allowed acceleration, such as IRS 16NW, 
have strong lower limits on their line-of-sight distances, which translate 
into significant constraints on the direction of the angular momentum
vector, $\vec{L}$, and can be equivalently expressed as constraints on 
inclination, $i$, and on the angle to the ascending node, $\Omega$ 
(Figure \ref{fig:pdfParams_example}, {\it middle}). 
Finally, even stars without significant limits on their 
line-of-sight distance from accelerations have some well
constrained orbital elements. In particular, $i$ and $\Omega$ 
are well constrained as a result of the precise measurements for the
stellar velocities and potential parameters.
Furthermore, if the star's total velocity is higher than the circular
velocity at the two-dimensional projected radius, then it is higher than the
circular velocity at {\it all} distances and only non-zero eccentricity 
orbits are allowed (Figure \ref{fig:pdfParams_example}, {\it bottom}).

The Monte Carlo analysis described above assumes that, in the absence
of an acceleration measurement, the acceleration should be drawn from a 
uniform probability distribution; or, in other words, we adopt a
uniform acceleration prior. For those stars that are
only in the extended sample, the Monte Carlo 
orbit analysis samples from this uniform acceleration prior 
ranging from the largest allowed acceleration by the
projected radius to the smallest allowed for the orbit to remain bound. 
For these stars and for stars in the primary sample with acceleration 
limits that are not significantly 
smaller than the maximum physically allowed acceleration,
the uniform-$a_\rho$ prior is an important assumption.
To test how sensitive our results are to this assumption, we performed
the same Monte Carlo analysis as detailed above using an 
alternative assumption that the prior acceleration distribution is
uniform in z,
which shifts the line-of-sight distance PDF to larger values when
compared with a uniform-$a_\rho$ prior. On a star-by-star basis, the
resulting orbital parameters are consistent within 1$\sigma$
for both priors, with one exception. The young star S0-14 has an 
eccentricity that is constrained to be higher than 0.93 (3$\sigma$)
with a uniform-$a_\rho$ prior, while with a uniform-z prior, all 
eccentricities are allowed within 3$\sigma$. S0-14 is distinguishable
from all other stars in our sample in that it has a total velocity of
only 50 km s$^{-1}$, as compared to 160-640 km s$^{-1}$ for the rest of 
the sample. 
Such a small velocity translates into a very large range of allowed 
line-of-sight distances which are not well sampled by a uniform-$a_\rho$ 
prior. S0-14's range of $i$ and $\Omega$ are not largely affected by the
choice of prior; therefore, we exclude S0-14 from our eccentricity analysis,
but we keep it in all other orbital analyses.

\begin{figure}
\epsscale{1.0}
\plotone{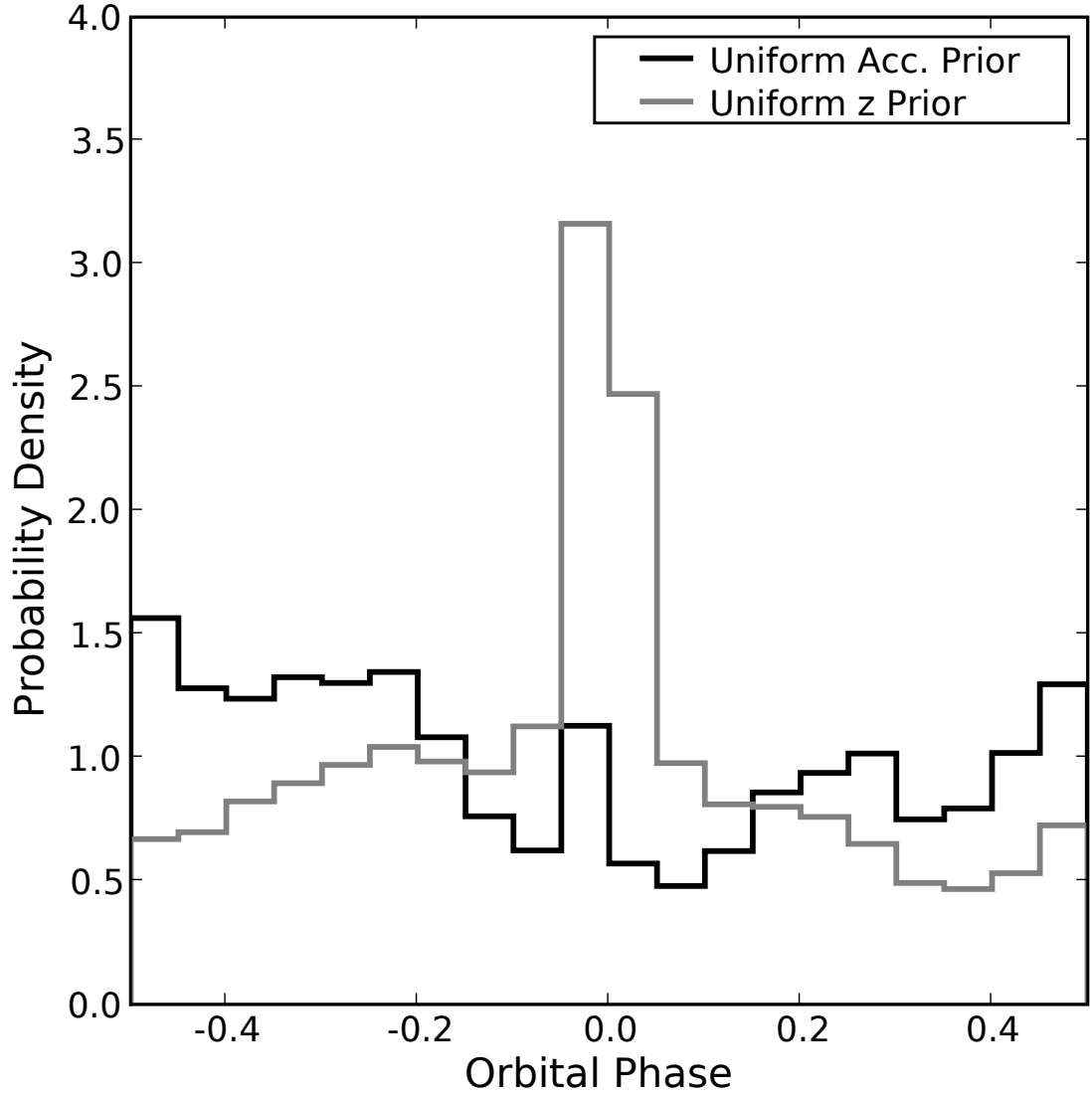}
\caption{The resulting distribution of orbital phases for all stars 
when assuming either a uniform acceleration prior ({\it black}) or
a uniform z prior ({\it gray}) and then imposing the measured
accelerations. The uniform z prior shows a strong bias
towards an orbital phase of 0, which corresponds to periapse; while
the uniform acceleration prior shows a more uniform distribution.
}
\label{fig:comparePriorsPhase}
\end{figure}

To distinguish between these two possible priors, we examine the resulting
distribution of orbital phases.
For a set of stars whose motion is dominated by the supermassive black hole
and that have been orbiting for more than a few
orbital time scales, the distribution of orbital phases should be 
uniform. 
The distribution of orbital phases for our sample is constructed by 
summing the orbital phase PDFs for all the stars. 
Figure \ref{fig:comparePriorsPhase} shows that 
while the uniform-$a_\rho$ prior produces a population that is 
uniformly distributed in orbital phase, the uniform-z prior produces 
a distribution that is strongly peaked at 0 (periapse) due to the
higher occurence of large line-of-sight distances that, for a given
velocity, creates an artificial bias towards periapse.
Such a strong bias towards periapse is unlikely to occur even if
some of the young stars reside in a gravitationally bound cluster,
such as IRS 13, where all the cluster members would have a similar
orbital phase.
Based on our assumption that the distribution of orbital phases
should be roughly uniform, we adopt a uniform-$a_\rho$ prior instead of
the uniform-z prior in the following sections.

\section{Orbit Results}
\label{sec:orbitResults}

\subsection{Detection of the Clockwise Disk}
\label{sec:diskDetect}

\begin{figure}
\centering
\includegraphics[scale=0.3,angle=90]{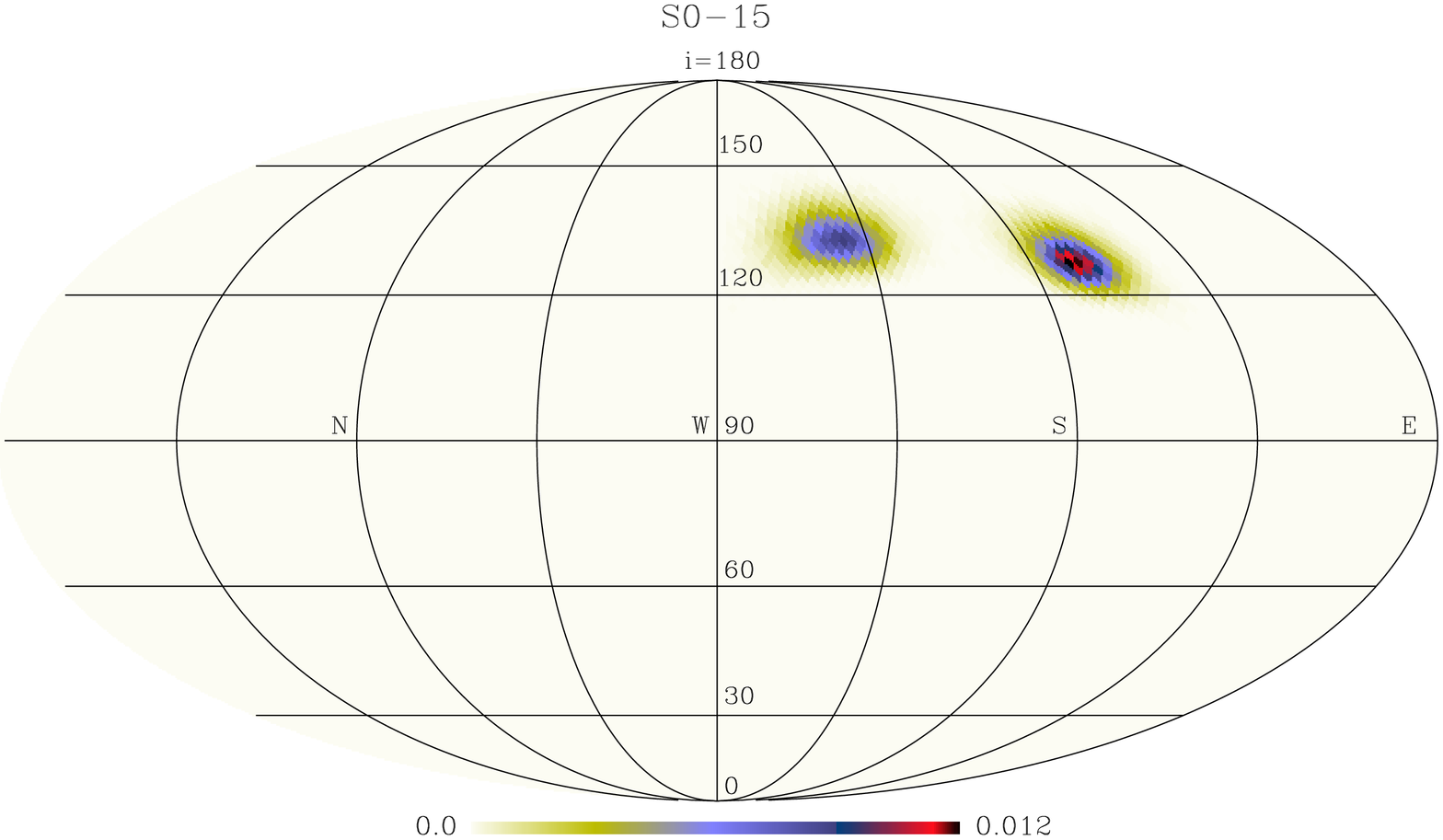} \\
\includegraphics[scale=0.3,angle=90]{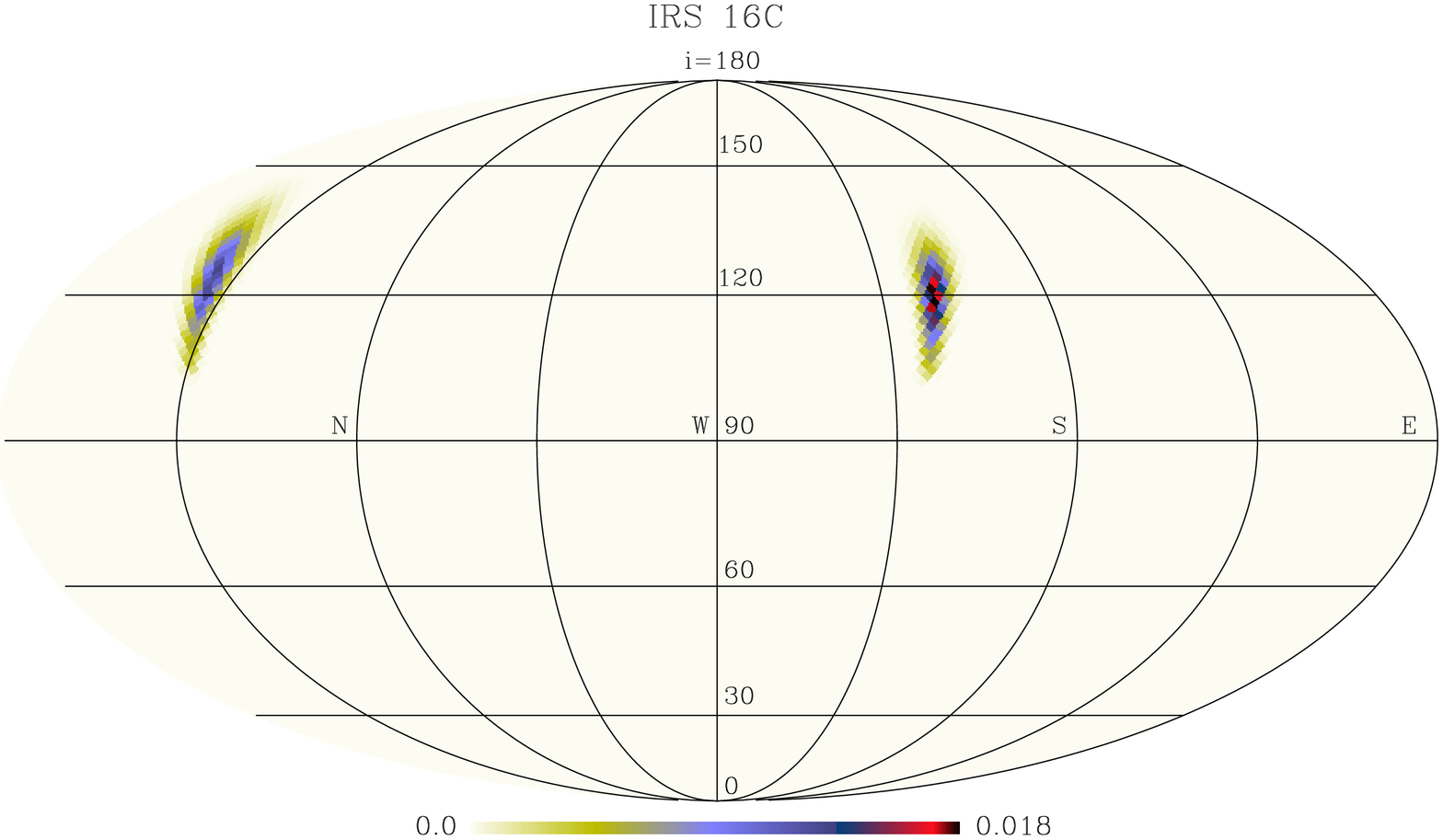} \\
\includegraphics[scale=0.3,angle=90]{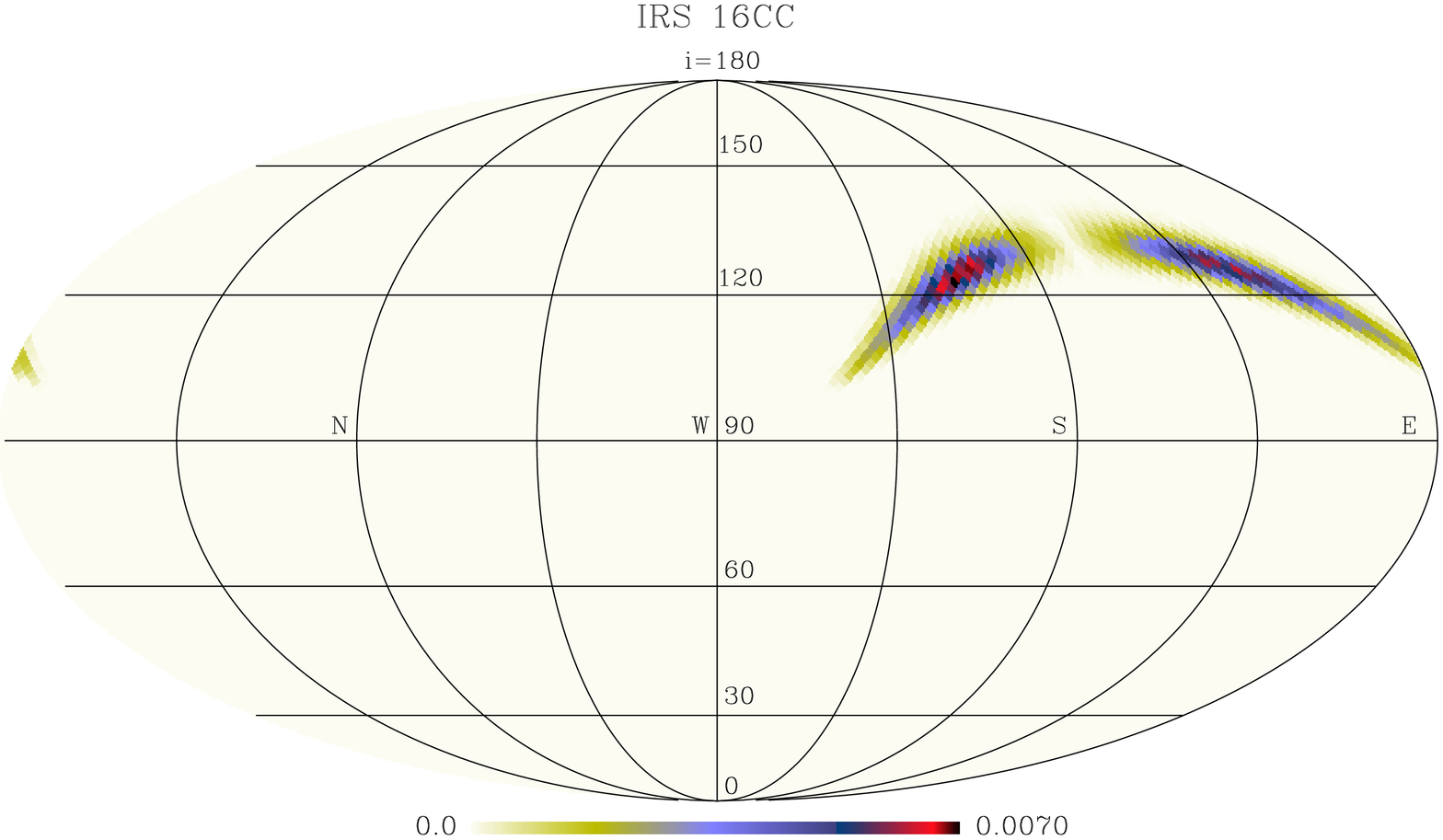}
\caption{
The orientation of three stars' orbital planes as described by the
probability distribution of the planes' normal vector projected
onto the sky as viewed from Sgr A*. 
Colors indicate the probability density for a star's
normal vector to point at each pixel on the sky.
The constraint on the stars' normal vectors are set
by ({\it top}: S0-15) a measured acceleration; ({\it middle}: IRS 16C) 
a significant acceleration limit; ({\it bottom}: IRS 16CC) the star's high 
velocity and assuming the orbit is bound.
}
\label{fig:iomap}
\end{figure}

\begin{figure}
\begin{center}
\includegraphics[scale=0.5,angle=90]{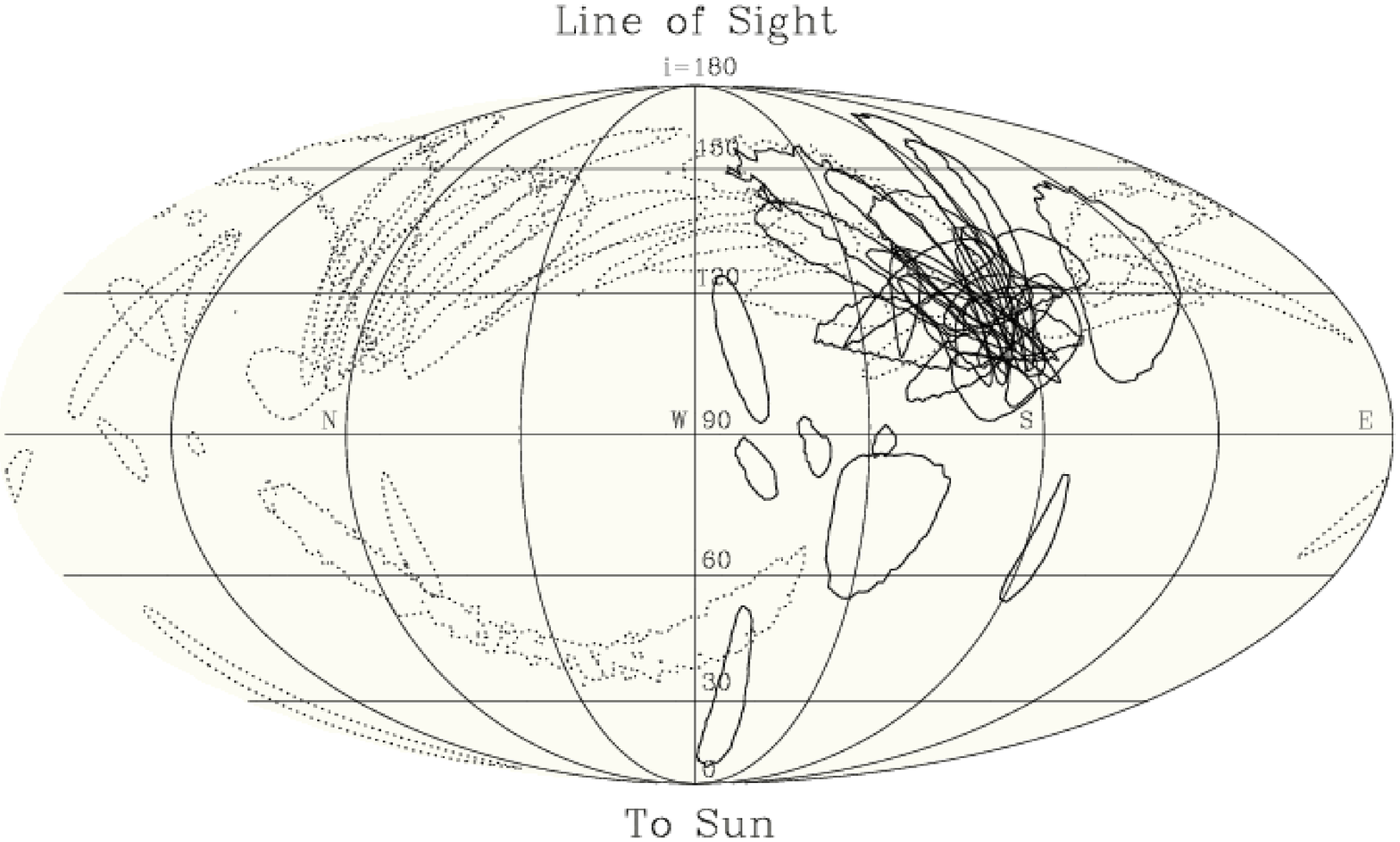} \\
\includegraphics[scale=0.5,angle=90]{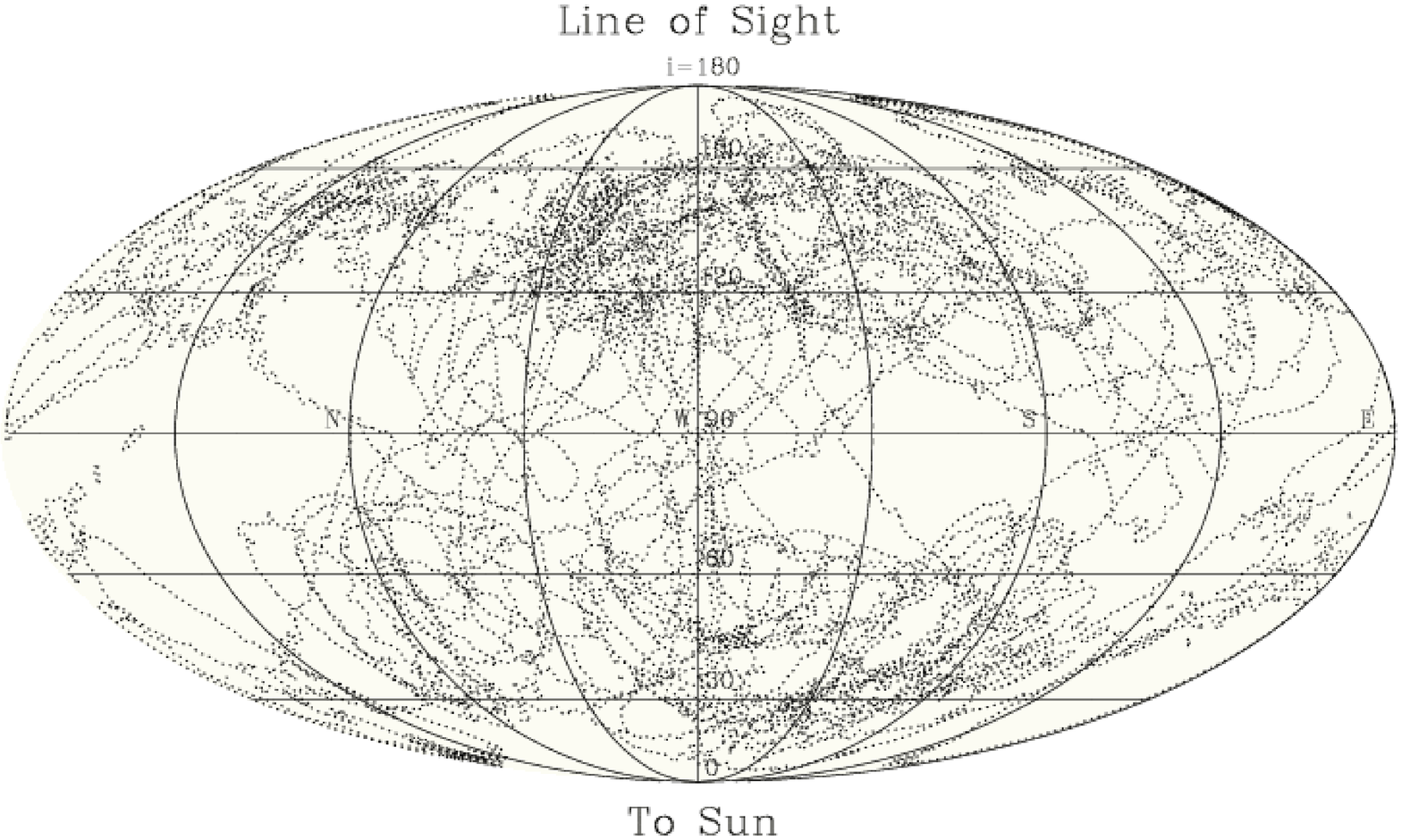}
\caption{The 1$\sigma$ contours of all stars' 
probability distribution functions for
the orientation of their orbital planes. This shows the distribution of
stellar orbit orientations around the sky. 
The primary sample is plotted on the {\it top} and
if there are degenerate solutions for a given star,
then one solution is plotted with a solid line and the other with a dashed 
line. Additional sources found only in the secondary sample are plotted on the 
{\it bottom} and are plotted with dashed lines as there are no acceleration
constraints and each star has a single solution with large uncertainties.
We note that the orientation of the projection shown in this figure is 
rotated by 180$^\circ$ with respect to that shown in earlier publications
\citep[e.g.][]{eisenhauer06,nayakshin06thick} in order to more easily
see the region around the proposed disks.
}
\label{fig:allPlanePDF}
\end{center}
\end{figure}

A large number of stars appear to share a common orbital plane
based on our analysis, which has no prior assumption about the 
existence of a disk.
The orientation of a star's orbital plane can be described by
a unit vector originating at Sgr A*'s position and pointing 
normal to the orbital plane ($\vec{n}$); and, this normal vector's
direction can be expressed by the inclination angle ($i$) and the angle to the 
ascending node ($\Omega$) using
\begin{equation}
\vec{n} = \left( \begin{array}{c} n_x \\ n_y \\ n_z \end{array} \right) = 
\left( \begin{array}{c} 
    \sin\, i\; \cos\, \Omega \\ 
    -\sin\, i\; \sin\, \Omega \\ 
    -\cos\, i\; \end{array} \right).
\end{equation}
The direction of each star's orbital plane normal vector is 
determined from the joint two-dimensional probability density 
function of $i$ and $\Omega$,
PDF($i$, $\Omega$), 
which is constructed by
binning the resulting $i$ and $\Omega$ values from the Monte Carlo
simulation in a two-dimensional histogram with equal solid angle bins using 
the HEALpix framework \citep{healpix}.
Figure \ref{fig:iomap} shows PDF($i$, $\Omega$)
projected onto the sky as viewed from Sgr A*
for the same three example stars shown in Figure \ref{fig:pdfParams_example}. 
Figure \ref{fig:allPlanePDF} shows, for all stars, 
the contours for the 68\% confidence region, which, 
on average, covers a solid angle of $SA_{\vec{n}} \sim$0.2 steradian (sr)
for the primary sample and 0.6 sr for stars found only in the extended sample, 
which have larger proper motion uncertainties.
Table \ref{tab:eccDisk} \& \ref{tab:eccDiskExtended}
list this solid angle, $SA_{\vec{n}}$, for each star in the primary and
extended samples.
The bound orbit assumption does not greatly impact the size of the 
$SA_{\vec{n}}$ because the orbital
parameters $i$ and $\Omega$ asymptote at large line-of-sight distances 
as can be seen in Figure \ref{fig:pdfParams_example}.
Stars with acceleration limits significantly smaller than $a_{{\rho}, max}$
have two isolated solutions because small line-of-sight distances (z)
are not permitted and at large line-of-sight distances the positive-z 
and negative-z solutions asymptote to
two different values of $\Omega$ (see Figure \ref{fig:pdfParams_example}).
Despite this degeneracy, the clockwise ($i$=90$^\circ$-180$^\circ$)
stars' normal vectors appear to cluster around a common point indicating 
that many of these stars lie on a common orbital plane. 

\begin{figure*}
\begin{center}
\includegraphics[scale=0.5,angle=90]{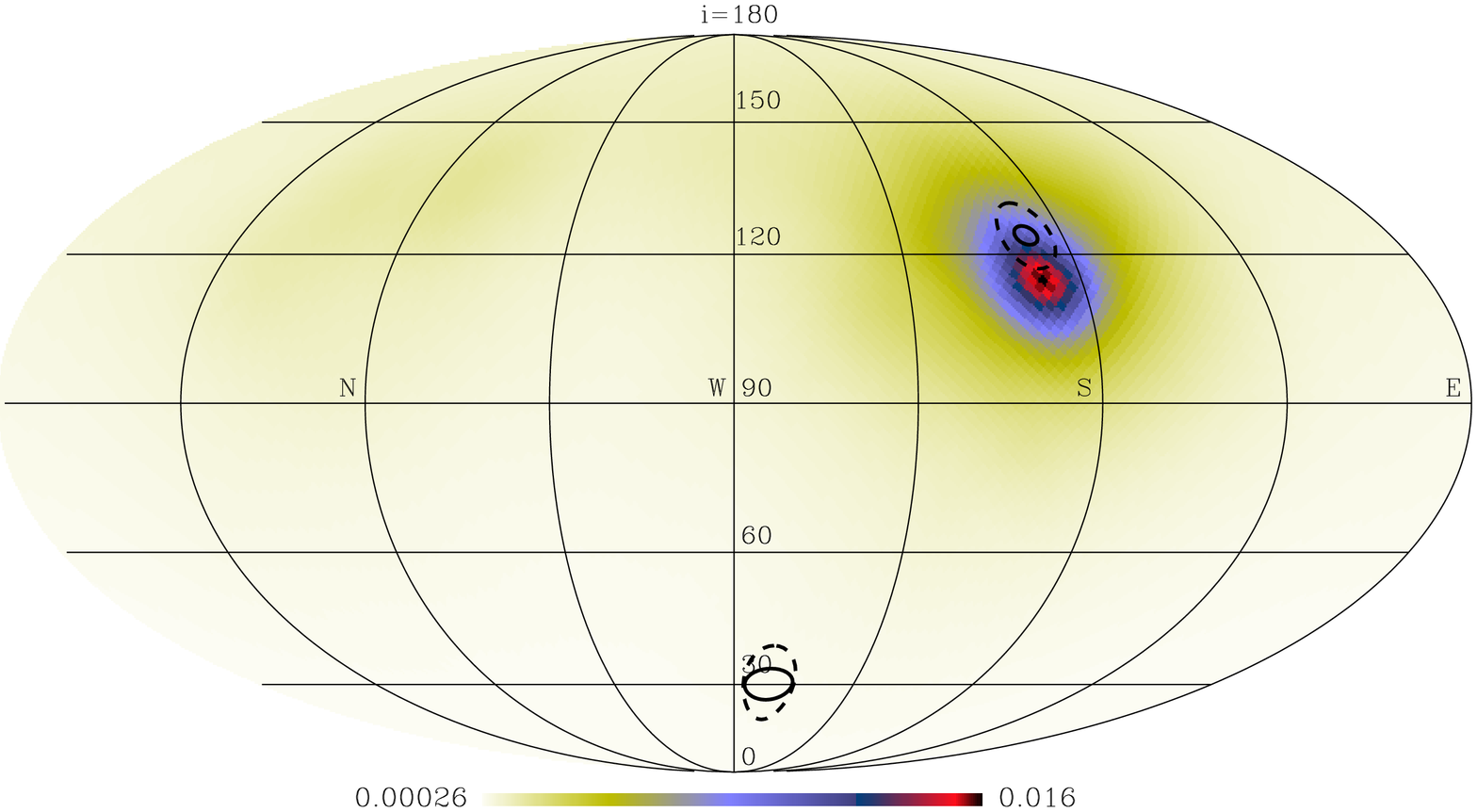} \\
\includegraphics[scale=0.5,angle=90]{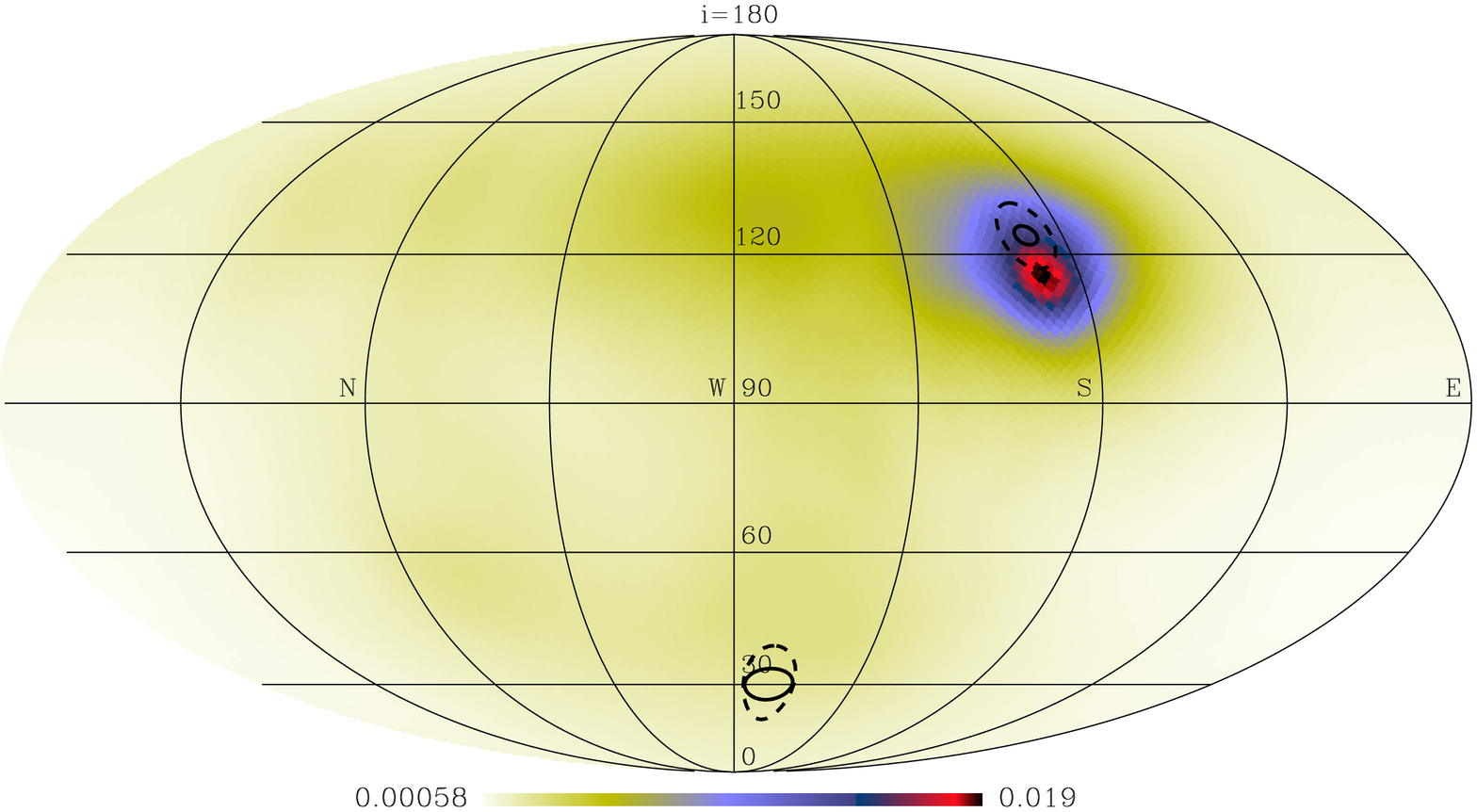}
\caption{The density of normal vectors to the orbital planes of the stars 
in our primary ({\it top}) and extended ({\it bottom}) samples. 
Densities are indicated in colors (stars deg$^{-2}$) on a linear scale and
the peak indicates an over-density of stars with similar orbital planes. 
Over-plotted in black are the candidate orbital
planes as proposed by \citet{levin03} and \citet{genzel03cusp} with
updated values from \citet{paumard06} for the candidate plane normal 
vector and uncertainties (solid black) and the disk thickness (dashed black)
shown as solid angles of 0.05 sr and 0.09 sr for the clockwise and
counter-clockwise disks respectively. 
}
\label{fig:orbitPlane}
\end{center}
\end{figure*}

The directions of the stars' normal vectors show a 
statistically significant clustering as measured by the 
the density of normal vectors in the sky as viewed from Sgr A*.
To quantify the density of normal vector directions, we use
a nearest neighbor density estimate, which is commonly used to identify
galaxy clusters \citep[e.g.,][]{nearestNeighbor}, and take the density
at each point on the sky to be 
\begin{equation}
\Sigma = \frac{k}{2\pi (1-\cos{\theta_k})} \textrm{stars sr}^{-1}
\end{equation}
where $\theta_k$ is the angle to the $k^{th}$ nearest star 
and $k$ is taken to be 6.
We calculate the expectation value for the density of normal vectors at
each point on the sky using the Monte Carlo simulation discussed earlier.
For each Monte Carlo trial, the sky is divided into 12288 equal area pixels
(0.001 sr) using a HEALpix grid and the 
density of normal vectors is calculated for each pixel.
These estimates are then averaged together over all the trials to provide
an average density per pixel on the sky.
The resulting average density of normal vectors is nearly the same 
for a choice of 4th, 5th, or 7th nearest neighbor.
Additionally, a similar analysis using a fixed 
aperture to calculate the density of normal vectors at each point on the
sky produced similar, but less smooth, results as the nearest neighbor
approach we adopt here. 
A peak in the density of normal vectors is detected at
$i = 115^\circ \pm 3^\circ$ and $\Omega = 100^\circ \pm 3^\circ$,
which provides direct evidence of a common orbital plane without any
prior assumptions (see Figure \ref{fig:orbitPlane}).
The uncertainty on the peak position is taken as the
half-width at half-maximum of the peak divided by the square-root of the 
number of stars that are candidate disk members, 
$\sqrt{N_{disk-stars}}$ (see below).
We also note that an analysis of the entire extended
sample produces a peak at the exact same position.
The mean density of normal vectors at the peak is 
0.016 stars deg$^{-2}$ with a negligible uncertainty on the mean value
($< 10^{-4}$ stars deg$^{-2}$).
The significance of the peak is determined by comparing the 
background density of normal vectors, which is defined by the 
average (0.001 stars deg$^{-2}$) and standard deviation (0.0008 stars deg$^{-2}$)
of all other pixels on the sky 
after first rejecting those pixels ($\sim$0.25 sr)
that are high outliers (more than three standard deviations).
The density peak is $\sim$19$\sigma$ above the observed 
background density.
A second comparison can be made to the density expected 
if the 32 stars in our sample were isotropically distributed over
4$\pi$ steradians.
The observed peak in the density is
$\gtrsim$20 times higher than this isotropic density. Thus we conclude
that there is a statistically significant common orbital plane of young stars.

The majority of the young stars that are orbiting in the clockwise direction
are likely to be orbiting in this common plane.
A comparison of each star's normal vector to the common plane's normal vector 
allows us to determine which stars are {\it not} on the common plane with 
high statistical significance. 
All other stars are then considered {\it candidate} members.
First, a preliminary estimate of the thickness of the common plane is
determined by defining the solid angle extent of the
plane, $SA_{plane}$, encompassed by the contour at which the  
density drops to half of the peak value. 
This corresponds to a region
with a solid angle of $SA_{plane} \sim$0.1 sr, which gives a half-opening 
angle of 0.2 radians (10$^\circ$) for a cone with the same $SA_{plane}$. 
Then, each star's probability density function, PDF($i$, $\Omega$),
is integrated over this region to determine the probability that the
star is a disk member.
The orientation of the stars' normal vectors have a wide range of 
uncertainties as expressed by the total solid angle covered by each star, 
so it is 
necessary to distinguish between those stars that have a low probability 
due to a large $\vec{n}$-uncertainty (i.e. large solid angle) 
vs. those stars that have a low probability
because they are significantly offset from the common plane.
Therefore, we normalize the above integrated probability by 
the probability at the peak of the star's PDF integrated over a region
that has the same total area as the common plane
\begin{eqnarray}
\mathrm{L}(\mathrm{not\;on\;plane}) & = & 1 - 
\frac{\int_{\mathrm{plane}} \mathrm{PDF}(i, \Omega)\;\; d\mathrm{SA}}
{\int_{\mathrm{peak}} \mathrm{PDF}(i, \Omega)\;\; d\mathrm{SA}} \\
\int_{\mathrm{plane}} d\mathrm{SA} & = & \int_{\mathrm{peak}} d\mathrm{SA}
\end{eqnarray}
where SA is the solid angle and L(not on plane) 
is the likelihood that the star is {\it not} on the common plane.
Those stars with likelihoods, L(not on plane), 
of greater than 0.9973 (equivalent to 3$\sigma$ for a gaussian distribution)
are flagged as non-members of the common plane.
The remaining set of stars are considered candidate members of the common plane.
Table \ref{tab:eccDisk} \& \ref{tab:eccDiskExtended}
list [1 - L(not on disk)] for each
star and Figure \ref{fig:PMimage} shows the positions of 
candidate members of the common plane in red and non-members in blue. 
Of the primary sample of 32 stars, 26 of which are orbiting in a clockwise 
sense on the plane-of-the-sky, we find that 22 are possible members of the 
common plane ($N_{disk-stars} = 22$).

The clockwise common plane that we measure is slightly offset from the 
clockwise planes proposed in earlier works.
Over-plotted in black on Figure \ref{fig:orbitPlane} is the candidate
orbital plane proposed by \citet{levin03} with
updated values from \citet{paumard06} for the candidate plane normal 
vector (solid black) and thickness (dashed black). 
The previously proposed plane was derived by minimizing a statistical metric,
K, in order to find the best-fit common orbital plane from the velocity
vectors of a sample of stars (see Appendix \ref{app:kmetric}). 
However, some stars are not members of the common plane and including 
them in the fit biases the result since they have extremely well 
measured velocities (S0-15, IRS 16C, S3-19).
For example, using the K metric
approach of \citet{levin03}, fitting all 26 clockwise stars in our 
primary sample gives i = 128$^\circ$ and $\Omega$ = 102$^\circ$ with K = 0.7,
which is closer to the disk found by \citet{paumard06} at
i = 127$^\circ$ and $\Omega$ = 99$^\circ$.
While fitting only the 22 stars that are consistent with the clockwise disk 
based on our orbit analysis gives i = 117$^\circ$ and $\Omega$ = 98$^\circ$ with
K = 0.2. Therefore, using the K metric to determine the common plane can 
produce biased results due to the inclusion of non-members. 
By combining position, velocity, and acceleration information in order to 
determine the orbital plane for each star, the direction of a common 
orbital plane can be estimated more robustly.

\begin{figure}
\epsscale{1.0}
\plottwo{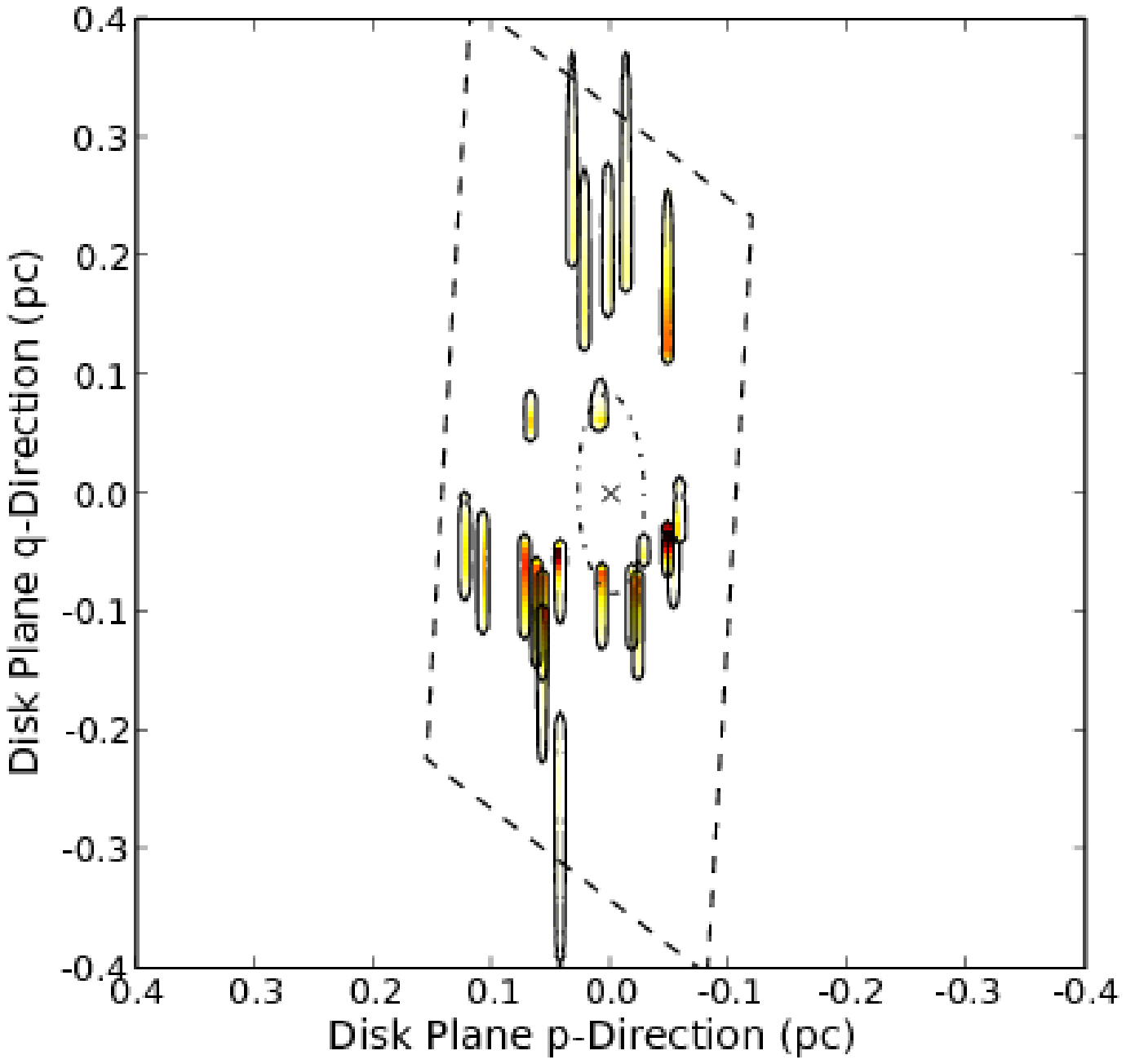}{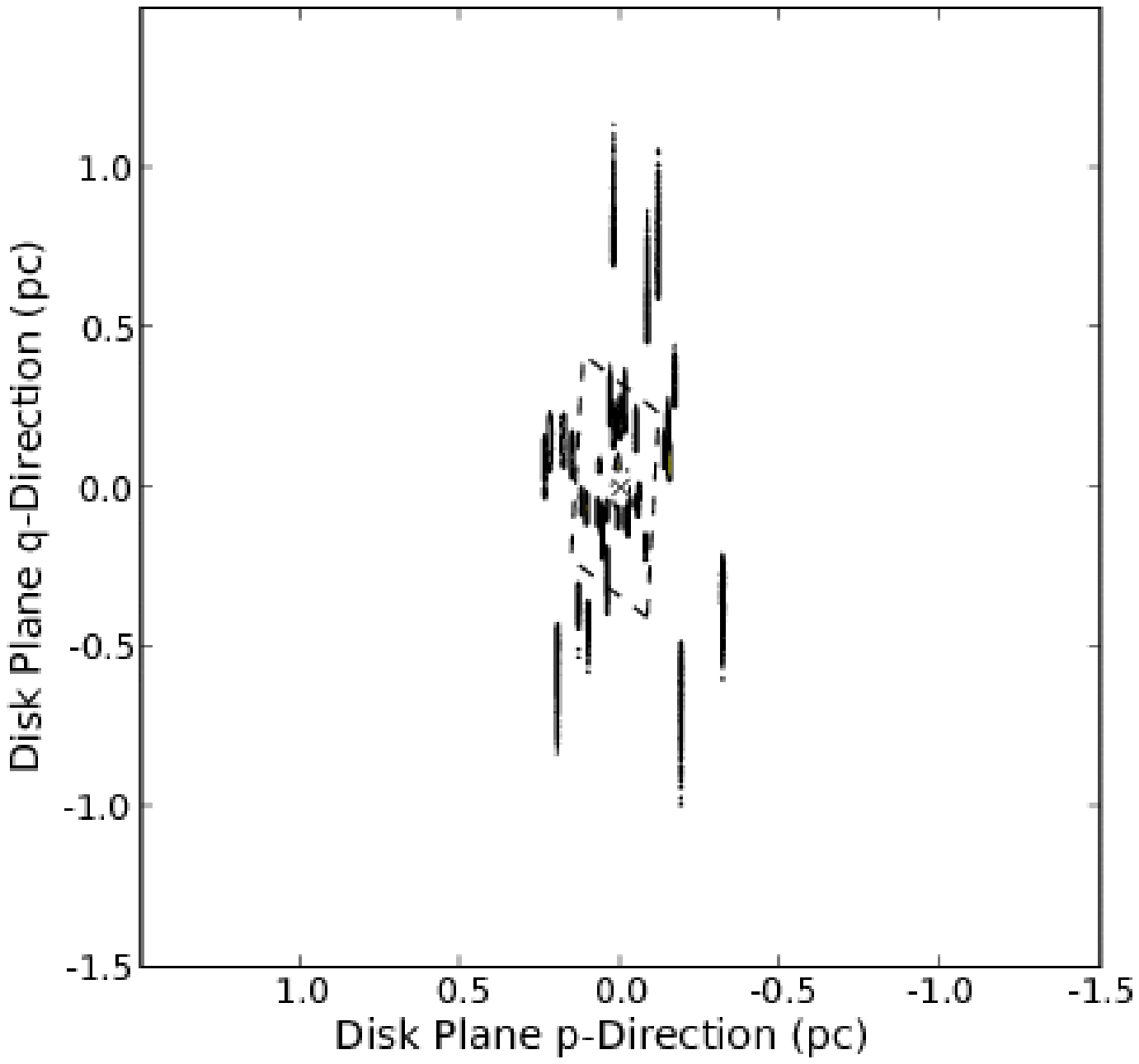}
\caption{
Positions for all candidate disk members in the disk plane from
our primary ({\it left}) and extended ({\it right}) samples.
The field of view for this study is projected onto the disk plane
and shows the outer ({\it dash}) and inner ({\it dot-dash}) boundaries.
For each disk candidate, a contour shows the star's position within 
the disk for all orbital solutions within 10$^\circ$ of the disk plane.
The color scale shows the probability density function for each star's 
position in the disk, normalized by the likelihood of disk membership. 
This normalization shows stars with a higher and lower likelihood of 
disk membership as {\it darker red} or {\it lighter yellow}, respectively.
}
\label{fig:diskVelRadius}
\end{figure}

The detected common orbital plane is composed of stars dispersed in a 
disk rather than in a single cluster as can be seen from the 
stars' positions within the common plane shown in 
Figure \ref{fig:diskVelRadius}. 
In this figure, the stars' positions have been 
converted into a disk coordinate system defined as 
[$\hat{p}$, $\hat{q}$, $\hat{n}$] where $\hat{n}$ is perpendicular 
to the disk plane, 
$\hat{p}$ is along the line of ascending nodes (where the plane of the sky 
intersects the disk plane), and $\hat{q} = \hat{n}\times\hat{p}$.
For each star, all orbital solutions that fall within 10$^\circ$ of the 
common orbital plane are combined to create
a probability distribution for the star's position in the disk, 
PDF($p$, $q$), which is shown in Figure \ref{fig:diskVelRadius} ({\it left}).
Each stars probability distribution is elongated in the $q$-direction
due to the large range of line-of-sight distances, $z$, that
are possible within the small range of possible disk inclinations for this
nearly edge-on plane of the disk. 
The thickness in the $p$ direction is largely set by the uncertainties
in the potential parameters ($M_\bullet$, R$_o$) and velocities.
The distribution of young stars within the plane shows a range of 
position angles on the plane, consistent with a stellar disk rather
than a stellar cluster.

\begin{figure}
\epsscale{1.0}
\plotone{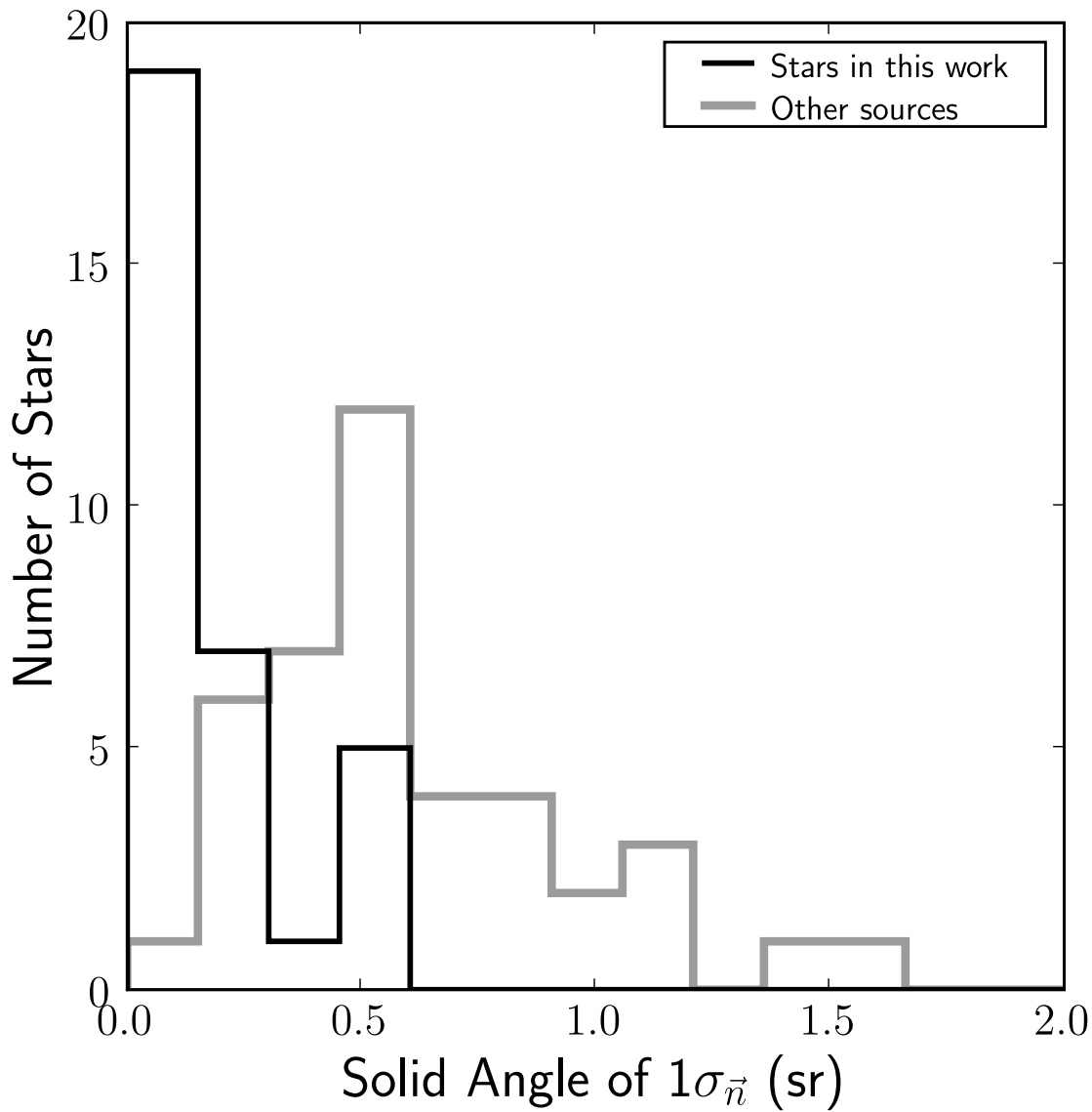}
\caption{The distribution of $\vec{n}$-uncertainties as expressed by
the area of the 1$\sigma$ region in which a star's normal vector can point.
The uncertainties are shown both for the sample in this work 
({\it black}) and for the stars with only three-dimensional position 
and two-dimensional velocity
information extracted from \citet{paumard06} that are used in the
search for a second disk ({\it gray}).}
\label{fig:histSolidAngle}
\end{figure}

The CW stellar disk is detected both in our analysis of the primary sample
and in a similar analysis of the entire extended sample.
The additional young stars in the extended sample have larger velocity 
uncertainties and no acceleration information, therefore the Monte Carlo 
orbit analysis samples from a prior probability distribution that is 
uniform in acceleration ranging from the largest allowed by the
projected radius to the smallest allowed for the orbit to 
remain bound. 
We note that even if we ignore the acceleration measurements for our primary
sample analysis, the CW stellar disk is still detected, although the 
significance is lowered from $\sim$19$\sigma$ to $\sim$8$\sigma$ above the 
background density of normal vectors. 
Thus the additional stars' orbits in the extended 
sample are still constrained (see Figure \ref{fig:histSolidAngle}),
even though they have larger uncertainties as compared to the stars in 
just the primary sample.
The density of normal vectors from the extended sample analysis shows a
peak within 1$^\circ$ of the disk's position from the primary sample.

The analysis of the extended sample shows that $\sim$50\% of the young
stars reside on the CW disk and there is no statistically significant 
change ($>3\sigma$) in the fractional number of disk stars at different radii.
For reference, the 73 young stars in the extended sample are
distributed on the plane of the sky with a surface density
that decreases with radius as $\rho^{-2.1 \pm 0.4}$.
Within a projected radius of 3'', the fraction of candidate
disk members is 72\% $\pm$ 9\% (18 out of 25) and
at projected radii larger than 3'', the fraction of candidate disk members
is 42\% $\pm$ 7\% (20 out of 48). 
Given the small number of known young stars, Poisson statistics indicate
that this change in the fraction of candidate disk members is only marginally 
statistically significant at the 2.6$\sigma$ level. 
Likewise, the projected surface density for the on-disk and off-disk 
populations shown no significant difference from each other or from that of
the total population.
Thus the number of candidate disk members does not change with radius and 
roughly half of the young stars reside on the CW disk.

The K-band luminosity function (KLF) of the young stars does not change 
significantly with radius or when considering stars on and off the
disk. To compare the KLF as a function of radius, the entire extended 
sample of young stars is divided into a near sample (r $<$ 3\farcs5)
and a far sample (r $\geq$ 3\farcs5) and the KLF is constructed for
each. A two-sample KS test yields a probability of 46\% that the
near and far samples have the same KLF. Similarly, the KLF is constructed
for stars on and off the disk and a two-sample KS test yields
a probability of 74\% that the on-disk and off-disk samples have the
same KLF. Finding more young stars will allow for a more detailed 
comparison of the KLF for different subsets within the young stars
population.

\subsection{Limits on Additional Stellar Disks}

In our primary sample, no common orbital plane is detected for the 
counter-clockwise population of stars; however, our sample is limited to 
six counter-clockwise
orbiting stars, only two of which (IRS 16NE, IRS 16NW) are claimed by
\citet{paumard06} to 
reside on the counter-clockwise disk.
Out of the 6 counter-clockwise stars in our primary sample, we find that only 
IRS 16NE and S2-66 could be consistent with the previously proposed 
counter-clockwise disk. 
The proposed counter-clockwise disk may have a larger
radial extent than is covered by our observations, 
so in order to fully explore whether
our lack of detection of a 2nd disk is due to our limited field-of-view, 
it is necessary to analyze the extended sample. 
As discussed in \S\ref{sec:orbitAnalysis}, 
the uniform acceleration prior adopted for this analysis tends to
overemphasize face-on orbital planes, making it easier to detect the 
proposed CCW disk, as \citet{paumard06} suggest it has an inclination 
of 24$^\circ$. 

Using the extended sample, our analysis of the density of normal vectors,
in the region of the proposed counter-clockwise disk, reveals no
significant over-density.
Of the 73 stars in the extended sample, at least 34 are not on the clockwise
disk and thus we compare the density observed in the region of the 
proposed counter-clockwise
disk to that expected for an isotropic distribution of 34 stars.
The observed density of normal vectors in the region of the 
counter-clockwise disk
is 2.4$\times$10$^{-3}$ stars deg$^{-2}$, which is only a factor of 3 above
what is expected for an isotropic distribution and is less than
1$\sigma$ above the background over the rest of the sky (excluding
the clockwise peak).
This density of normal vectors corresponds to only 3 stars 
within 19$^\circ$ of the putative CCW disk, 
where 19$^\circ$ is the disk thickness proposed
by \citet{paumard06}, and is consistent
with random fluctuations of an isotropic distribution having the
$\vec{n}$-uncertainties shown in Figure \ref{fig:histSolidAngle}. 
%
%
We estimate that this analysis is capable of revealing, at the 3$\sigma$ level, 
a stellar disk with more than 7 stars within a solid angle cone of radius =
19$^\circ$ at the location of the proposed CCW disk;
thus the proposed CCW disk containing 17 stars as suggested by
\citet{paumard06} should have been detected with this approach.

There are several principle differences between our analysis and 
that in earlier works. First, previous works make the {\it a priori}
assumption that a disk exists through the use of the statistical metric, 
K, and the results were not compared to a null hypothesis (i.e. no disk)
to establish the statistical significance of a disk detection.
Furthermore, the K metric used in previous works suffers from a bias which 
is described in Appendix \ref{app:kmetric}. 
The primary goal of our methodology
is to minimize the number of {\it a priori} assumptions and to 
fully quantify the significance of any disk detected as compared to the null
hypothesis that there is no disk. Therefore, we choose to search for disks
using all the young stars rather than first trimming out stars based on
projected angular momentum criteria or radii.
Also, we determine the range of allowed orbital orientations for each
star individually rather than searching for a disk from a statistical
sample of young stars. In this fashion, we utilize not only the 
direction information for a velocity vector, as has been used previously,
but also the physical relationships between the magnitude of the velocity 
and the positional information. This method allows for no disk to be
detected, while the previously used statistical tests assumed a disk
model and, therefore, must be compared
to the no-disk hypothesis using simulations of isotropic populations.
Without the simulations, the significance of any disk detection via
the K metric cannot be fully quantified. The resulting distribution
of orbits from our analysis is consistent with the 
hypothesis of a single, clockwise disk plus a more randomly
distributed population.

\subsection{Properties of the Clockwise Disk}
\label{sec:diskProperties}

We now examine, in detail, the properties of the detected clockwise disk.
With the identification of a single stellar disk and a 
candidate list of disk members, we investigate the following:
(1) the thickness of the disk,
(2) the radial profile of the disk,
(3) the azimuthal isotropy of the disk,
(4) the eccentricities of stars in the disk,
and (5) the luminosity function of the stars in the disk.
These properties are critical for distinguishing between {\it in situ}
and infalling cluster formation scenarios, as well as for understanding the
dynamical evolution of the young stars both on and off the disk.

The observed disk of young stars has a significant intrinsic thickness;
however, the vertical velocity dispersion is less than previously determined.
To measure the thickness of the disk, 
the dispersion of the velocities out of the plane (along the $\vec{n}$ 
direction) is calculated from all candidate disk members by projecting
each star's three-dimensional velocity vector along the
disk's normal vector to give $v_{\vec{n}}$.
The measurement uncertainties in both $\vec{v}$ and $\vec{n}$
are propagated through this coordinate transformation.
The intrinsic velocity dispersion is calculated using
\begin{eqnarray}
\sigma^2_{\vec{n},intrinsic} & \;=\; & 
\sigma^2_{\vec{n},measured} \;-\; \sigma^2_{\vec{n},bias} \\
\sigma^2_{\vec{n},intrinsic} & \;=\; & 
\left( \frac{1}{N_{disk-stars} - 1} \right) \left(
\displaystyle\sum_{i=0}^{N_{disk-stars}} v^2_{\vec{n},i} \;-\; 
\displaystyle\sum_{i=0}^{N_{disk-stars}} error^2(v_{\vec{n},i})
\right)
\end{eqnarray}
where the bias term, $\sigma_{\vec{v},bias}$, is 19 km s$^{-1}$ and accounts for 
added dispersion as a result of uncertainties in the measurements.
The resulting intrinsic velocity dispersion is 28 $\pm$ 6 km s$^{-1}$, which 
is significantly different from zero, thus a finite thickness is required. 
However, this velocity dispersion is a factor of 2 smaller than 
that found using the previously proposed disk solution of \citet{paumard06} 
and is slightly smaller than the value reported in \citet{beloborodov06} 
due to our improved identification of candidate disk members. 
The disk's
thickness can be expressed as the ratio of the vertical scale height to
radius, $h/r = \sigma_{\vec{n},intrinsic} / <|\vec{v}|>$, and is 0.08 $\pm$ 0.02.
Following a similar analysis to \citet{beloborodov06}, but with the above
relationship between $h/r$ and the velocity dispersion, the disk thickness 
can also be described using a gaussian distribution of inclination angles
about the disk plane with a standard deviation of
$\Delta\theta$ and is related to the scale height of the disk by
$h/r \sim \sqrt{1/2} \Delta\theta$. This yields a dispersion angle of 
$\Delta\theta = 7^\circ \pm 2^\circ$ for the young stellar disk. This
more rigorous determination of the disk thickness is consistent with the
thickness we derived in \S\ref{sec:diskDetect} from the half-width at 
half-maximum
of the peak in the density of normal vectors; thus the selection of the
candidate disk members is likely robust.
In comparison, the previously proposed disk solutions
yield a disk thickness of $h/r = 0.2$ ($\Delta\theta = 14^\circ$)
and $h/r = 0.1$ ($\Delta\theta = 9^\circ$) 
for \citet{paumard06} and \citet{beloborodov06}, respectively. 
We caution that all of these conversions from velocity dispersion to disk 
scale height and dispersion angle assume circular orbits and an isothermal 
disk structure. 
From our analysis, we note that the out-of-the-plane velocity dispersion 
shows no statistically significant variation with radius in the disk both 
for the primary (difference is $1\sigma \sim$ 7 km s$^{-1}$) and 
the full extended samples (difference is $1\sigma \sim$ 14 km s$^{-1}$).
Therefore, the observations are consistent with a thin disk of
uniform velocity dispersion at all radii.

\begin{figure}
\epsscale{1.0}
\plotone{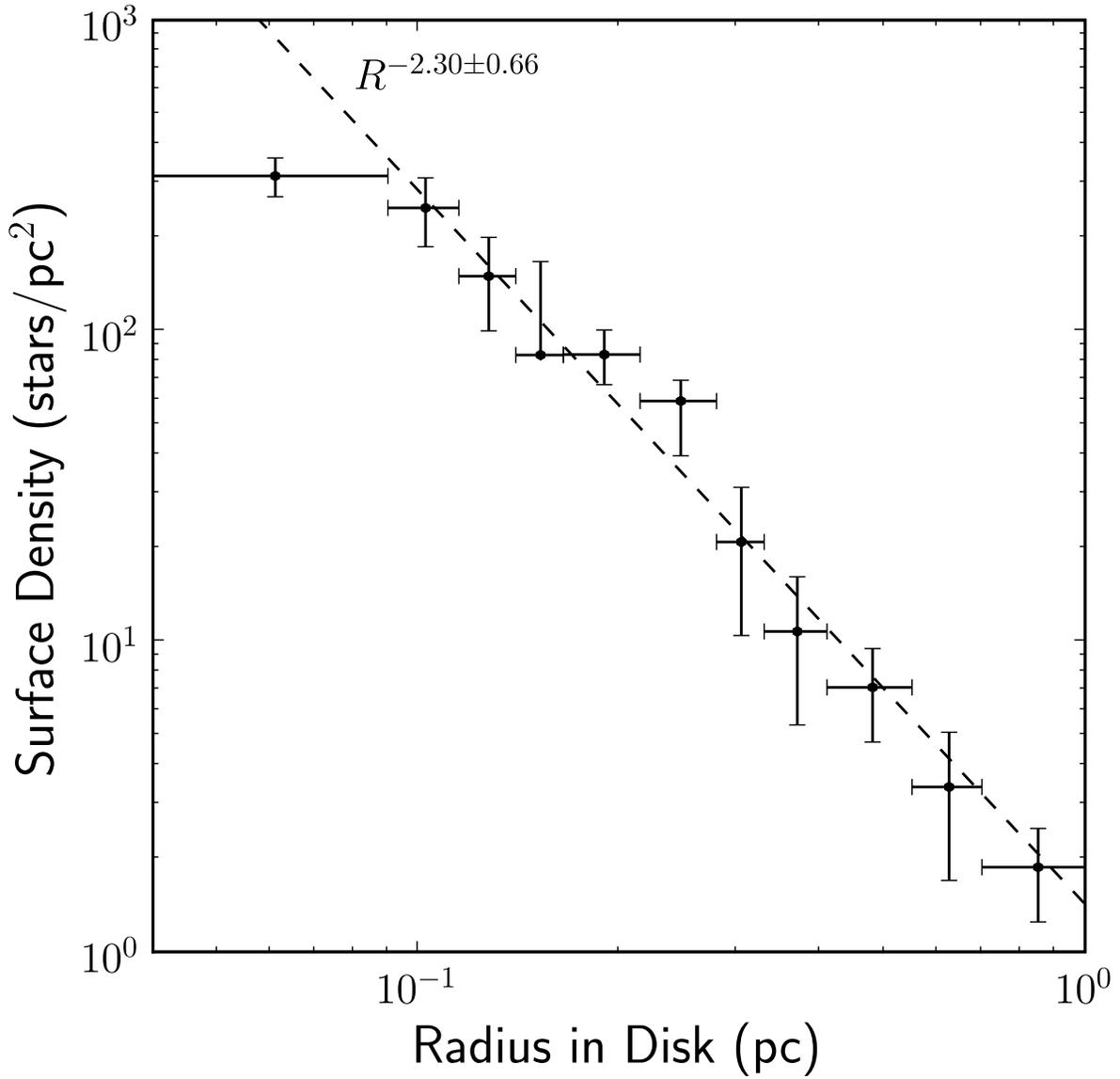}
\caption{
The radial distribution of stars within the disk plane for the 
extended sample. The best fit line is shown ({\it dashed}) and was
constructed by excluding the first data point and the last three data
points where field of view limitations may affect the distribution.
}
\label{fig:diskRadialDist}
\end{figure}

The surface density of stars in the disk falls off rapidly as a 
function of radius. In order to extend the radial coverage, we
consider the entire extended sample in this analysis.
The young stars that are candidate disk members have constraints
on their three-dimensional radii if we limit their orbital solutions
to those close to the disk plane. Thus the disk's surface
density can be determined as a function of three-dimensional radius
rather than just the projected two-dimensional radius as discussed at the 
end of \S\ref{sec:diskDetect}.
The distribution for each star's position within the disk plane, 
PDF($p$, $q$), is constructed from orbits that are within 10$^\circ$ 
of the disk and is shown in Figure \ref{fig:diskVelRadius}.
Then the disk's surface density at each radius is computed
numerically by sampling the PDF($p$, $q$) 10$^5$ times for all the candidate 
disk members and constructing a radial histogram for each trial. 
The radial histograms are combined for all the trials to find the peak
and 68\% confidence bounds for the expected number of stars at each radius.
This is converted into an azimuthally integrated
surface density by dividing by the area of a ring
at each radius. This method of constructing the surface density captures
both the measurement error in the individual stars and the finite 
thickness of the disk, which has not been incorporated into previous
estimates. The resulting azimuthally averaged surface density on the 
disk is shown for the extended sample in Figure \ref{fig:diskRadialDist} 
and has a best-fit power-law profile of $r^{-2.3\pm0.7}$. 
This is consistent with the previous results \citep{paumard06}, 
but our analysis accounts for the uncertainty in each stars
line-of-sight distance due to the finite disk thickness and, therefore
yields a larger uncertainty on the power-law index.

Visual examination of the stars' positions in the disk plane 
(Figure \ref{fig:diskVelRadius}) suggests there may be 
some anisotropy as evidenced by the clustering of stars on the lower 
part of the disk; however, this over-density is only marginally
statistically significant based on the following analysis.
In order to search for non-uniformities, 
we compare the observed stellar surface density of the
extended sample within the disk plane 
with the surface density expected for an azimuthally symmetric disk.
The observed stellar surface density is measured by 
sampling from all stars' PDF($p$, $q$) for 10$^5$ trials and calculating the 
stellar surface density over a grid of points in the disk plane 
for each trial. For each point on the
disk plane, the surface densities from all trials are combined, yielding
the most probable surface density with uncertainties.
The resulting two-dimensional map of observed surface densities is then 
compared to the expected surface densities for an azimuthally symmetric 
disk by subtracting the two values and dividing by the uncertainties.
This produces a surface density excess map that shows the significance
of any excess.
The disk shows a marginally significant ($\sim 3\sigma$) over-density 
on the front side (q $<$ 0) of the disk and a corresponding under-density on 
the back side (q $>$ 0).

\begin{figure}
\epsscale{0.8}
\plotone{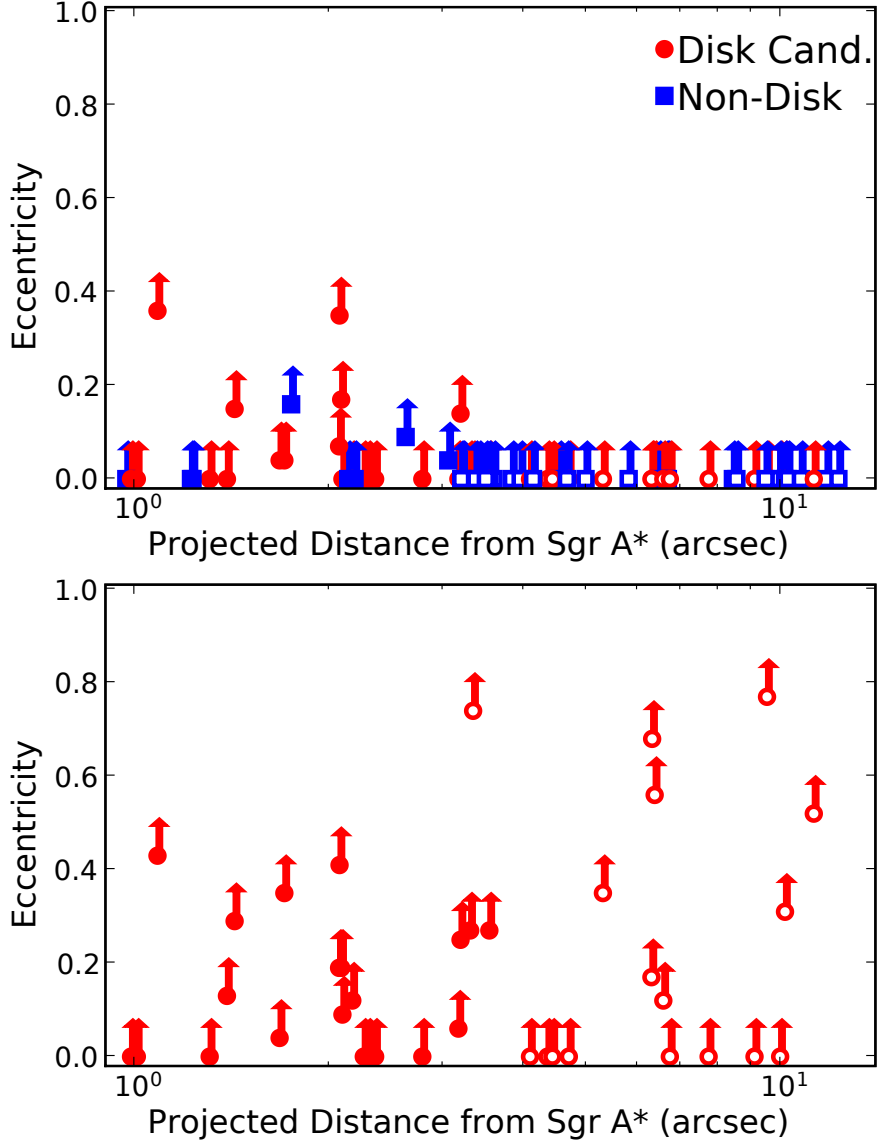}
\caption{The distribution of eccentricity lower limits
as determined from individual stellar
orbits, excluding S0-14. The top panel shows 
the 99.7\% confidence lower limit from all possible orbital solutions
for candidate disk members ({\it red circles}) and non-disk members
({\it blue squares}). Stars from the primary sample ({\it filled}) and 
stars added in the extended sample ({\it unfilled}) are both shown. 
Sources in only the 
extended sample have less constrained eccentricities due to their larger
velocity uncertainties. The bottom panel
shows the candidate disk members 99.7\% confidence lower limits after
restricting the orbital solutions to those with 
normal vectors within 10$^\circ$ of the disk. By assuming disk membership,
the range of eccentricities is more restricted for the candidate disk
members. 
}
\label{fig:eccentricity}
\end{figure}

A few candidate disk stars show evidence for eccentric orbits. 
To determine whether any of the stars' eccentricities are consistent
with a circular orbit, the six-dimensional probability density function for 
the orbital parameters is marginalized and re-expressed as a PDF for
the eccentricity vector (see Appendix \ref{app:orbits}),
PDF($e_x$, $e_y$, $e_z$). The magnitude of this
vector is the orbital eccentricity and the direction
points along the semi-major axis towards the periapse position. 
The PDF for the eccentricity vector cannot be further marginalized to 
produce a PDF of the eccentricity magnitude without introducing a bias
due to the positive, definite nature of a vector magnitude. This is the
same bias term as described in the velocity dispersion analysis; however,
unlike the velocities, the eccentricity distributions are strongly 
non-gaussian and the bias term cannot be easily accounted for in the
marginalization.
The peak of PDF($e_x$, $e_y$, $e_z$) gives the unbiased orbital 
eccentricity and the 99.7\% confidence interval of the three-dimensional 
distribution is used to determine the range for the 
one-dimensional eccentricity. 
Tables \ref{tab:eccDisk} and \ref{tab:eccDiskExtended}
show the 99.7\% confidence range of the eccentricities for all stars 
in the primary and extended samples. Also, 
Figure \ref{fig:eccentricity} shows the eccentricity 99.7\% confidence
lower limit
for the candidate disk members in red, non-disk members in blue, and excludes
S0-14 (see \S\ref{sec:orbitAnalysis}).
When considering all possible orbital solutions, the resulting eccentricity 
ranges show that 2 candidate disk members from the primary sample
have 99.7\% confidence eccentricity lower limits of greater than 0.2.
Restricting the possible orbital solutions to only those 
having normal vectors oriented within 10$^\circ$ of the disk normal
vector increases the number to 8 candidate disk members
with 99.7\% confidence eccentricity lower limits larger than 0.2. 

We find high-eccentricity
stars in the disk, similar to the analysis of \citet{beloborodov06} in
which they assumed an infinitely thin disk. However, our analysis 
incorporates the finite thickness of the disk and places statistical
errors on the eccentricities for individual stars.

\begin{figure}
\epsscale{1.0}
\plotone{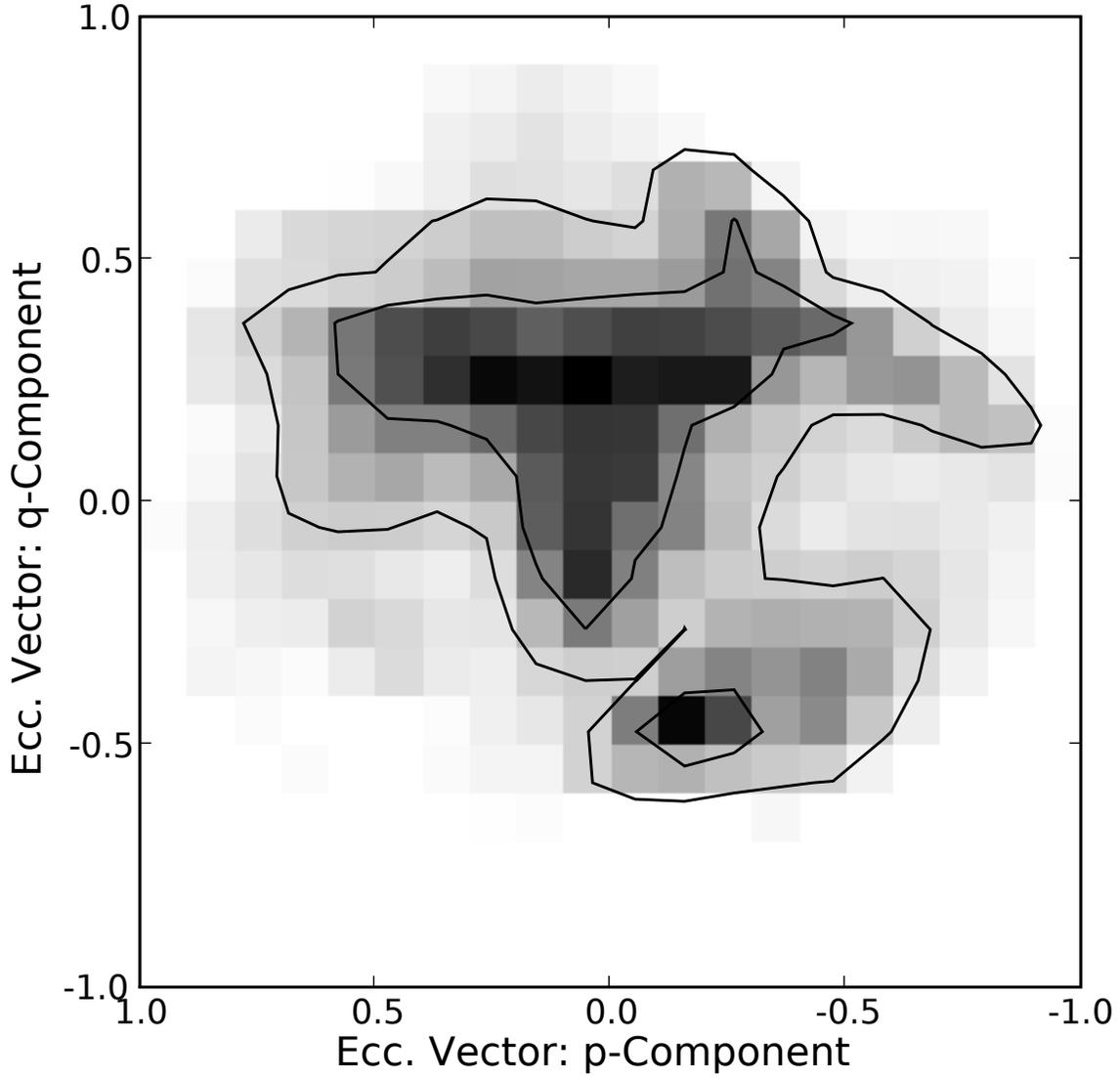}
\caption{Combined probability distribution for the candidate
disk stars' eccentricity vectors. The eccentricity vectors for 
orbital solutions with normal vectors within 10$^\circ$ of 
the disk's normal vector are projected onto the disk plane. 
The 1$\sigma$ and 2$\sigma$ confidence-level contours are shown in 
black.
}
\label{fig:eccVector}
\end{figure}

The average eccentricity of the entire population is not yet well 
constrained.
The eccentricity for the stellar disk is determined using the 
eccentricity vector. 
For each candidate disk member, orbital solutions are selected 
whose normal vectors point within 10$^\circ$ of the disk normal vector.
These orbital solutions are combined for all the disk stars by
averaging their PDFs to create
a combined probability distribution for all stars' eccentricity vectors, 
which is then projected into the disk plane and plotted in two dimensions
(Figure \ref{fig:eccVector}). 
This two-dimensional probability distribution gives an unbiased estimate
of the eccentricity magnitude
and shows that while the characteristic disk eccentricity peaks
at e=0.22, it is consistent with e=0.0 $-$ 0.8 at the 1$\sigma$ level, 
reflecting the large eccentricity uncertainties for the majority of the 
candidate disk members.

\section{Discussion}
\label{sec:discussion}

The kinematic analysis of the young stars in the central parsec
around our Galaxy's supermassive black hole has implications for the recent 
star formation history in this region.
Our first attempt at determining individual orbits for young stars that 
reside outside the central arcsecond shows definitive evidence for
the clockwise-rotating disk that was suggested
by \citet{levin03} and was subsequently refined by 
\citet{genzel03cusp} and \citet{paumard06}.
Our results do not show a statistically significant second disk.
The presence of a single stellar disk eliminates the need to 
invoke two distinct starburst events occuring roughly 6 Myr ago and 
greatly simplifies the demands on both {\it in situ} and infalling 
cluster scenarios. For instance, in the self-gravitating gas disk scenario,
the detection of only a single stellar disk lifts the requirement
for a second disk to rapidly build up gas, fragment, 
and form stars within 1-2 Myr of the formation of the first disk.
Likewise, for the infalling cluster 
scenario, the presence of only one stellar disk means that the 
frequency of such infall events is half that required for
the existence of two disks. 
On the strength of our confirming only one stellar disk, we consider 
whether all of the young stars within the central parsec may have formed 
in a single burst of star formation. 

Such a scenario must explain not only the observed
clockwise stellar disk, oriented at
$i\sim$115$^\circ$ and $\Omega\sim$100$^\circ$, but also the 
presence of roughly half of the young stars from our extended
sample on more isotropically distributed orbits out of the disk.
In the single starburst scenario, the out-of-the-disk stars could either
be generated during the formation process or could intially be in 
the disk and then perturbed through subsequent dynamical evolution.
Self-relaxation of the disk has not had sufficient time to produce
the out-of-the-plane population \citep{alexander07imf,cuadra08}, 
but other mechanisms have been 
proposed such as scattering by an inward-migrating IMBH \citep{yu07}. 
Currently, our results show that the
on-disk and off-disk populations of young stars look very similar
outside the central arcsecond (0.04 pc)
both in terms of the K-band luminosity function and the surface density 
profiles that decreases at larger projected radii as $\propto r^{-2}$.
However, the number of young stars in the disk drops at radii smaller 
than 0.08 pc; and at radii of $\lesssim$0.04 pc, none of the observed young 
S-stars are in the disk \citep{ghez05orbits,eisenhauer06}. 
This drop in the number of disk stars at
small radii may be the result of resonant relaxation or other dynamical
processes if the central arcsecond S-stars are a continuation of the
disk population \citep{hopman06}. 
Thus, if dynamical evolution produced the off-disk population, then
the dynamical process must not be a strong function of radius
beyond 0.08 pc.

Our distributions show that a potential problem with the single starburst 
scenario is the presence of the apparent massive star cluster, IRS 13, located 
$\sim$4'' from the supermassive black hole \citep{maillard04irs13,schodel05}. 
The cluster's orbit is not in the disk plane
and, given the proposed mass of IRS 13 ($>$10$^3$ \msun), it is
unlikely that it could have been ejected from the disk. 
However, the definition of IRS 13 as a cluster and the derived
mass is based on observations of only 3-4 bright stars and
is complicated by enhanced dust and gas emission in the vicinity. 
More data are needed to determine the total mass of IRS 13 and its
relationship to the disk stars.

Our results also have implications for the star formation mechanism.
For both infalling cluster and {\it in situ} formation
scenarios, we consider whether the observed characteristics 
of the young stellar disk can be explained. 
We observe a stellar disk with an out-of-the-disk velocity dispersion  
of 28 $\pm$ 6 km s$^{-1}$. Additionally, if we consider only orbital 
solutions within the disk (disk prior), we find that at least 8 of the 22 
candidate disk stars have 99.7\% confidence lower limits on the 
eccentricity of greater than 0.2.
Therefore, any formation scenario should explain not only a single thin 
stellar disk but also allow for non-circular stellar orbits of 
some stars in the disk.

First, for the infalling star cluster formation scenario, 
some of the disk properties we observe are well explained
and others appear difficult to reconcile with this model.
For instance, eccentric orbits are easily produced.
Stars that are stripped from a cluster as
it spirals in should have a similar inclination and 
eccentricity as the cluster itself. Therefore, an infalling cluster
with an initially eccentric orbit will produce a disk of stars with
similarly eccentric orbits \citep{berukoff06}. 
Previous studies have observed co-moving clumps of stars, 
such as IRS 16SW \citep{lu05irs16sw} and IRS 13 \citep{schodel05}, 
that appeared to support the infalling cluster formation scenario as 
they could be the remnant core of the dissipated cluster. 
We tentatively observe evidence for an over-density of stars on the 
front half of the disk at the position of the IRS 16SW co-moving group.
However this over-density may be explained by the effects of 
extinction that reduces the number of young stars identified 
on the back half of the disk at a given magnitude.
The extinction is highly variable throughout the region
and the back half of the disk is behind a patch of higher exctinction 
\citep[$\Delta$A$_K$ = 0.3 - 1.4; ][]{scoville03,schodel07}.
Thus the apparent overdensity on the front half of the disk,
corresponding to the IRS 16SW co-moving group, can perhaps be ascribed to
differential extinction.
More data are needed to confirm the observed disk asymmetry and to determine 
whether the cause is extinction.
Our results yield a steep radial profile for the young stars in 
the disk, as also found by \citet{paumard06}, which appears to be 
inconsistent with the flatter profile expected for an infalling cluster
\citep[$r^{-0.75}$, ][]{berukoff06}. 
We note that mass segregation is observed in massive star
clusters that are only a few million years old 
\citep{hillenbrand98,fischer98,stolte06}. Any mass segregation that existed
prior to the cluster's dissolution may impact the observed radial profile 
as the massive stars would have resided preferentially in the cluster core 
and would therefore have been deposited at the smallest radii. 
Thus, the massive stars O stars that we observe today 
may have a steeper radial profile than the entire young star population.
Additionally, the lack of X-ray emission from pre-main-sequence stars 
\citep{nayakshinSunyaev06} is not well explained by an infalling cluster 
model.
A larger and deeper survey for young stars over the central $\sim$5 pc could
definitively rule out this scenario if the tidal tails of the disrupted
clusters are not detected.

Some theories of {\it in situ} star formation
take place in a circular gas disk. Such a gas disk
can be built up from a steady inward migration of
material or from many small cloud-infall events and the disk 
would circularize prior to becoming massive enough to form stars from 
self-gravity ($>10^4$ \msun). Such a formation scenario would most 
likely produce a steep radial profile in agreement with our observations.
Our observations of over 30\% of the
candidate disk members with eccentricities greater than 0.2 appears to be
inconsistent with an initially circular disk of stars and a normal
initial mass function.
A disk of stars on initially circular orbits and with a normal IMF
will relax over 6 Myr and produce a thermal distribution of 
eccentricities with an rms eccentricity of 0.15 or less \citep{alexander07imf}.
For such a disk, only 4 out of 22 stars should have eccentricities higher than 
0.2, compared with the 8 out of 22 observed when a disk prior
is imposed on the primary sample.
Therefore, in order for the disk to have been initially circular with 
a normal-IMF, some additional dynamical processing other than self-relaxation
is needed. Other possibilities are that the initial mass function may
have been top-heavy, the binary fraction may have been extremely high,
or IMBHs could have formed, all resulting in faster relaxation
to higher eccentricities, but these are not 
sufficient to explain the out-of-the-disk population of young stars
\citep{alexander07imf,cuadra08}. The gas disk formation scenario may be
modified \citep{alexander07imf,cuadra08} to accommodate the observed high 
stellar eccentricities and out-of-the plane population by 
building up a massive gas disk in a single cloud infall or a cloud-cloud
collision event, in which the clouds are on eccentric orbits
\citep{sanders98,vollmerDuschl01,nayakshin07sims}.
The gas disk would then have a high eccentricity for a short period
of time during which stars might form \citep{alexander08,bonnell08}. 
The cloud-cloud collision scenario may yield both a thin stellar disk and
a more distributed population of stars at larger radii
with a range of angular momenta as a result of the complex interactions 
and shocks during the collision. 
It is also conceivable that a cloud-cloud
collision scenario might give rise to out-of-the-disk clumps of gas
that could form a cluster such as IRS 13. 
Refined estimates of the eccentricity and inclination distributions
of the young stars and more detailed theoretical
analysis are needed to investigate the viability of this scenario.

\section{Conclusions}

In summary, the advent of laser guide star adaptive optics has allowed
us to retroactively improve our 11 year astrometric data set used for
monitoring stars orbiting our Galactic Center. This has increased our
proper motion precision, with resulting uncertainties of $\sim$3 km s$^{-1}$,
and allowed us, for the first time, to make measurements of and place
limits on accelerations for stars outside the central arcsecond
out to a radius of 3\farcs5, with typical 3$\sigma$ acceleration 
limits of -0.19 mas yr$^{-2}$.
By combining our improved stellar positions and proper motions with radial
velocity information from the literature, we compute orbits
for individual young stars proposed to lie in stellar disks orbiting
the supermassive black hole.
The orbits for the young stars confirm only a single disk
of young stars at a high inclination rotating in a clockwise sense and
there is no statistically significant evidence for a second disk.
Stars within the well-defined, clockwise disk
have an out-of-the-disk velocity dispersion of 28 $\pm$ 6 km s$^{-1}$ 
and several stars have high eccentricities. These disk properties
suggest that star formation may have occurred in a single event, rather
than the two events previously needed to explain two stellar disks; however,
there are open questions as to how $\sim$50\% of all young stars can be
perturbed out of the disk plane and whether 
the apparent compact cluster, IRS 13, which is not part of the stellar
disk, requires a separate star formation or dynamical event.
Future directions include (1) obtaining new LGSAO 
data sets with improved astrometry to measure accelerations for the
young stars at all radii and (2) identifying new young stars within
the central parsec in order to better constrain the orbital 
properties of these stars and to study in detail the 
distribution of eccentricities and semi-major axes for stars both in and 
out of the disk.

\acknowledgements

Support for this work was provided by NSF grant AST-0406816, and
the NSF Science \& Technology Center for AO, managed by UCSC
(AST-9876783).
Additional support for J.R.L. was provided by
a NSF Graduate Research Fellowship. 
We would like to thank Brad Hansen
and the anonymous referee for helpful comments.
The W.M.~Keck Observatory is operated as a scientific
partnership among the California Institute of Technology, the University
of California and the National Aeronautics and Space Administration.
The Observatory was made possible by the generous financial support of
the W.M.~Keck Foundation.

{\it Facilities:} \facility{Keck:II (NIRC2)}, \facility{Keck:I (NIRC)}

%
%
\begin{appendix}

\section{NIRC Speckle Distortion}
\label{app:speckDistort}

\begin{figure}
\epsscale{0.7}
\plotone{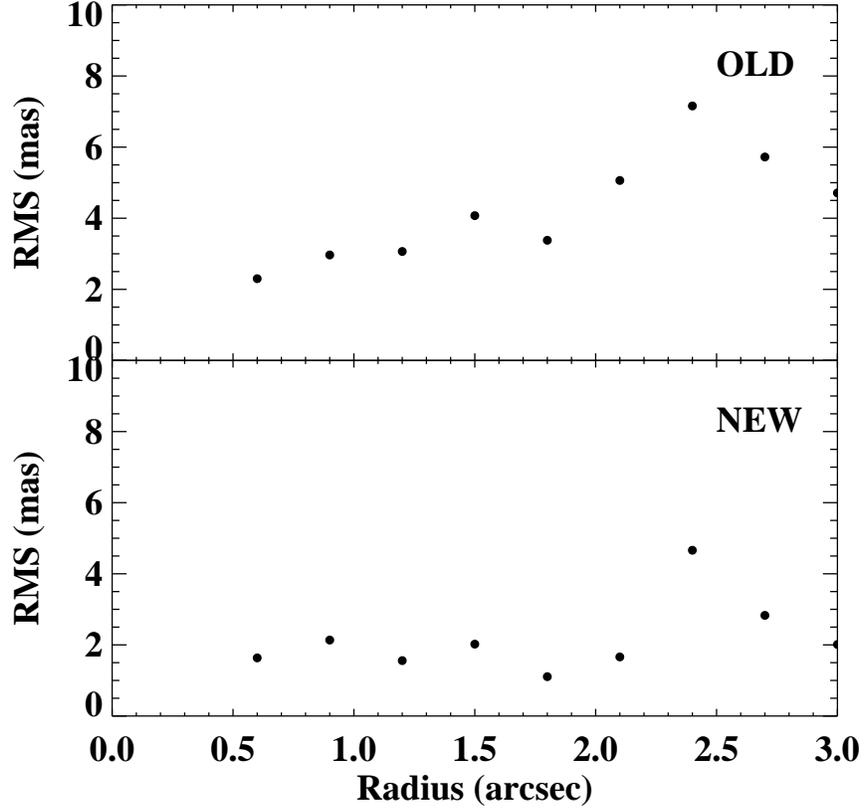}
\caption{
The improvement in positional accuracy at large radii as a result
of correcting geometric distortion in speckle data sets. 
To characterize the systematic positional uncertainty, we take
each star at each epoch and calculate the residual positional offset,
which is defined as the difference between the 
measured position and the position as determined by the best fit velocity
($x = x_o + v*\Delta t$).  Then the RMS of the residuals is calculated
across all epochs for each star. All stars' resulting RMS values are sorted
by the distance between the star and Sgr A* (which was at the center of
the images) and then averaged over radius bins of 0\farcs3. 
The radial trend is shown for 
data prior to the new distortion correction (top) and after the new 
distortion correction (bottom). 
}
\label{fig:distort}
\end{figure}

In the speckle data sets, optical distortions, introduced by the NIRC
reimager, are small near the center of the field-of-view where 
Sgr A* was positioned, but grow to dominate the positional 
uncertainties for stars located more than $\sim$0\farcs5 from Sgr A*
(see Figure \ref{fig:distort} and \S\ref{sec:images}).
Now, utilizing images of the Galactic Center obtained with NIRC2,
which has optical distortions characterized at the $\sim$2 mas level
\citep{ghez08},
we can, for the first time, similarly quantify and correct 
the optical distortions in the NIRC reimager speckle data sets.

\begin{figure}
\begin{center}
\epsscale{1.0}
\plottwo{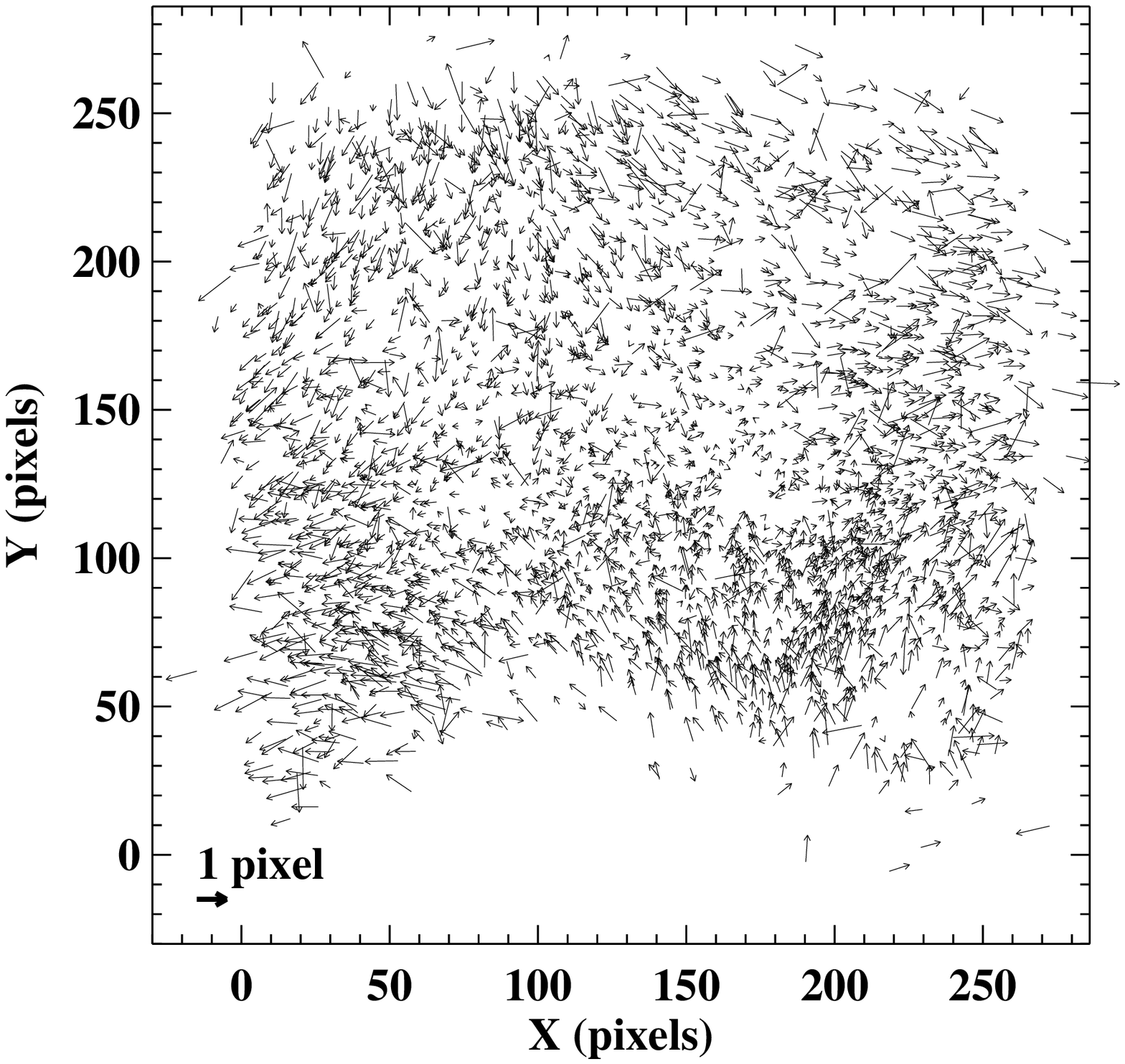}{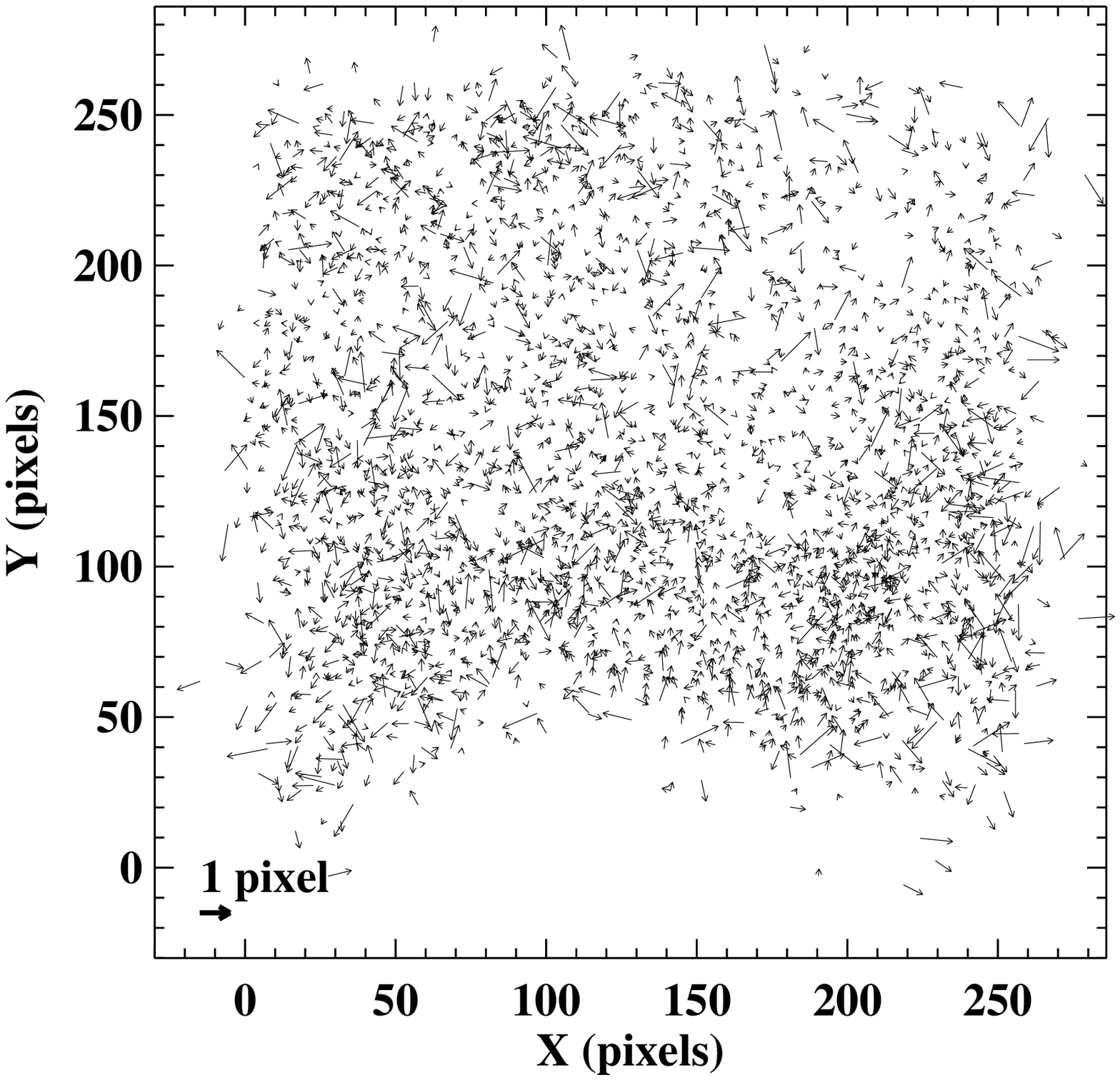}
\caption{Map of the positional differences between stars 
observed near-simultaneously with NIRC and NIRC2. 
The maps are plotted in the original NIRC detector coordinates
and show residuals before {\it (left; a)} and after {\it (right; b)}
the NIRC-reimager distortion solution.
}
\label{fig:distort2d}
\end{center}
\end{figure}

\begin{figure}
\epsscale{1.0}
\plotone{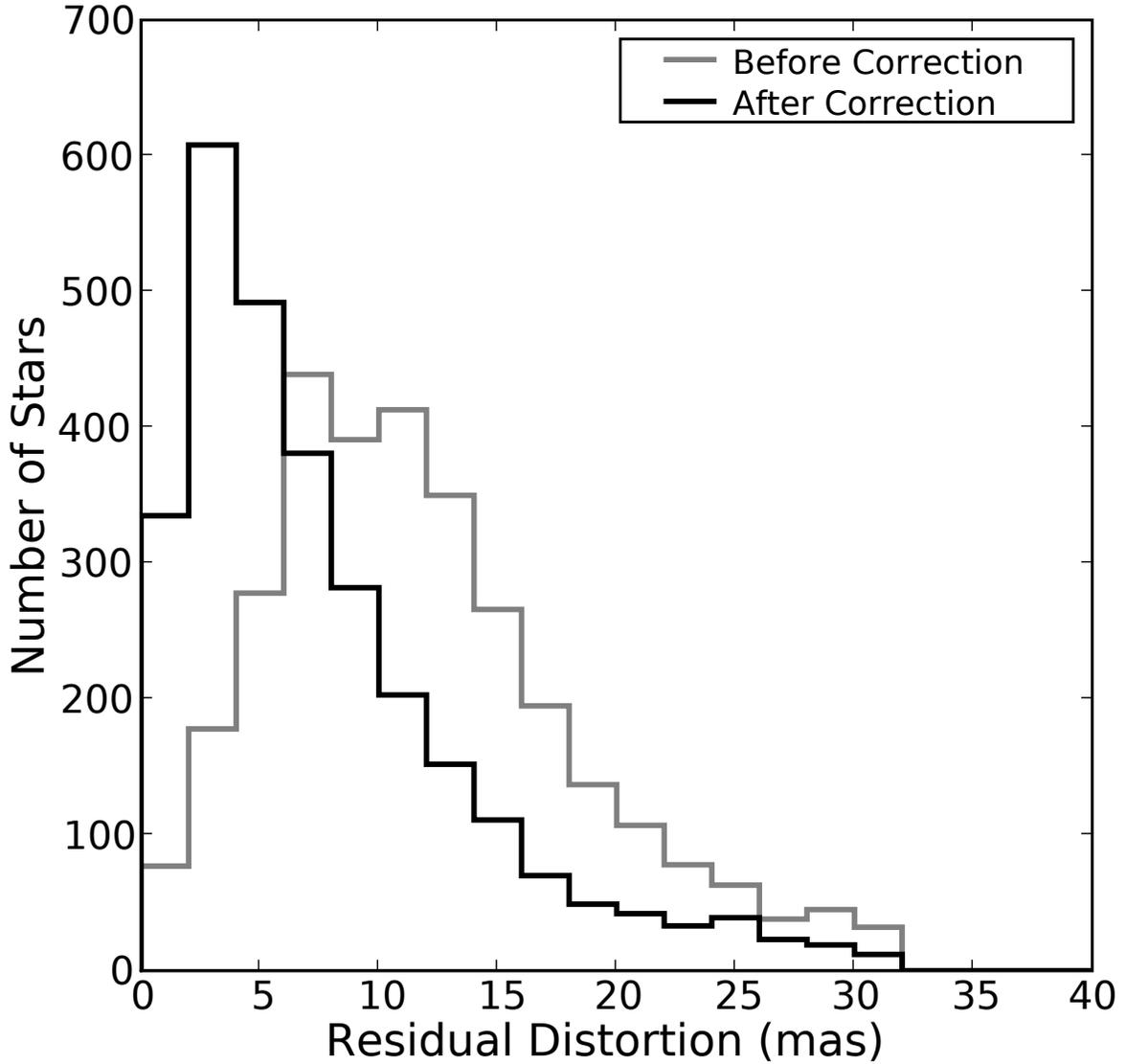}
\caption{Distribution of the residuals before ({\it gray}) and
after ({\it black}) correcting for the NIRC image converter
distortion. Residuals are calculated by comparing a star's position in
each NIRC image stack to the position in the LGS AO/NIRC2 image. 
These residuals are further reduced in the final image because the
stacks are dithered small amounts on the detector and residual
distortion can be averaged out if it is randomly oriented over
the scale of the dither.
}
\label{fig:speck_distort}
\end{figure}

Images of the Galactic Center were obtained with both NIRC and NIRC2 on
consecutive nights during July 2004 
and the NIRC2 images were used as a reference coordinate system.
The individual NIRC speckle exposure times are only 
0.1 seconds and have insufficient signal-to-noise to detect more than the
brightest 5 stars. Exposures were obtained in sets of 100 and each set is
combined to produce a single image in which approximatly 100 stars are 
detected. It is assumed that the images are mostly stationary on the
NIRC detector during each set of exposures. For each stacked image, the stars'
positions are compared to those in the NIRC2 image and the offsets are 
mapped into NIRC detector coordinates 
(see Figure \ref{fig:distort2d}, {\it left}). 
In this fashion, a distortion map is built up from many stacks of images
which are dithered and rotated such that stars fall on many different 
positions on the detector. 
The distortion solution was obtained by fitting the distortion map with 
polynomials of the form
\begin{equation}
(x^\prime + 128) = a_0 + a_1(x - 128) + a_2(y - 128)
\end{equation}
\begin{equation}
(y^\prime + 128) = b_0 + b_1(x - 128) + b_2(y - 128)
\end{equation}
where the best-fit distortion parameters are listed in Table 
\ref{tab:distort}. 
The new distortion solution improves the RMS residual errors
per stack by a factof of 3 to $\sim$3 mas
(Figure \ref{fig:speck_distort}), which is further reduced in the 
final image by averaging the dithered stacks.
Higher-order polynomial terms did not sufficiently improve the 
fit to warrant inclusion. The above solution is applied 
after the initial application of the standard NIRC distortion correction. 
The map of positional differences between stars in the NIRC and NIRC2 
images before and after the NIRC-reimager distortion correction is shown
in Figure \ref{fig:distort2d} ({\it right}). 
The resulting radial dependence on the RMS
positional uncertainty is greatly improved and is shown in 
Figure \ref{fig:distort}, which plots many stars' RMS residual offset from 
their best-fit proper motions across all epochs. In the final analysis
of the speckle data, the relative astrometric uncertainty is $\sim$2 mas.

\section{Analytic Orbit Equations}
\label{app:orbits}

The orbit of a star in a known point source potential can be derived from a 
single measurement of a star's orbital state vector.
At epoch t$_{ref}$, the orbital state vector is usually
described by the star's position, $\vec{r}$, and velocity, $\vec{v}$,
relative to the central mass. For the analysis in this paper, 
the state vector is estimated using measurements of the three-dimensional
velocity, $\vec{v} = [v_x, v_y, v_z]$, and the
projected position, $\vec{r}_{2D} = [x, y]$, and $z$ is derived from
the radial acceleration on the plane of the sky. For brevity, we have
removed the $ref$ subscript notation and all of the above variables are 
measured at t$_{ref}$.
Orbital trajectories are then inferred from conservation of
energy, specific angular momentum, and eccentricity 
($\epsilon$, $\vec{h}$, $\vec{e}$),
which are related by $\vec{e} \cdot \vec{h} = 0$ and 
$|e|^2 - 1 = 2\,\epsilon\,h^2 / GM$ giving 5 constants of motion 
plus an undetermined reference time. 
Equivalently, the orbital trajectory can
be expressed using the standard Keplerian orbital elements:
period ($P$), eccentricity ($e$), time of periapse passage (T$_{\circ}$),
inclination ($i$), position angle of the ascending node ($\Omega$), 
and the longitude of periapse 
\citep[$\omega$; see Equations \ref{eqn:i}, \ref{eqn:e}, \ref{eqn:w}, 
\ref{eqn:o}, \ref{eqn:p}, \ref{eqn:t0} and][for detailed descriptions of these orbital parameters]{ghez05orbits}.
The 3D position and velocity state vectors can be used to calculate
the orbit of the star around the black hole (by algebraic manipulation of 
Kepler's Laws). 
Here we present the analytic expressions used to compute the orbital elements
from the state vectors.

Orbit determination for the young stars in our sample is tractable because
the mass and position of the black hole are determined by independent means,
namely the well determined orbits of stars much closer to the black hole.
The coordinate system is set such that Sgr A* resides at the origin,
$\hat{x}$ and $\hat{y}$ increase with right ascension and declination,
and $\hat{z}$ increases with the line-of-sight distance from the Earth
to Sgr A* with z=0 at the location of the black hole. 
Combining the two state vectors, $\vec{r}$ and $\vec{v}$, and the
black hole mass, there are three intermediate vectors that describe the
geometry of the orbit both in three-dimensions and projected onto the plane
of the sky. These are (1) the specific angular momentum vector, 
$\vec{h}$, which points normal to the plane of the orbit, (2) the
eccentricity vector, $\vec{e}$, which points in the direction of periapse, 
and (3) the ascending node vector, $\vec{\Omega}$, which points to where
the star passes through the plane of the sky moving away from us, 
and are given by
\begin{eqnarray}
\vec{h} & = & \vec{r} \times \vec{v} \\
\vec{e} & = & \frac{\vec{v} \times \vec{h}}{GM} - \frac{\vec{r}}{|\vec{r}|} \\ 
\vec{\Omega} & = & \vec{h} \times \hat{z}.
\end{eqnarray}
The semi-major axis can also be calculated as an intermediate quantity
\begin{eqnarray}
a & = & \left ( \frac{2}{|\vec{r}|} - \frac{|\vec{v}|^2}{GM} \right )^{-1}.
\end{eqnarray}
Then the five standard orbital parameters that describe the shape and 
period of the orbit are then
\begin{eqnarray}
i & = & \arccos{ \left ( \frac{-\vec{h} \cdot \hat{z}}{|\vec{h}|} \right )}
\label{eqn:i} \\ 
e & = & |\vec{e}| 
\label{eqn:e} \\ 
\omega & = & \arccos \left ( \frac{(\hat{z} \times \vec{h}) \cdot \vec{e}}
  {|\hat{z} \times \vec{h}||\vec{e}|} \right ) \quad
  (\textrm{if } \vec{e} \cdot \hat{z} < 0 \textrm{ then } 
  \omega = 2\pi - \omega) 
\label{eqn:w} \\ 
\Omega & = & \arctan \left ( \frac{\vec{\Omega} \cdot \hat{x}}
  {\vec{\Omega} \cdot \hat{y}} \right )
\label{eqn:o} \\
\left ( \frac{P}{[yr]} \right ) & = & \sqrt{ \left ( \frac{a}{[AU]} \right )^3 
  \left ( \frac{[M_\sun]}{M} \right ) } 
\label{eqn:p} \\ 
\end{eqnarray}
where i = 0 if the orbit is in the plane of the sky and $\Omega$ is 
measured East ($\hat{x}$) of North ($\hat{y}$).
The remaining orbital parameter is the epoch of 
periapse passage and can be computed in a number of different ways. 
We first compute several intermediate quantities of interest such as the
Thiele-Innes constants (A,B,C,F,G,H), and the eccentric anomaly as shown below:
\begin{eqnarray}
A & = & a(\cos \omega \cos \Omega - \sin \omega \sin \Omega \cos i) \\ 
B & = & a(\cos \omega \sin \Omega + \sin \omega \cos \Omega \cos i) \\ 
F & = & a(-\sin \omega \cos \Omega - \cos \omega \sin \Omega \cos i) \\ 
G & = & a (-\sin \omega \sin \Omega + \cos \omega \cos \Omega \cos i) \\ 
\cos E & = & \frac{Gr_y - Fr_x}{AG - BF} + e \\ 
\sin E & = & \frac{Ar_x - Br_y}{AG - BF} \frac{1}{\sqrt{ 1 - e^2 }} \\ 
E & = & \arctan \left( \frac{\sin E}{\cos E} \right ).
\end{eqnarray}
And finally the epoch of periapse passage are calculated from 
these intermediate quantities using
\begin{eqnarray}
T_o & = & t_{ref} - \frac{P}{2\pi} (E - e \sin E).
\label{eqn:t0}
\end{eqnarray}

\section{K Metric}
\label{app:kmetric}

The previously proposed planes were derived by minimizing a metric that
\citet{levin03} call $\chi^2$, but we call K, and which is defined as
\begin{equation}
K = \frac{1}{N-1}\displaystyle\sum^N_{i=1} 
\frac{(\vec{n}\cdot\vec{v_i})^2}{(n_x\sigma_{v_{x,i}})^2 + 
(n_y\sigma_{v_{y,i}})^2 + (n_z\sigma_{v_{z,i}})^2}
\end{equation}
where $N$ is the number of stars, $\vec{v}_i$ is the velocity of each star,
$\sigma_{v_{x,i}}$, $\sigma_{v_{y,i}}$, $\sigma_{v_{z,i}}$ are the velocity
uncertainties for each star, and $\vec{n}$ is the normal vector to the disk 
plane that is found in the fitting process.
This metric is used to find, statistically, the best-fit common orbital
plane from the velocity vectors of a sample of stars.
The K metric suffers from several shortcomings. 
First, the K metric is described as a $\chi^2$ metric; however, 
standard $\chi^2$ minimization takes the form of 
(data - model)$^2$/(data errors)$^2$ where the data errors have no 
dependency on the model parameters. The K metric includes the model
parameters in the data-error term and does not necessarily have an 
expectation value of 1 for normal errors. The appropriate function
to minimize in order to find the best-fit common orbital plane 
can be derived from maximum likelihood theory if we assume that the
likelihood function is given by
\begin{equation}
L = \prod_{i=1}^N \frac{1}{\sqrt{2 \pi \sigma_i^2}} 
\exp{ \left [ - \frac{(\vec{n} \cdot \vec{v}_i)^2}{2 \sigma_i^2} \right ]}
\end{equation}
where $\sigma_i$ depends on the disk model parameters that are 
being sought by
\begin{equation}
\sigma_i^2 = (n_x \sigma_{v_{x,i}})^2 + 
(n_y\sigma_{v_{y,i}})^2 + (n_z\sigma_{v_{z,i}})^2.
\end{equation}
Standard practice is then to take the logarithm of the likelihood, $L$, and 
minimize the resulting function in Equation \ref{eq:minfunc} in order to find 
the best fit disk model parameters. The above likelihood function then becomes
\begin{eqnarray}
\ln L & = & -\frac{N}{2} \ln (2\pi) - \sum_{i=1}^N \ln \sigma_i 
+ \sum_{i=1}^N -\frac{ (\vec{n} \cdot \vec{v}_i)^2 }{ 2 \sigma_i^2 } \\
-2 \ln L & = & N \ln (2 \pi) + 2 \sum_{i=1}^N \ln \sigma_i +
\sum_{i=1}^N \frac{ (\vec{n} \cdot \vec{v}_i)^2 }{\sigma_i^2} \label{eq:minfunc}
\end{eqnarray}
and the first two terms are constant and do not factor into finding an 
extremum in the above equation. The third term on the right-hand side is the 
K metric previously used to determing the disk parameters. However, the
second term on the right-hand side also depends on the free parameters 
in $\vec{n}$ and must be included in the minimization process. 
This extra term that has not
previously been included in the disk fitting process has the full form
\begin{equation}
\ln{ \sqrt{(n_x\sigma_{v_{x,i}})^2 + (n_y\sigma_{v_{y,i}})^2 + 
(n_z\sigma_{v_{z,i}})^2}}
\end{equation}
and standard chi-squared probability functions cannot be applied. 
Second, even when accounting for the extra term, the metric can still
introduce substantial bias. In particular, radial
velocity uncertainties, $\sigma_{v_{z,i}}$, are larger than the proper motion
errors by a factor of 2 on average in previous publications. During 
K-minimization, this over-weights solutions with a larger $n_z$ resulting
in a bias against edge-on planes. Finally, in order to properly 
evaluate the probability of obtaining a given 
value of the K-metric by random chance, one must perform simulations of
an isotropic distribution of stars. However, such simulations are extremely 
sensitive to the input distribution of semi-major axes and eccentricities
which are not yet well constrained by observations.
Thus, when utilizing such statistical tests for finding a common orbital 
plane, it is difficult to compare to the null hypothesis -- an isotropic
distribution of stars -- and to quantify the significance of a disk.

\end{appendix}

\clearpage

%
%

\begin{deluxetable}{llrrrrrrrl}
\tabletypesize{\scriptsize}
\tablewidth{0pt}
\tablecaption{List of Observations}
\tablehead{
  \colhead{Date\tablenotemark{a}} &
  \colhead{Filter\tablenotemark{b}} &
  \colhead{t$_{exp,i}$} &
  \colhead{Frames\tablenotemark{c}} &
  \colhead{FWHM} &
  \colhead{Strehl} &
  \colhead{Number} &
  \colhead{K$_{turnover}$\tablenotemark{d}} &
  \colhead{Pos. Error\tablenotemark{e}} &
  \colhead{Data Source\tablenotemark{f}} \\
  \colhead{} &
  \colhead{} &
  \colhead{(sec)} &
  \colhead{Used} &
  \colhead{(mas)} &
  \colhead{} &
  \colhead{of Stars} &
  \colhead{(mag)} &
  \colhead{(mas)} &
  \colhead{}
}

\startdata
1995.439 & K  & 0.12 & 1562 & 58 & 0.06 & 124 & 15.2 & 1.1 & speckle; (ref. 1) \\
1996.485 & K  & 0.13 &  857 & 60 & 0.03 &  71 & 13.5 & 1.7 & speckle; (ref. 1) \\
1997.367 & K  & 0.13 & 1834 & 61 & 0.05 & 116 & 15.2 & 1.1 & speckle; (ref. 1) \\
1998.251 & K  & 0.15 & 1645 & 62 & 0.04 &  81 & 12.9 & 1.4 & speckle; (ref. 2) \\
1998.366 & K  & 0.14 & 2096 & 69 & 0.05 & 120 & 15.1 & 1.2 & speckle; (ref. 2) \\
1998.505 & K  & 0.14 &  936 & 63 & 0.07 & 101 & 15.6 & 1.7 & speckle; (ref. 2) \\
1998.590 & K  & 0.14 & 1914 & 62 & 0.06 & 139 & 15.5 & 0.9 & speckle; (ref. 2) \\
1998.771 & K  & 0.14 & 1085 & 56 & 0.07 & 111 & 15.4 & 1.1 & speckle; (ref. 2) \\
1999.333 & K  & 0.14 & 1848 & 72 & 0.08 & 136 & 15.5 & 1.1 & speckle; (ref. 2) \\
1999.559 & K  & 0.14 & 2092 & 57 & 0.10 & 141 & 15.6 & 0.8 & speckle; (ref. 2) \\
2000.305 & K  & 0.14 & 1471 & 56 & 0.03 &  62 & 13.5 & 1.6 & speckle; (ref. 3) \\
2000.381 & K  & 0.14 & 2180 & 56 & 0.09 & 142 & 15.6 & 0.9 & speckle; (ref. 3) \\
2000.548 & K  & 0.14 & 1572 & 63 & 0.07 & 132 & 15.6 & 1.2 & speckle; (ref. 3) \\
2000.797 & K  & 0.14 & 1506 & 60 & 0.04 &  77 & 14.0 & 1.8 & speckle; (ref. 3) \\
2001.351 & K  & 0.14 & 1979 & 56 & 0.07 & 137 & 15.5 & 0.9 & speckle; (ref. 3) \\
2001.572 & K  & 0.14 & 1687 & 57 & 0.12 & 141 & 15.6 & 1.1 & speckle; (ref. 3) \\
2002.309 & K  & 0.14 & 1957 & 67 & 0.06 & 137 & 15.5 & 1.0 & speckle; (ref. 3) \\
2002.391 & K  & 0.14 & 1433 & 60 & 0.09 & 141 & 15.5 & 0.8 & speckle; (ref. 3) \\
2002.547 & K  & 0.14 & 1137 & 63 & 0.06 & 115 & 14.3 & 1.7 & speckle; (ref. 3) \\
2003.303 & K  & 0.14 & 1815 & 62 & 0.04 & 119 & 15.2 & 1.2 & speckle; (ref. 3) \\
2003.554 & K  & 0.14 & 1713 & 65 & 0.07 & 134 & 15.7 & 1.5 & speckle; (ref. 3) \\
2003.682 & K  & 0.14 & 1780 & 65 & 0.07 & 130 & 15.3 & 1.1 & speckle; (ref. 3) \\
2004.327 & K  & 0.14 & 1444 & 63 & 0.09 & 136 & 15.6 & 1.0 & speckle; (ref. 4) \\
2004.564 & K  & 0.14 & 2156 & 60 & 0.07 & 143 & 15.5 & 1.1 & speckle; (ref. 4) \\
2004.567 & K' &    9 &   12 & 60 & 0.31 & 145 & 15.8 & 1.0 & LGSAO; (ref. 5) \\
2004.660 & K  & 0.14 & 1300 & 59 & 0.08 & 114 & 15.2 & 1.3 & speckle; (ref. 4) \\
2005.312 & K  & 0.14 & 1677 & 60 & 0.07 & 132 & 15.3 & 1.0 & speckle; (ref. 6) \\
2005.495\tablenotemark{g} & K$_{CO}$, K$_{cont}$  & 36, 59.5 & 10 & 61 & 0.32 & 146 & 15.7 & 1.2 & LGSAO; (new) \\
2005.566 & K  & 0.14 & 1825 & 62 & 0.05 & 113 & 15.1 & 1.7 & speckle; (ref. 6) \\
\enddata
\label{tab:obs}

\tablenotetext{a}{Dates are computed as the weighted average of UT dates from
  the individual exposures.}
\tablenotetext{b}{Filters used include 
  K ($\lambda_{o}$=2.2 \micron, $\Delta\lambda$=0.4 \micron),
  K' ($\lambda_{o}$=2.12 \micron, $\Delta\lambda$=0.35 \micron),
  K$_{CO}$ ($\lambda_{o}$=2.289 \micron, $\Delta\lambda$=0.048 \micron), and 
  K$_{cont}$ ($\lambda_{o}$=2.270 \micron, $\Delta\lambda$=0.030 \micron).
}
\tablenotetext{c}{The number of frames used in the final combined image.}
\tablenotetext{d}{The turnover of the number of stars at a given
  magnitude provides a rough estimate of the completeness limit.}
\tablenotetext{e}{The average positional uncertainty due to centroiding
  in each epoch is estimated from a set of 25 stars detected in all 
  epochs and brighter than K$<$13. The two LGSAO epochs positional errors
  include and additional term of 0.88 mas to account for residual distortion.}
\tablenotetext{f}{Data originally reported in 
  (1) \citet{ghez98pm}
  (2) \citet{ghez00nat}
  (3) \citet{ghez05orbits}
  (4) \citet{lu05irs16sw}
  (5) \citet{ghez05lgs}
  (6) \citet{rafelski07}.
}
\tablenotetext{g}{Five exposures were taken in each of two narrow-band
filters with different exposure times, but similar sensitivity and 
astrometric precision. All frames from both filters were combined in order
to extract astrometric measurements from this data set.}
\end{deluxetable}

\begin{deluxetable}{lrrrrrrrrrrrl}
\rotate
\tabletypesize{\scriptsize}
\tablewidth{0pt}
\tablecaption{Proper Motions for Young Stars\label{tab:yng_pm_table}}
\tablehead{
  \colhead{Name} &
  \colhead{K} &
  \colhead{N$_{epochs}$} &
  \colhead{Epoch} &
  \colhead{Radius} &
  \colhead{$\Delta RA$ \tablenotemark{a}} &
  \colhead{$\Delta DEC$ \tablenotemark{a}} &
  \colhead{$v_{ra}$} &
  \colhead{$v_{dec}$} &
  \colhead{$v_z$ \tablenotemark{b}} &
  \colhead{$a_\rho$} &
  \colhead{$a_{tan}$} &
  \colhead{AltName \tablenotemark{b}} \\
  \colhead{} &
  \colhead{(mag)} &
  \colhead{} &
  \colhead{(year)} &
  \colhead{(arcsec)} &
  \colhead{(arcsec)} &
  \colhead{(arcsec)} &
  \colhead{(mas/yr)} &
  \colhead{(mas/yr)} &
  \colhead{(km/s)} &
  \colhead{(mas/yr$^2$)} &
  \colhead{(mas/yr$^2$)} &
  \colhead{}
}

\startdata
       S0-14 & 13.7 &  29 & 2001.290 &  0.82 &  -0.770 &  -0.270 &    1.62 $\pm$   0.06 &   -0.46 $\pm$   0.07 &   -14 $\pm$  40 &  0.05 $\pm$ 0.05 &  0.09 $\pm$ 0.05 & E14 \\
       S0-15 & 13.7 &  29 & 2001.680 &  0.97 &  -0.930 &   0.280 &   -5.32 $\pm$   0.07 &  -10.23 $\pm$   0.08 &  -424 $\pm$  70 & -0.21 $\pm$ 0.05 &  0.10 $\pm$ 0.06 & E16 \\
        S1-3 & 12.1 &  29 & 2001.980 &  0.98 &   0.440 &   0.879 &  -13.83 $\pm$   0.05 &    2.00 $\pm$   0.05 &    68 $\pm$  40 & -0.10 $\pm$ 0.03 & -0.08 $\pm$ 0.03 & E15 \\
        S1-2 & 14.9 &  26 & 2001.860 &  1.01 &  -0.025 &  -1.007 &   11.70 $\pm$   0.13 &   -0.65 $\pm$   0.13 &    26 $\pm$  30 &  0.01 $\pm$ 0.08 &  0.17 $\pm$ 0.08 & E17 \\
        S1-8 & 14.2 &  29 & 2001.680 &  1.08 &  -0.651 &  -0.865 &    7.64 $\pm$   0.10 &   -4.63 $\pm$   0.10 &  -364 $\pm$  40 &  0.04 $\pm$ 0.06 &  0.06 $\pm$ 0.06 & E18 \\
    IRS 16NW & 10.1 &  29 & 2001.560 &  1.22 &   0.029 &   1.221 &    6.30 $\pm$   0.06 &    0.87 $\pm$   0.06 &   -44 $\pm$  20 &  0.04 $\pm$ 0.04 & -0.05 $\pm$ 0.04 & E19 \\
     IRS 16C &  9.8 &  29 & 2001.570 &  1.23 &   1.121 &   0.497 &   -8.74 $\pm$   0.05 &    7.42 $\pm$   0.05 &   125 $\pm$  30 & -0.04 $\pm$ 0.03 & -0.11 $\pm$ 0.03 & E20 \\
       S1-12 & 13.8 &  28 & 2001.500 &  1.30 &  -0.837 &  -1.000 &    9.93 $\pm$   0.07 &   -1.88 $\pm$   0.07 &   -24 $\pm$  30 &  0.04 $\pm$ 0.06 & -0.00 $\pm$ 0.06 & E21 \\
       S1-14 & 12.8 &  29 & 2001.380 &  1.39 &  -1.355 &  -0.302 &    4.01 $\pm$   0.06 &   -6.79 $\pm$   0.07 &  -434 $\pm$  50 & -0.11 $\pm$ 0.05 & -0.02 $\pm$ 0.05 & E22 \\
    IRS 16SW & 10.0 &  29 & 2001.490 &  1.43 &   1.051 &  -0.966 &    6.80 $\pm$   0.05 &    2.22 $\pm$   0.06 &   320 $\pm$  40 & -0.08 $\pm$ 0.04 &  0.03 $\pm$ 0.04 & E23 \\
       S1-21 & 13.3 &  17 & 2001.190 &  1.68 &  -1.669 &   0.141 &    3.52 $\pm$   0.09 &   -3.84 $\pm$   0.09 &  -344 $\pm$  50 &  0.03 $\pm$ 0.07 &  0.02 $\pm$ 0.07 & E24 \\
       S1-22 & 12.7 &  29 & 2001.200 &  1.70 &  -1.631 &  -0.493 &    6.95 $\pm$   0.07 &   -1.70 $\pm$   0.08 &  -224 $\pm$  50 & -0.09 $\pm$ 0.06 & -0.08 $\pm$ 0.06 & E25 \\
       S1-24 & 11.6 &  29 & 2001.420 &  1.75 &   0.718 &  -1.591 &    1.13 $\pm$   0.07 &   -6.37 $\pm$   0.08 &   206 $\pm$  30 &  0.02 $\pm$ 0.05 &  0.08 $\pm$ 0.05 & E26 \\
        S2-4 & 12.3 &  29 & 2001.480 &  2.07 &   1.452 &  -1.476 &    6.69 $\pm$   0.08 &    2.51 $\pm$   0.08 &   286 $\pm$  20 & -0.03 $\pm$ 0.06 & -0.06 $\pm$ 0.06 & E28 \\
    IRS 16CC & 10.6 &  25 & 2000.840 &  2.07 &   1.999 &   0.550 &   -1.88 $\pm$   0.06 &    5.48 $\pm$   0.06 &   241 $\pm$  25 & -0.02 $\pm$ 0.04 & -0.09 $\pm$ 0.04 & E27 \\
        S2-6 & 12.1 &  29 & 2001.290 &  2.09 &   1.594 &  -1.345 &    6.80 $\pm$   0.05 &    1.66 $\pm$   0.06 &   216 $\pm$  20 & -0.02 $\pm$ 0.04 & -0.02 $\pm$ 0.04 & E30 \\
        S2-7 & 14.1 &  27 & 2002.350 &  2.09 &   0.979 &   1.849 &   -6.15 $\pm$   0.11 &    1.01 $\pm$   0.11 &   -94 $\pm$  50 &  0.04 $\pm$ 0.08 & -0.07 $\pm$ 0.08 & E29 \\
     IRS 29N & 10.3 &  29 & 2001.410 &  2.14 &  -1.595 &   1.423 &    5.26 $\pm$   0.08 &   -4.41 $\pm$   0.08 &  -190 $\pm$  90 & -0.02 $\pm$ 0.06 & -0.09 $\pm$ 0.06 & E31 \\
  IRS 16SW-E & 11.0 &  29 & 2001.430 &  2.17 &   1.846 &  -1.141 &    4.83 $\pm$   0.06 &    2.98 $\pm$   0.06 &   366 $\pm$  70 & -0.03 $\pm$ 0.04 & -0.03 $\pm$ 0.04 & E32 \\
     IRS 33N & 11.4 &  29 & 2001.630 &  2.19 &  -0.048 &  -2.189 &    1.72 $\pm$   0.12 &   -5.15 $\pm$   0.12 &    68 $\pm$  20 & -0.01 $\pm$ 0.08 &  0.18 $\pm$ 0.08 & E33 \\
       S2-17 & 10.9 &  29 & 2001.660 &  2.26 &   1.271 &  -1.871 &    7.51 $\pm$   0.09 &   -0.51 $\pm$   0.09 &   100 $\pm$  20 & -0.07 $\pm$ 0.07 &  0.04 $\pm$ 0.07 & E34 \\
       S2-16 & 11.9 &  29 & 2001.410 &  2.30 &  -0.992 &   2.073 &   -8.07 $\pm$   0.08 &   -0.29 $\pm$   0.09 &  -100 $\pm$  70 &  0.08 $\pm$ 0.06 & -0.01 $\pm$ 0.06 & E35 \\
       S2-19 & 12.6 &  28 & 2001.770 &  2.35 &   0.446 &   2.310 &   -7.30 $\pm$   0.09 &    0.55 $\pm$   0.09 &    41 $\pm$  20 &  0.02 $\pm$ 0.06 & -0.05 $\pm$ 0.06 & E36 \\
       S2-66 & 14.8 &  21 & 2003.490 &  2.62 &  -1.457 &   2.173 &    3.25 $\pm$   0.46 &   -1.57 $\pm$   0.46 &  -114 $\pm$  30 &  0.64 $\pm$ 0.22 & -0.18 $\pm$ 0.22 & E37 \\
       S2-74 & 13.3 &  24 & 2002.670 &  2.78 &   0.179 &   2.779 &   -7.63 $\pm$   0.17 &    1.14 $\pm$   0.17 &    36 $\pm$  20 & -0.01 $\pm$ 0.09 & -0.08 $\pm$ 0.09 & E38 \\
    IRS 16NE &  9.0 &  28 & 2000.990 &  3.06 &   2.868 &   1.053 &    3.11 $\pm$   0.06 &  -10.94 $\pm$   0.06 &   -10 $\pm$  20 & -0.06 $\pm$ 0.04 & -0.11 $\pm$ 0.04 & E39 \\
        S3-5 & 12.2 &  29 & 2001.030 &  3.17 &   2.938 &  -1.183 &    1.44 $\pm$   0.08 &    3.44 $\pm$   0.08 &   327 $\pm$ 100 & -0.01 $\pm$ 0.06 & -0.13 $\pm$ 0.06 & E40 \\
     IRS 33E & 10.6 &  16 & 2003.890 &  3.20 &   0.665 &  -3.126 &    5.38 $\pm$   0.49 &    0.04 $\pm$   0.50 &   170 $\pm$  20 &  0.02 $\pm$ 0.18 & -0.36 $\pm$ 0.18 & E41 \\
       S3-19 & 12.5 &  17 & 2003.700 &  3.21 &  -1.591 &  -2.785 &    6.40 $\pm$   0.49 &    1.65 $\pm$   0.50 &  -114 $\pm$  50 & -0.28 $\pm$ 0.19 & -0.15 $\pm$ 0.18 & E43 \\
       S3-25 & 14.1 &  18 & 2003.030 &  3.30 &   1.452 &   2.963 &   -5.86 $\pm$   0.35 &   -0.84 $\pm$   0.37 &  -114 $\pm$  40 & -0.12 $\pm$ 0.14 &  0.06 $\pm$ 0.14 & E44 \\
       S3-30 & 12.9 &  25 & 2003.120 &  3.40 &   1.668 &  -2.963 &   -3.14 $\pm$   0.27 &    5.07 $\pm$   0.28 &    91 $\pm$  30 & -0.21 $\pm$ 0.19 & -0.27 $\pm$ 0.18 & E47 \\
       S3-10 & 12.4 &  26 & 2001.780 &  3.54 &   3.345 &  -1.143 &   -1.78 $\pm$   0.12 &    4.11 $\pm$   0.13 &   281 $\pm$  20 & -0.14 $\pm$ 0.08 & -0.17 $\pm$ 0.09 & E50 \\
\enddata
\tablecomments{All uncertainties are 1$\sigma$ relative
errors and do not include errors in the plate scale, 
location of Sgr A*, or position angle.}
\tablenotetext{a}{Positions as determined from polynomial
fitting have relative errors of 
$\sim$0.4 mas.}
\tablenotetext{b}{Radial velocities and alternate names
obtained from Paumard et al. (2006).}
\end{deluxetable}

\begin{deluxetable}{lrrrrrrl}
\tablecaption{Orbital Eccentricity and Disk Membership}
\tabletypesize{\scriptsize}
\tablewidth{0pt}

\tablehead{
  \colhead{Name} &
  \colhead{SA$_{\vec{n}}$} &
  \colhead{1 - L(not on disk)} &
  \multicolumn{2}{c}{Eccentricity (All Solutions)} &
  \multicolumn{2}{c}{Eccentricity (Disk Solutions)} &
  \colhead{Direction} \\
  \colhead{} &
  \colhead{(steradians)} &
  \colhead{} &
  \colhead{Peak} &
  \colhead{3$\sigma$ Range} &
  \colhead{Peak} &
  \colhead{3$\sigma$ Range} &
  \colhead{} 
}

\startdata
\multicolumn{8}{c}{CANDIDATE DISK MEMBERS} \\
          S2-16 &  0.47 &     7.89e-01 & 0.60 & 0.00 $-$ 1.00 & 0.21 & 0.00 $-$ 1.00 &   CW \\ 
      irs16SW-E &  0.18 &     6.76e-01 & 0.37 & 0.00 $-$ 1.00 & 0.37 & 0.12 $-$ 1.00 &   CW \\ 
          S1-14 &  0.12 &     5.80e-01 & 0.33 & 0.00 $-$ 1.00 & 0.33 & 0.13 $-$ 1.00 &   CW \\ 
           S2-6 &  0.15 &     5.76e-01 & 0.79 & 0.17 $-$ 1.00 & 0.30 & 0.17 $-$ 0.60 &   CW \\ 
           S3-5 &  0.20 &     5.62e-01 & 0.64 & 0.00 $-$ 1.00 & 0.53 & 0.06 $-$ 1.00 &   CW \\ 
        irs16SW &  0.14 &     5.38e-01 & 0.78 & 0.04 $-$ 1.00 & 0.41 & 0.29 $-$ 0.91 &   CW \\ 
          S1-12 &  0.10 &     4.56e-01 & 0.41 & 0.00 $-$ 1.00 & 0.33 & 0.00 $-$ 0.61 &   CW \\ 
           S2-4 &  0.13 &     4.23e-01 & 0.69 & 0.10 $-$ 1.00 & 0.32 & 0.21 $-$ 0.94 &   CW \\ 
           S1-8 &  0.09 &     4.06e-01 & 0.62 & 0.37 $-$ 1.00 & 0.57 & 0.45 $-$ 1.00 &   CW \\ 
           S1-2 &  0.10 &     3.66e-01 & 0.26 & 0.00 $-$ 1.00 & 0.19 & 0.00 $-$ 0.80 &   CW \\ 
          S2-17 &  0.21 &     3.63e-01 & 0.77 & 0.00 $-$ 1.00 & 0.40 & 0.00 $-$ 0.56 &   CW \\ 
          S3-10 &  0.08 &     3.21e-01 & 0.16 & 0.00 $-$ 1.00 & 0.67 & 0.24 $-$ 0.81 &   CW \\ 
           S2-7 &  0.54 &     3.02e-01 & 0.76 & 0.00 $-$ 1.00 & 0.55 & 0.07 $-$ 0.67 &   CW \\ 
          S3-25 &  0.50 &     2.64e-01 & 0.76 & 0.00 $-$ 1.00 & 0.61 & 0.27 $-$ 1.00 &   CW \\ 
          S2-74 &  0.24 &     2.01e-01 & 0.45 & 0.00 $-$ 1.00 & 0.15 & 0.00 $-$ 1.00 &   CW \\ 
          S1-21 &  0.14 &     1.67e-01 & 0.92 & 0.00 $-$ 1.00 & 0.46 & 0.04 $-$ 0.79 &   CW \\ 
          S2-19 &  0.22 &     1.60e-01 & 0.57 & 0.00 $-$ 1.00 & 0.18 & 0.00 $-$ 0.46 &   CW \\ 
        irs16CC &  0.17 &     1.49e-01 & 0.62 & 0.34 $-$ 1.00 & 0.54 & 0.40 $-$ 0.66 &   CW \\ 
         irs33E &  0.31 &     1.42e-01 & 0.49 & 0.11 $-$ 1.00 & 0.50 & 0.27 $-$ 1.00 &   CW \\ 
           S1-3 &  0.06 &     1.32e-01 & 0.34 & 0.00 $-$ 1.00 & 0.09 & 0.00 $-$ 1.00 &   CW \\ 
          S1-22 &  0.27 &     1.02e-01 & 0.92 & 0.00 $-$ 1.00 & 0.68 & 0.34 $-$ 0.83 &   CW \\ 
          S0-14 &  0.13 &     5.35e-02 &  $-$ &         $-$ &  $-$ &         $-$ &   CW \\ 
 & & & & & & & \\
\multicolumn{8}{c}{STARS NOT IN THE DISK} \\
          S0-15 &  0.12 &     1.55e-03 & 0.30 & 0.00 $-$ 1.00 &  $-$ &         $-$ &   CW \\ 
         irs16C &  0.07 &     5.28e-04 & 0.50 & 0.00 $-$ 1.00 &  $-$ &         $-$ &   CW \\ 
         irs33N &  0.14 & $<$ 1.00e-05 & 0.97 & 0.00 $-$ 1.00 &  $-$ &         $-$ &   CW \\ 
          S1-24 &  0.06 & $<$ 1.00e-05 & 0.98 & 0.13 $-$ 1.00 &  $-$ &         $-$ &   CCW \\ 
          S3-19 &  0.57 & $<$ 1.00e-05 & 0.80 & 0.00 $-$ 1.00 &  $-$ &         $-$ &   CW \\ 
          S3-30 &  0.03 & $<$ 1.00e-05 & 0.99 & 0.00 $-$ 1.00 &  $-$ &         $-$ &   CCW \\ 
        irs16NE &  0.14 & $<$ 1.00e-05 & 0.19 & 0.04 $-$ 1.00 &  $-$ &         $-$ &   CCW \\ 
        irs16NW &  0.08 & $<$ 1.00e-05 & 0.70 & 0.00 $-$ 1.00 &  $-$ &         $-$ &   CCW \\ 
         irs29N &  0.01 & $<$ 1.00e-05 & 0.99 & 0.00 $-$ 1.00 &  $-$ &         $-$ &   CCW \\ 
          S2-66 &  0.53 & $<$ 1.00e-05 & 0.97 & 0.08 $-$ 1.00 &  $-$ &         $-$ &   CCW \\ 
\enddata
\tablenotetext{a}{No eccentricity is reported for S0-14
since the uniform-$a_\rho$ prior is not appropriate
for this very low velocity star.}
\label{tab:eccDisk}

\end{deluxetable}

\begin{deluxetable}{lrrrrrrl}
\tablecaption{Orbital Eccentricity and Disk Membership for Stars Added to 
the Extended Sample}
\tabletypesize{\scriptsize}
\tablewidth{0pt}

\tablehead{
  \colhead{Name} &
  \colhead{SA$_{\vec{n}}$} &
  \colhead{1 - L(not on disk)} &
  \multicolumn{2}{c}{Eccentricity (All Solutions)} &
  \multicolumn{2}{c}{Eccentricity (Disk Solutions)} &
  \colhead{Direction} \\
  \colhead{} &
  \colhead{(steradians)} &
  \colhead{} &
  \colhead{Peak} &
  \colhead{3$\sigma$ Range} &
  \colhead{Peak} &
  \colhead{3$\sigma$ Range} &
  \colhead{} 
}

\startdata
\multicolumn{8}{c}{CANDIDATE DISK MEMBERS} \\
        paumE57 &  0.41 &     8.57e-01 & 0.29 & 0.00 $-$ 1.00 & 0.34 & 0.00 $-$ 0.91 &   CW \\ 
         irs34W &  0.21 &     5.12e-01 & 0.20 & 0.00 $-$ 1.00 & 0.20 & 0.00 $-$ 1.00 &   CW \\ 
        paumE72 &  1.50 &     4.31e-01 & 0.81 & 0.00 $-$ 1.00 & 0.56 & 0.00 $-$ 1.00 &   CW \\ 
        paumE73 &  0.98 &     2.41e-01 & 0.02 & 0.00 $-$ 1.00 & 0.96 & 0.00 $-$ 1.00 &   CW \\ 
        irs34NW &  0.39 &     2.33e-01 & 0.04 & 0.00 $-$ 1.00 & 0.07 & 0.00 $-$ 1.00 &   CW \\ 
         AFNWNW &  1.52 &     2.06e-01 & 0.99 & 0.00 $-$ 1.00 & 0.96 & 0.00 $-$ 1.00 &   CW \\ 
        paumE69 &  0.51 &     1.34e-01 & 0.02 & 0.00 $-$ 1.00 & 0.80 & 0.12 $-$ 1.00 &   CW \\ 
         irs9SW &  0.60 &     1.10e-01 & 0.05 & 0.00 $-$ 1.00 & 0.35 & 0.00 $-$ 1.00 &   CW \\ 
        paumE54 &  0.36 &     1.03e-01 & 0.08 & 0.00 $-$ 1.00 & 0.17 & 0.00 $-$ 0.68 &   CW \\ 
          irs1E &  0.88 &     7.35e-02 & 0.77 & 0.00 $-$ 1.00 & 0.93 & 0.56 $-$ 1.00 &   CW \\ 
          irs9W &  0.57 &     6.29e-02 & 0.02 & 0.00 $-$ 1.00 & 0.67 & 0.17 $-$ 1.00 &   CW \\ 
        paumE87 &  0.65 &     5.09e-02 & 0.09 & 0.00 $-$ 1.00 & 0.94 & 0.52 $-$ 1.00 &   CW \\ 
        irs15SW &  0.35 &     2.64e-02 & 0.02 & 0.00 $-$ 1.00 & 0.94 & 0.31 $-$ 1.00 &   CW \\ 
             AF &  0.23 &     1.22e-02 & 0.11 & 0.00 $-$ 1.00 & 0.99 & 0.77 $-$ 1.00 &   CCW \\ 
          irs1W &  0.59 &     9.61e-03 & 0.18 & 0.00 $-$ 1.00 & 0.87 & 0.35 $-$ 1.00 &   CW \\ 
         irs7SW &  0.06 &     8.14e-03 & 0.99 & 0.00 $-$ 1.00 & 0.95 & 0.68 $-$ 1.00 &   CW \\ 
          S3-26 &  0.22 &     3.32e-03 & 1.00 & 0.00 $-$ 1.00 & 0.92 & 0.74 $-$ 1.00 &   CW \\ 
 & & & & & & & \\
\multicolumn{8}{c}{STARS NOT IN THE DISK} \\
           AFNW &  1.08 &     1.31e-03 & 0.95 & 0.00 $-$ 1.00 &  $-$ &         $-$ &   CW \\ 
        irs15NE &  1.07 &     8.92e-04 & 0.99 & 0.00 $-$ 1.00 &  $-$ &         $-$ &   CCW \\ 
        paumE78 &  0.61 &     4.06e-04 & 0.98 & 0.00 $-$ 1.00 &  $-$ &         $-$ &   CCW \\ 
         irs9SE &  0.89 &     7.12e-05 & 0.02 & 0.00 $-$ 1.00 &  $-$ &         $-$ &   CCW \\ 
        irs7E2? &  0.56 &     6.80e-05 & 0.99 & 0.00 $-$ 1.00 &  $-$ &         $-$ &   CW \\ 
        paumE84 &  0.22 &     4.09e-05 & 0.98 & 0.00 $-$ 1.00 &  $-$ &         $-$ &   CW \\ 
        paumE89 &  0.94 & $<$ 1.00e-05 & 0.02 & 0.00 $-$ 1.00 &  $-$ &         $-$ &   CCW \\ 
        irs13E1 &  0.50 & $<$ 1.00e-05 & 0.98 & 0.00 $-$ 1.00 &  $-$ &         $-$ &   CCW \\ 
        paumE86 &  0.32 & $<$ 1.00e-05 & 0.06 & 0.00 $-$ 1.00 &  $-$ &         $-$ &   CCW \\ 
        paumE82 &  0.79 & $<$ 1.00e-05 & 0.02 & 0.00 $-$ 1.00 &  $-$ &         $-$ &   CCW \\ 
        paumE75 &  0.42 & $<$ 1.00e-05 & 0.99 & 0.00 $-$ 1.00 &  $-$ &         $-$ &   CW \\ 
        paumE64 &  0.67 & $<$ 1.00e-05 & 0.62 & 0.00 $-$ 1.00 &  $-$ &         $-$ &   CW \\ 
        paumE62 &  0.58 & $<$ 1.00e-05 & 0.02 & 0.00 $-$ 1.00 &  $-$ &         $-$ &   CCW \\ 
        paumE60 &  0.21 & $<$ 1.00e-05 & 0.99 & 0.00 $-$ 1.00 &  $-$ &         $-$ &   CCW \\ 
        paumE55 &  0.61 & $<$ 1.00e-05 & 0.95 & 0.00 $-$ 1.00 &  $-$ &         $-$ &   CCW \\ 
        paumE53 &  0.49 & $<$ 1.00e-05 & 0.97 & 0.00 $-$ 1.00 &  $-$ &         $-$ &   CW \\ 
        paumE52 &  0.26 & $<$ 1.00e-05 & 0.09 & 0.00 $-$ 1.00 &  $-$ &         $-$ &   CCW \\ 
        paumE42 &  0.59 & $<$ 1.00e-05 & 0.04 & 0.00 $-$ 1.00 &  $-$ &         $-$ &   CCW \\ 
          irs7W &  0.33 & $<$ 1.00e-05 & 0.98 & 0.00 $-$ 1.00 &  $-$ &         $-$ &   CCW \\ 
         irs7SE &  0.81 & $<$ 1.00e-05 & 0.07 & 0.00 $-$ 1.00 &  $-$ &         $-$ &   CCW \\ 
    irs7E1(ESE) &  1.18 & $<$ 1.00e-05 & 0.05 & 0.00 $-$ 1.00 &  $-$ &         $-$ &   CCW \\ 
        irs13E4 &  0.60 & $<$ 1.00e-05 & 0.70 & 0.00 $-$ 1.00 &  $-$ &         $-$ &   CCW \\ 
       irs13E3b &  0.53 & $<$ 1.00e-05 & 0.64 & 0.00 $-$ 1.00 &  $-$ &         $-$ &   CCW \\ 
        irs13E2 &  0.59 & $<$ 1.00e-05 & 0.60 & 0.00 $-$ 1.00 &  $-$ &         $-$ &   CCW \\ 
\enddata
\label{tab:eccDiskExtended}

\end{deluxetable}


\begin{deluxetable}{rrr}
\tabletypesize{\scriptsize}
\tablewidth{0pt}
\tablecaption{NIRC Reimager Distortion Coefficients}

\tablehead{
  \colhead{i} &
  \colhead{X (a$_i$)} &
  \colhead{Y (b$_i$)}
}
\startdata
0 &  1.713$\times$10$^{-2}$ & -2.654$\times$10$^{-2}$ \\
1 &  9.957$\times$10$^{-1}$ & -1.759$\times$10$^{-3}$ \\
2 & -3.371$\times$10$^{-3}$ &  1.004
\enddata
\label{tab:distort}
\end{deluxetable}

\clearpage

%
%

\figsetstart
\figsetnum{7}
\figsettitle{Orbital Parameters}

\figsetend

\clearpage

\begin{thebibliography}{78}
\expandafter\ifx\csname natexlab\endcsname\relax\def\natexlab#1{#1}\fi

\bibitem[{{Alexander} {et~al.}(2008){Alexander}, {Armitage}, {Cuadra}, \&
  {Begelman}}]{alexander08}
{Alexander}, R.~D., {Armitage}, P.~J., {Cuadra}, J., \& {Begelman}, M.~C. 2008,
  \apj, 674, 927

\bibitem[{{Alexander} {et~al.}(2007){Alexander}, {Begelman}, \&
  {Armitage}}]{alexander07imf}
{Alexander}, R.~D., {Begelman}, M.~C., \& {Armitage}, P.~J. 2007, \apj, 654,
  907

\bibitem[{{Alexander}(2005)}]{alexander05review}
{Alexander}, T. 2005, \physrep, 419, 65

\bibitem[{{Alexander} \& {Morris}(2003)}]{alexander03}
{Alexander}, T., \& {Morris}, M. 2003, \apjl, 590, L25

\bibitem[{{Allen} {et~al.}(1990){Allen}, {Hyland}, \& {Hillier}}]{allen90}
{Allen}, D.~A., {Hyland}, A.~R., \& {Hillier}, D.~J. 1990, \mnras, 244, 706

\bibitem[{{Babu} \& {Feigelson}(1996)}]{astrostats}
{Babu}, G.~J., \& {Feigelson}, E.~D. 1996, {Astrostatistics} (Astrostatistics
  by G.J.~Babu and E.D.~Feigelson.~London: Chapman and Hall, 1996.)

\bibitem[{{Beloborodov} {et~al.}(2006){Beloborodov}, {Levin}, {Eisenhauer},
  {Genzel}, {Paumard}, {Gillessen}, \& {Ott}}]{beloborodov06}
{Beloborodov}, A.~M., {Levin}, Y., {Eisenhauer}, F., {Genzel}, R., {Paumard},
  T., {Gillessen}, S., \& {Ott}, T. 2006, \apj, 648, 405

\bibitem[{{Bender} {et~al.}(2005){Bender}, {Kormendy}, {Bower}, {Green},
  {Thomas}, {Danks}, {Gull}, {Hutchings}, {Joseph}, {Kaiser}, {Lauer},
  {Nelson}, {Richstone}, {Weistrop}, \& {Woodgate}}]{bender05}
{Bender}, R., {Kormendy}, J., {Bower}, G., {Green}, R., {Thomas}, J., {Danks},
  A.~C., {Gull}, T., {Hutchings}, J.~B., {Joseph}, C.~L., {Kaiser}, M.~E.,
  {Lauer}, T.~R., {Nelson}, C.~H., {Richstone}, D., {Weistrop}, D., \&
  {Woodgate}, B. 2005, \apj, 631, 280

\bibitem[{{Berukoff} \& {Hansen}(2006)}]{berukoff06}
{Berukoff}, S.~J., \& {Hansen}, B.~M.~S. 2006, \apj, 650, 901

\bibitem[{{Blum} {et~al.}(1995){Blum}, {Depoy}, \& {Sellgren}}]{blum95heI}
{Blum}, R.~D., {Depoy}, D.~L., \& {Sellgren}, K. 1995, \apj, 441, 603

\bibitem[{{Bonnell} \& {Rice}(2008)}]{bonnell08}
{Bonnell}, I.~A., \& {Rice}, W.~K.~M. 2008, Science, 321 (5892), 1060

\bibitem[{{Cuadra} {et~al.}(2008){Cuadra}, {Armitage}, \&
  {Alexander}}]{cuadra08}
{Cuadra}, J., {Armitage}, P.~J., \& {Alexander}, R.~D. 2008, \mnras, 388, L64

\bibitem[{{Davies} {et~al.}(1998){Davies}, {Blackwell}, {Bailey}, \&
  {Sigurdsson}}]{davies98}
{Davies}, M.~B., {Blackwell}, R., {Bailey}, V.~C., \& {Sigurdsson}, S. 1998,
  \mnras, 301, 745

\bibitem[{{Davies} \& {King}(2005)}]{davies05}
{Davies}, M.~B., \& {King}, A. 2005, \apjl, 624, L25

\bibitem[{{Diolaiti} {et~al.}(2000){Diolaiti}, {Bendinelli}, {Bonaccini},
  {Close}, {Currie}, \& {Parmeggiani}}]{starfinder}
{Diolaiti}, E., {Bendinelli}, O., {Bonaccini}, D., {Close}, L., {Currie}, D.,
  \& {Parmeggiani}, G. 2000, \aaps, 147, 335

\bibitem[{{Dressler}(1980)}]{nearestNeighbor}
{Dressler}, A. 1980, \apj, 236, 351

\bibitem[{{Eckart} \& {Genzel}(1996)}]{eckart96}
{Eckart}, A., \& {Genzel}, R. 1996, \nat, 383, 415

\bibitem[{{Eckart} {et~al.}(2002){Eckart}, {Genzel}, {Ott}, \&
  {Sch{\"o}del}}]{eckart02}
{Eckart}, A., {Genzel}, R., {Ott}, T., \& {Sch{\"o}del}, R. 2002, \mnras, 331,
  917

\bibitem[{{Eisenhauer} {et~al.}(2005){Eisenhauer}, {Genzel}, {Alexander},
  {Abuter}, {Paumard}, {Ott}, {Gilbert}, {Gillessen}, {Horrobin}, {Trippe},
  {Bonnet}, {Dumas}, {Hubin}, {Kaufer}, {Kissler-Patig}, {Monnet},
  {Str{\"o}bele}, {Szeifert}, {Eckart}, {Sch{\"o}del}, \&
  {Zucker}}]{eisenhauer06}
{Eisenhauer}, F., {Genzel}, R., {Alexander}, T., {Abuter}, R., {Paumard}, T.,
  {Ott}, T., {Gilbert}, A., {Gillessen}, S., {Horrobin}, M., {Trippe}, S.,
  {Bonnet}, H., {Dumas}, C., {Hubin}, N., {Kaufer}, A., {Kissler-Patig}, M.,
  {Monnet}, G., {Str{\"o}bele}, S., {Szeifert}, T., {Eckart}, A.,
  {Sch{\"o}del}, R., \& {Zucker}, S. 2005, \apj, 628, 246

\bibitem[{{Fischer} {et~al.}(1998){Fischer}, {Pryor}, {Murray}, {Mateo}, \&
  {Richtler}}]{fischer98}
{Fischer}, P., {Pryor}, C., {Murray}, S., {Mateo}, M., \& {Richtler}, T. 1998,
  \aj, 115, 592

\bibitem[{{Fruchter} \& {Hook}(2002)}]{drizzle}
{Fruchter}, A.~S., \& {Hook}, R.~N. 2002, \pasp, 114, 144

\bibitem[{{Genzel} {et~al.}(2000){Genzel}, {Pichon}, {Eckart}, {Gerhard}, \&
  {Ott}}]{genzel00}
{Genzel}, R., {Pichon}, C., {Eckart}, A., {Gerhard}, O.~E., \& {Ott}, T. 2000,
  \mnras, 317, 348

\bibitem[{{Genzel} {et~al.}(2003){Genzel}, {Sch{\" o}del}, {Ott}, {Eisenhauer},
  {Hofmann}, {Lehnert}, {Eckart}, {Alexander}, {Sternberg}, {Lenzen}, {Cl{\'
  e}net}, {Lacombe}, {Rouan}, {Renzini}, \& {Tacconi-Garman}}]{genzel03cusp}
{Genzel}, R., {Sch{\" o}del}, R., {Ott}, T., {Eisenhauer}, F., {Hofmann}, R.,
  {Lehnert}, M., {Eckart}, A., {Alexander}, T., {Sternberg}, A., {Lenzen}, R.,
  {Cl{\' e}net}, Y., {Lacombe}, F., {Rouan}, D., {Renzini}, A., \&
  {Tacconi-Garman}, L.~E. 2003, \apj, 594, 812

\bibitem[{{Genzel} {et~al.}(1996){Genzel}, {Thatte}, {Krabbe}, {Kroker}, \&
  {Tacconi-Garman}}]{genzel96}
{Genzel}, R., {Thatte}, N., {Krabbe}, A., {Kroker}, H., \& {Tacconi-Garman},
  L.~E. 1996, \apj, 472, 153

\bibitem[{{Gerhard}(2001)}]{gerhard01}
{Gerhard}, O. 2001, \apjl, 546, L39

\bibitem[{{Ghez} {et~al.}(2003){Ghez}, {Duch{\^ e}ne}, {Matthews}, {Hornstein},
  {Tanner}, {Larkin}, {Morris}, {Becklin}, {Salim}, {Kremenek}, {Thompson},
  {Soifer}, {Neugebauer}, \& {McLean}}]{ghez03spec}
{Ghez}, A.~M., {Duch{\^ e}ne}, G., {Matthews}, K., {Hornstein}, S.~D.,
  {Tanner}, A., {Larkin}, J., {Morris}, M., {Becklin}, E.~E., {Salim}, S.,
  {Kremenek}, T., {Thompson}, D., {Soifer}, B.~T., {Neugebauer}, G., \&
  {McLean}, I. 2003, \apjl, 586, L127

\bibitem[{{Ghez} {et~al.}(2005{\natexlab{a}}){Ghez}, {Hornstein}, {Lu},
  {Bouchez}, {Le Mignant}, {van Dam}, {Wizinowich}, {Matthews}, {Morris},
  {Becklin}, {Campbell}, {Chin}, {Hartman}, {Johansson}, {Lafon}, {Stomski}, \&
  {Summers}}]{ghez05lgs}
{Ghez}, A.~M., {Hornstein}, S.~D., {Lu}, J.~R., {Bouchez}, A., {Le Mignant},
  D., {van Dam}, M.~A., {Wizinowich}, P., {Matthews}, K., {Morris}, M.,
  {Becklin}, E.~E., {Campbell}, R.~D., {Chin}, J.~C.~Y., {Hartman}, S.~K.,
  {Johansson}, E.~M., {Lafon}, R.~E., {Stomski}, P.~J., \& {Summers}, D.~M.
  2005{\natexlab{a}}, \apj, 635, 1087

\bibitem[{{Ghez} {et~al.}(1998){Ghez}, {Klein}, {Morris}, \&
  {Becklin}}]{ghez98pm}
{Ghez}, A.~M., {Klein}, B.~L., {Morris}, M., \& {Becklin}, E.~E. 1998, \apj,
  509, 678

\bibitem[{{Ghez} {et~al.}(2000){Ghez}, {Morris}, {Becklin}, {Tanner}, \&
  {Kremenek}}]{ghez00nat}
{Ghez}, A.~M., {Morris}, M., {Becklin}, E.~E., {Tanner}, A., \& {Kremenek}, T.
  2000, \nat, 407, 349

\bibitem[{{Ghez} {et~al.}(2005{\natexlab{b}}){Ghez}, {Salim}, {Hornstein},
  {Tanner}, {Lu}, {Morris}, {Becklin}, \& {Duch{\^e}ne}}]{ghez05orbits}
{Ghez}, A.~M., {Salim}, S., {Hornstein}, S.~D., {Tanner}, A., {Lu}, J.~R.,
  {Morris}, M., {Becklin}, E.~E., \& {Duch{\^e}ne}, G. 2005{\natexlab{b}},
  \apj, 620, 744

\bibitem[{{Ghez} {et~al.}(2008){Ghez}, {Salim}, {Weinberg}, {Lu}, {Do}, {Dunn},
  {Matthews}, {Morris}, {Yelda}, {Becklin}, {Kremenek}, {Milosavljevic}, \&
  {Naiman}}]{ghez08}
{Ghez}, A.~M., {Salim}, S., {Weinberg}, N.~N., {Lu}, J.~R., {Do}, T., {Dunn},
  J.~K., {Matthews}, K., {Morris}, M., {Yelda}, S., {Becklin}, E.~E.,
  {Kremenek}, T., {Milosavljevic}, M., \& {Naiman}, J. 2008, \apj, submitted

\bibitem[{{Goodman}(2003)}]{goodman03}
{Goodman}, J. 2003, \mnras, 339, 937

\bibitem[{{G{\'o}rski} {et~al.}(2005){G{\'o}rski}, {Hivon}, {Banday},
  {Wandelt}, {Hansen}, {Reinecke}, \& {Bartelmann}}]{healpix}
{G{\'o}rski}, K.~M., {Hivon}, E., {Banday}, A.~J., {Wandelt}, B.~D., {Hansen},
  F.~K., {Reinecke}, M., \& {Bartelmann}, M. 2005, \apj, 622, 759

\bibitem[{{G{\"u}rkan} \& {Rasio}(2005)}]{gurkan05}
{G{\"u}rkan}, M.~A., \& {Rasio}, F.~A. 2005, \apj, 628, 236

\bibitem[{{Hansen} \& {Milosavljevi{\' c}}(2003)}]{hansen03}
{Hansen}, B.~M.~S., \& {Milosavljevi{\' c}}, M. 2003, \apjl, 593, L77

\bibitem[{{Hillenbrand} \& {Hartmann}(1998)}]{hillenbrand98}
{Hillenbrand}, L.~A., \& {Hartmann}, L.~W. 1998, \apj, 492, 540

\bibitem[{{Hopman} \& {Alexander}(2006)}]{hopman06}
{Hopman}, C., \& {Alexander}, T. 2006, \apj, 645, 1152

\bibitem[{{Hornstein}(2007)}]{sethThesis}
{Hornstein}, S.~D. 2007, PhD thesis, UCLA

\bibitem[{{Kim} {et~al.}(2004){Kim}, {Figer}, \& {Morris}}]{kim04}
{Kim}, S.~S., {Figer}, D.~F., \& {Morris}, M. 2004, \apjl, 607, L123

\bibitem[{{Kim} \& {Morris}(2003)}]{kim03}
{Kim}, S.~S., \& {Morris}, M. 2003, \apj, 597, 312

\bibitem[{{Kolykhalov} \& {Syunyaev}(1980)}]{kolykhalov80}
{Kolykhalov}, P.~I., \& {Syunyaev}, R.~A. 1980, Soviet Astronomy Letters, 6,
  357

\bibitem[{{Krabbe} {et~al.}(1991){Krabbe}, {Genzel}, {Drapatz}, \&
  {Rotaciuc}}]{krabbe91}
{Krabbe}, A., {Genzel}, R., {Drapatz}, S., \& {Rotaciuc}, V. 1991, \apjl, 382,
  L19

\bibitem[{{Krabbe} {et~al.}(1995){Krabbe}, {Genzel}, {Eckart}, {Najarro},
  {Lutz}, {Cameron}, {Kroker}, {Tacconi-Garman}, {Thatte}, {Weitzel},
  {Drapatz}, {Geballe}, {Sternberg}, \& {Kudritzki}}]{krabbe95}
{Krabbe}, A., {Genzel}, R., {Eckart}, A., {Najarro}, F., {Lutz}, D., {Cameron},
  M., {Kroker}, H., {Tacconi-Garman}, L.~E., {Thatte}, N., {Weitzel}, L.,
  {Drapatz}, S., {Geballe}, T., {Sternberg}, A., \& {Kudritzki}, R. 1995,
  \apjl, 447, L95

\bibitem[{{Lee}(1996)}]{lee96gcmergers}
{Lee}, H.~M. 1996, in IAU Symposium, Vol. 169, Unsolved Problems of the Milky
  Way, ed. L.~{Blitz} \& P.~J. {Teuben}, 215

\bibitem[{{Levin}(2007)}]{levin06}
{Levin}, Y. 2007, \mnras, 374, 515

\bibitem[{{Levin} \& {Beloborodov}(2003)}]{levin03}
{Levin}, Y., \& {Beloborodov}, A.~M. 2003, \apjl, 590, L33

\bibitem[{{Lin} \& {Pringle}(1987)}]{linPringle87}
{Lin}, D.~N.~C., \& {Pringle}, J.~E. 1987, \mnras, 225, 607

\bibitem[{{Lu} {et~al.}(2005){Lu}, {Ghez}, {Hornstein}, {Morris}, \&
  {Becklin}}]{lu05irs16sw}
{Lu}, J.~R., {Ghez}, A.~M., {Hornstein}, S.~D., {Morris}, M., \& {Becklin},
  E.~E. 2005, \apjl, 625, L51

\bibitem[{{Maillard} {et~al.}(2004){Maillard}, {Paumard}, {Stolovy}, \&
  {Rigaut}}]{maillard04irs13}
{Maillard}, J.~P., {Paumard}, T., {Stolovy}, S.~R., \& {Rigaut}, F. 2004, \aap,
  423, 155

\bibitem[{{Matthews} {et~al.}(1996){Matthews}, {Ghez}, {Weinberger}, \&
  {Neugebauer}}]{NIRCs}
{Matthews}, K., {Ghez}, A.~M., {Weinberger}, A.~J., \& {Neugebauer}, G. 1996,
  \pasp, 108, 615

\bibitem[{{Matthews} \& {Soifer}(1994)}]{NIRC}
{Matthews}, K., \& {Soifer}, B.~T. 1994, Experimental Astronomy, 3, 77

\bibitem[{{McMillan} \& {Portegies Zwart}(2003)}]{mcmillan03}
{McMillan}, S.~L.~W., \& {Portegies Zwart}, S.~F. 2003, \apj, 596, 314

\bibitem[{{Milosavljevi{\'c}} \& {Loeb}(2004)}]{milosav04}
{Milosavljevi{\'c}}, M., \& {Loeb}, A. 2004, \apjl, 604, L45

\bibitem[{{Morris}(1993)}]{morris93}
{Morris}, M. 1993, \apj, 408, 496

\bibitem[{{Morris} \& {Serabyn}(1996)}]{morris96}
{Morris}, M., \& {Serabyn}, E. 1996, \araa, 34, 645

\bibitem[{{Najarro} {et~al.}(1997){Najarro}, {Krabbe}, {Genzel}, {Lutz},
  {Kudritzki}, \& {Hillier}}]{najarro97}
{Najarro}, F., {Krabbe}, A., {Genzel}, R., {Lutz}, D., {Kudritzki}, R.~P., \&
  {Hillier}, D.~J. 1997, \aap, 325, 700

\bibitem[{{Nayakshin} \& {Cuadra}(2005)}]{nayakshinCuadra05}
{Nayakshin}, S., \& {Cuadra}, J. 2005, \aap, 437, 437

\bibitem[{{Nayakshin} {et~al.}(2007){Nayakshin}, {Cuadra}, \&
  {Springel}}]{nayakshin07sims}
{Nayakshin}, S., {Cuadra}, J., \& {Springel}, V. 2007, \mnras, 379, 21

\bibitem[{{Nayakshin} {et~al.}(2006){Nayakshin}, {Dehnen}, {Cuadra}, \&
  {Genzel}}]{nayakshin06thick}
{Nayakshin}, S., {Dehnen}, W., {Cuadra}, J., \& {Genzel}, R. 2006, \mnras, 366,
  1410

\bibitem[{{Nayakshin} \& {Sunyaev}(2005)}]{nayakshinSunyaev06}
{Nayakshin}, S., \& {Sunyaev}, R. 2005, \mnras, 364, L23

\bibitem[{{Ott}(2003)}]{ott03thesis}
{Ott}, T. 2003, PhD thesis, Max-Planck-Institut f{\" u}r Extraterrestrische
  Physik

\bibitem[{{Paumard} {et~al.}(2006){Paumard}, {Genzel}, {Martins}, {Nayakshin},
  {Beloborodov}, {Levin}, {Trippe}, {Eisenhauer}, {Ott}, {Gillessen}, {Abuter},
  {Cuadra}, {Alexander}, \& {Sternberg}}]{paumard06}
{Paumard}, T., {Genzel}, R., {Martins}, F., {Nayakshin}, S., {Beloborodov},
  A.~M., {Levin}, Y., {Trippe}, S., {Eisenhauer}, F., {Ott}, T., {Gillessen},
  S., {Abuter}, R., {Cuadra}, J., {Alexander}, T., \& {Sternberg}, A. 2006,
  \apj, 643, 1011

\bibitem[{{Portegies Zwart} {et~al.}(2003){Portegies Zwart}, {McMillan}, \&
  {Gerhard}}]{pzwart03irs16}
{Portegies Zwart}, S.~F., {McMillan}, S.~L.~W., \& {Gerhard}, O. 2003, \apj,
  593, 352

\bibitem[{{Rafelski} {et~al.}(2007){Rafelski}, {Ghez}, {Hornstein}, {Lu}, \&
  {Morris}}]{rafelski07}
{Rafelski}, M., {Ghez}, A.~M., {Hornstein}, S.~D., {Lu}, J.~R., \& {Morris}, M.
  2007, \apj, 659, 1241

\bibitem[{{Sanders}(1992)}]{sanders92}
{Sanders}, R.~H. 1992, \nat, 359, 131

\bibitem[{{Sanders}(1998)}]{sanders98}
---. 1998, \mnras, 294, 35

\bibitem[{{Sch{\" o}del} {et~al.}(2005){Sch{\" o}del}, {Eckart}, {Iserlohe},
  {Genzel}, \& {Ott}}]{schodel05}
{Sch{\" o}del}, R., {Eckart}, A., {Iserlohe}, C., {Genzel}, R., \& {Ott}, T.
  2005, \apjl, 625, L111

\bibitem[{{Sch{\" o}del} {et~al.}(2003){Sch{\" o}del}, {Ott}, {Genzel},
  {Eckart}, {Mouawad}, \& {Alexander}}]{schodel03}
{Sch{\" o}del}, R., {Ott}, T., {Genzel}, R., {Eckart}, A., {Mouawad}, N., \&
  {Alexander}, T. 2003, \apj, 596, 1015

\bibitem[{{Sch{\"o}del} {et~al.}(2007){Sch{\"o}del}, {Eckart}, {Alexander},
  {Merritt}, {Genzel}, {Sternberg}, {Meyer}, {Kul}, {Moultaka}, {Ott}, \&
  {Straubmeier}}]{schodel07}
{Sch{\"o}del}, R., {Eckart}, A., {Alexander}, T., {Merritt}, D., {Genzel}, R.,
  {Sternberg}, A., {Meyer}, L., {Kul}, F., {Moultaka}, J., {Ott}, T., \&
  {Straubmeier}, C. 2007, \aap, 469, 125

\bibitem[{{Sch{\"o}del} {et~al.}(2002){Sch{\"o}del}, {Ott}, {Genzel},
  {Hofmann}, {Lehnert}, {Eckart}, {Mouawad}, {Alexander}, {Reid}, {Lenzen},
  {Hartung}, {Lacombe}, {Rouan}, {Gendron}, {Rousset}, {Lagrange}, {Brandner},
  {Ageorges}, {Lidman}, {Moorwood}, {Spyromilio}, {Hubin}, \&
  {Menten}}]{schodel02}
{Sch{\"o}del}, R., {Ott}, T., {Genzel}, R., {Hofmann}, R., {Lehnert}, M.,
  {Eckart}, A., {Mouawad}, N., {Alexander}, T., {Reid}, M.~J., {Lenzen}, R.,
  {Hartung}, M., {Lacombe}, F., {Rouan}, D., {Gendron}, E., {Rousset}, G.,
  {Lagrange}, A.-M., {Brandner}, W., {Ageorges}, N., {Lidman}, C., {Moorwood},
  A.~F.~M., {Spyromilio}, J., {Hubin}, N., \& {Menten}, K.~M. 2002, \nat, 419,
  694

\bibitem[{{Scoville} {et~al.}(2003){Scoville}, {Stolovy}, {Rieke},
  {Christopher}, \& {Yusef-Zadeh}}]{scoville03}
{Scoville}, N.~Z., {Stolovy}, S.~R., {Rieke}, M., {Christopher}, M., \&
  {Yusef-Zadeh}, F. 2003, \apj, 594, 294

\bibitem[{{Shlosman} \& {Begelman}(1989)}]{shlosman89}
{Shlosman}, I., \& {Begelman}, M.~C. 1989, \apj, 341, 685

\bibitem[{{Stolte} {et~al.}(2006){Stolte}, {Brandner}, {Brandl}, \&
  {Zinnecker}}]{stolte06}
{Stolte}, A., {Brandner}, W., {Brandl}, B., \& {Zinnecker}, H. 2006, \aj, 132,
  253

\bibitem[{{Tamblyn} {et~al.}(1996){Tamblyn}, {Rieke}, {Hanson}, {Close},
  {McCarthy}, \& {Rieke}}]{tamblyn96}
{Tamblyn}, P., {Rieke}, G.~H., {Hanson}, M.~M., {Close}, L.~M., {McCarthy},
  D.~W., \& {Rieke}, M.~J. 1996, \apj, 456, 206

\bibitem[{{van Dam} {et~al.}(2006){van Dam}, {Bouchez}, {Le Mignant},
  {Johansson}, {Wizinowich}, {Campbell}, {Chin}, {Hartman}, {Lafon}, {Stomski},
  \& {Summers}}]{vanDam06}
{van Dam}, M.~A., {Bouchez}, A.~H., {Le Mignant}, D., {Johansson}, E.~M.,
  {Wizinowich}, P.~L., {Campbell}, R.~D., {Chin}, J.~C.~Y., {Hartman}, S.~K.,
  {Lafon}, R.~E., {Stomski}, Jr., P.~J., \& {Summers}, D.~M. 2006, \pasp, 118,
  310

\bibitem[{{Vollmer} \& {Duschl}(2001)}]{vollmerDuschl01}
{Vollmer}, B., \& {Duschl}, W.~J. 2001, \aap, 377, 1016

\bibitem[{{Wizinowich} {et~al.}(2006){Wizinowich}, {Le Mignant}, {Bouchez},
  {Campbell}, {Chin}, {Contos}, {van Dam}, {Hartman}, {Johansson}, {Lafon},
  {Lewis}, {Stomski}, {Summers}, {Brown}, {Danforth}, {Max}, \&
  {Pennington}}]{wizinowich06}
{Wizinowich}, P.~L., {Le Mignant}, D., {Bouchez}, A.~H., {Campbell}, R.~D.,
  {Chin}, J.~C.~Y., {Contos}, A.~R., {van Dam}, M.~A., {Hartman}, S.~K.,
  {Johansson}, E.~M., {Lafon}, R.~E., {Lewis}, H., {Stomski}, P.~J., {Summers},
  D.~M., {Brown}, C.~G., {Danforth}, P.~M., {Max}, C.~E., \& {Pennington},
  D.~M. 2006, \pasp, 118, 297

\bibitem[{{Yu} {et~al.}(2007){Yu}, {Lu}, \& {Lin}}]{yu07}
{Yu}, Q., {Lu}, Y., \& {Lin}, D.~N.~C. 2007, \apj, 666, 919

\end{thebibliography}
\end{document}